\documentclass[]{article}
\input epsf

\voffset= - 1.5in
\hoffset= - 1.0in         

\catcode`\@=11
\def\marginnote#1{}

\newcount\hour
\newcount\minute
\newtoks\amorpm
\hour=\time\divide\hour by60
\minute=\time{\multiply\hour by60 \global\advance\minute by-\hour}
\edef\standardtime{{\ifnum\hour<12 \global\amorpm={am}%
        \else\global\amorpm={pm}\advance\hour by-12 \fi
        \ifnum\hour=0 \hour=12 \fi
        \number\hour:\ifnum\minute<10 0\fi\number\minute\the\amorpm}}
\edef\militarytime{\number\hour:\ifnum\minute<10 0\fi\number\minute}

\def\draftlabel#1{{\@bsphack\if@filesw {\let\thepage\relax
   \xdef\@gtempa{\write\@auxout{\string
      \newlabel{#1}{{\@currentlabel}{\thepage}}}}}\@gtempa
   \if@nobreak \ifvmode\nobreak\fi\fi\fi\@esphack}
        \gdef\@eqnlabel{#1}}
\def\@eqnlabel{}
\def\@vacuum{}
\def\draftmarginnote#1{\marginpar{\raggedright\scriptsize\tt#1}}

\def\draft{\oddsidemargin -.5truein
        \def\@oddfoot{\sl preliminary draft \hfil
        \rm\thepage\hfil\sl\today\quad\militarytime}
        \let\@evenfoot\@oddfoot \overfullrule 3pt
        \let\label=\draftlabel
        \let\marginnote=\draftmarginnote
   \def\@eqnnum{(\theequation)\rlap{\kern\marginparsep\tt\@eqnlabel}%
\global\let\@eqnlabel\@vacuum}  }

\makeatletter
\newdimen\normalarrayskip              
\newdimen\minarrayskip                 
\normalarrayskip\baselineskip
\minarrayskip\jot
\newif\ifold             \oldtrue            \def\new{\oldfalse}
\def\arraymode{\ifold\relax\else\displaystyle\fi} 
\def\eqnumphantom{\phantom{(\theequation)}}     
\def\@arrayskip{\ifold\baselineskip\z@\lineskip\z@
     \else
     \baselineskip\minarrayskip\lineskip2\minarrayskip\fi}
\def\@arrayclassz{\ifcase \@lastchclass \@acolampacol \or
\@ampacol \or \or \or \@addamp \or
   \@acolampacol \or \@firstampfalse \@acol \fi
\edef\@preamble{\@preamble
  \ifcase \@chnum
     \hfil$\relax\arraymode\@sharp$\hfil
     \or $\relax\arraymode\@sharp$\hfil
     \or \hfil$\relax\arraymode\@sharp$\fi}}
\def\@array[#1]#2{\setbox\@arstrutbox=\hbox{\vrule
     height\arraystretch \ht\strutbox
     depth\arraystretch \dp\strutbox
     width\z@}\@mkpream{#2}\edef\@preamble{\halign
\noexpand\@halignto
\bgroup \tabskip\z@ \@arstrut \@preamble \tabskip\z@ \cr}%
\let\@startpbox\@@startpbox \let\@endpbox\@@endpbox
  \if #1t\vtop \else \if#1b\vbox \else \vcenter \fi\fi
  \bgroup \let\par\relax
  \let\@sharp##\let\protect\relax
  \@arrayskip\@preamble}
\def\eqnarray{\stepcounter{equation}%
              \let\@currentlabel=\theequation
              \global\@eqnswtrue
              \global\@eqcnt\z@
              \tabskip\@centering
              \let\\=\@eqncr
 \halign to \displaywidth\bgroup
    \eqnumphantom\@eqnsel\hskip\@centering
    $\displaystyle \tabskip\z@ {##}$%
    \global\@eqcnt\@ne \hskip 2\arraycolsep
         $\displaystyle\arraymode{##}$\hfil
    \global\@eqcnt\tw@ \hskip 2\arraycolsep
         $\displaystyle\tabskip\z@{##}$\hfil
         \tabskip\@centering
    &{##}\tabskip\z@\cr}
\begingroup\ifx\undefined\newsymbol \else\def\input#1 \endgroup\fi

\newfont{\hr}{msbm10}
\newfont{\ams}{msam10}
\font\teneufm=cmmib10
\font\seveneufm=cmmib7
\font\fiveeufm=cmmib5
\def\bfit#1{{\textfont1=\teneufm\scriptfont1=\seveneufm
\scriptscriptfont1=\fiveeufm
\mathchoice{\hbox{$\displaystyle#1$}}{\hbox{$\textstyle#1$}}
{\hbox{$\scriptstyle#1$}}{\hbox{$\scriptscriptstyle#1$}}}}

\font\numbers=cmss12
\font\upright=cmu10 scaled\magstep1
\def\stroke{\vrule height8pt width0.4pt depth-0.1pt}
\def\topfleck{\vrule height8pt width0.5pt depth-5.9pt}
\def\botfleck{\vrule height2pt width0.5pt depth0.1pt}
\def\Zmath{\vcenter{\hbox{\numbers\rlap{\rlap{Z}\kern 0.8pt\topfleck}\kern 2.2pt
                   \rlap Z\kern 6pt\botfleck\kern 1pt}}}
\def\Qmath{\vcenter{\hbox{\upright\rlap{\rlap{Q}\kern
                   3.8pt\stroke}\phantom{Q}}}}
\def\Nmath{\vcenter{\hbox{\upright\rlap{I}\kern 1.7pt N}}}
\def\Cmath{\vcenter{\hbox{\upright\rlap{\rlap{C}\kern
                   3.8pt\stroke}\phantom{C}}}}
\def\Rmath{\vcenter{\hbox{\upright\rlap{I}\kern 1.7pt R}}}
\def\Z{\ifmmode\Zmath\else$\Zmath$\fi}
\def\Q{\ifmmode\Qmath\else$\Qmath$\fi}
\def\N{\ifmmode\Nmath\else$\Nmath$\fi}
\def\C{\ifmmode\Cmath\else$\Cmath$\fi}
\def\R{\ifmmode\Rmath\else$\Rmath$\fi}

\def\d{\partial}

\def\bea{\begin{eqnarray}}
\def\eea{\end{eqnarray}}
\def\nn{\nonumber}
\def\beq{\begin{equation}}
\def\eeq{\end{equation}}
\def\ba{\beq\new\begin{array}{c}}
\def\ea{\end{array}\eeq}
\def\be{\ba}
\def\ee{\ea}
\def\stackreb#1#2{\mathrel{\mathop{#2}\limits_{#1}}}
\def\Tr{{\rm Tr}}
\def\res{{\rm res}}
\def\Bf#1{\mbox{\boldmath $#1$}}

\def\bphi{{\Bf\phi}}
\def\bPhi{{\Bf\Phi}}

\def\bsigma{{\bfit\sigma}}

\def\bfeta{{\Bf\eta}}
\def\Im{{\rm Im}}

\def\rank{{\rm rank}}

\def\2{{1\over 2}}
\def\N2{${\cal N}=2$}
\def\4N{${\cal N}=4$}
\def\1N{${\cal N}=1$}
\def\F{{\cal F}}

\renewcommand{\theequation}{\thesection.\arabic{equation}}

\begin{document}


\begin{flushright}
LPTENS-02/62\\
IHES/P/02/91\\
FIAN/TD-16/02\\
ITEP/TH-61/02\\
\end{flushright}
\vspace{2.0 cm}

\setcounter{footnote}0
\begin{center}
{\Large\bf
STRING THEORY OR FIELD THEORY?
\footnote{Based on contribution to {\em Physics Uspekhi},
{\bf 172}, 2002}}\\
\vspace{1.0 cm}
{\large A.Marshakov}\\
\vspace{0.6 cm}
{\em Laboratoire de Physique Th\'eorique de l' \'Ecole
Normale Sup\'erieure,\footnote{Unit\'e mixte de Recherche
8549 du
Centre National de la Recherche Scientifique et de l'Ecole Normale
Sup\'erieure}\\
24 rue Lhomond, Paris, France,\\
 Institut des Hautes \'Etudes Scientifiques,\\
Bures-sur-Yvette, France}\\
\vspace{0.2 cm}
and\\
\vspace{0.2 cm}
{\em
Theory Department, P.N.Lebedev Physics Institute,\\
Institute of Theoretical and Experimental Physics\\ Moscow, Russia
\footnote{Permanent address}}\\
\vspace{0.3 cm}
{e-mail:\ \ mars@lpi.ru,\ \ mars@gate.itep.ru,\ \ andrei@ihes.fr}
\end{center}
\begin{quotation}
\begin{small}

The status of string theory is reviewed, and major recent
developments - especially those in going beyond perturbation
theory in the string theory and quantum field theory frameworks -
are discussed. This analysis helps better understand the role and
place of string theory in the modern picture of the physical
world. Even though quantum field theory describes a wide range of
experimental phenomena, it is emphasized that there are some
insurmountable problems inherent in it - notably the
impossibility to formulate the quantum theory of gravity on its
basis - which prevent it from being a fundamental physical theory
of the world of microscopic distances. It is this task, the
creation of such a theory, which string theory, currently far
from completion, is expected to solve. In spite of its somewhat
vague current form, string theory has already led to a number of
serious results and greatly contributed to progress in the
understanding of quantum field theory. It is these developments
which are our concern in this review.
\end{small}
\end{quotation}

\newpage
\tableofcontents
\newpage
\setcounter{section}{0}
\setcounter{footnote}{0}

\section{Introduction}

The 20th century may be considered as a century of success
({\em uspekhi})
for physics.
Absolutely new physical ideas about the world which surrounds us
have greatly affected every human being and indeed the whole of
mankind, especially those people in power. This is shown by the wide
spread use of radio and television, man going into space,
and -- perhaps chiefly -- by explosions of atomic and hydrogen bombs. Thus,
originally found "with pen and paper" electrodynamics, the theories of relativity
and quantum mechanics have completely proved their worth.

Probing further into the "deep secrets of the world" in an attempt to
understand the very small -- subatomic and
subnuclear -- structure of our world, has not proved straightforward.
The absence of an
experimental base, or at the very least, big problems with experiments directed
to check any statement about energies more than~100~GeV, has led to the situation
where theoretical physics has relied more and more upon its "internal
beauty". In other words, it develops, in a fashion similar to mathematics,
mostly based on its own logic.
As a result of such developments, one had by the end of the 20th century a situation
quite rare for physics before. This search for "internal harmony" among
theoretical physicists distanced them quite far from the desires of experimentalists, at
least in the field of elementary particle physics. The so called
Standard Model (unifying theory of electromagnetic and weak interaction
based on the Weinberg-Salam model and chromodynamics) appears to be almost
completely satisfactory
\footnote{Precise checks of the predictions of the Standard Model have not
found any contradictions between the theory and experiment, coming out of
three standard deviations, what is quite satisfactory since, as L.B.Okun
reminded me, Landau and Fermi suggested always multiply the errors of
experiment by $\pi$. The latest data can be found in the report by
M.Gr\"unewald (Talk at LEP Physics Jamboree, CERN, July 10, 2001) available
at http://www.cern/ch/LEPEWWG.}
from the point of view of all known experiments. Already for about thirty years
theoreticians look for a "nice fundamental" theory, which reproduce the
Standard Model at large distances or energies of the order of W-boson mass
(roughly, the same 100~GeV). Despite obvious weaknesses of the arguments about
"beauty" as a foundation for theoretical physics, the majority
of interested people including myself can say that
the Standard Model is not satisfactory only from the point of view of this
principle. Moreover, already within the framework of the Standard Model a
few ideas
were used (spontaneous breaking of the gauge symmetry or the Higgs
effect), which are not yet confirmed by experiment but were rather chosen
among all possible options only due to their beauty and simplicity. In this
way, the Standard Model W-bosons become massive due to interaction with the
condensating scalar field, in complete analogy with the Landau-Ginzburg
mechanism in the condensed matter physics, though the
excitations of this scalar field have never been seen in nature.

Hence, in this review we will try to discuss the theory, which cannot be
verified by experimental particle physics. In this sense this hypothetical
theory is somehow more close to gravity than to elementary particle physics,
where after the appearance of General Relativity "internal beauty"
plays the role of the main physical principle. In the theory of gravity,
which is responsible mainly for the physics of the {\em macro}world,
the separation from
experiment (or, better to say, lack of experimental base for fixing the
parameters of the theory) has always allowed the possibility of using some extra
purely "internal" theoretical principles. It turns out, that such a situation
permeates also more and more into the physics of the {\em micro}world.

A natural requirement to such a hypothetical theory would be an explanation of
"everything" including gravity (which is definitely beyond the Standard
Model), i.e. the formulation of all four interactions -- electromagnetic,
strong, weak and gravitational -- starting from some unique principles.
This review contains an attempt to formulate these general principles and to
demonstrate that they could lead to some progress not only in understanding
of quantum theory of gravity, but also to some absolutely new perspective
on the well-known problems in gauge theories, being the base of the
Standard Model. It is certainly clear that there cannot be any "uniqueness
theorem" for such hypothetical fundamental physical theory and therefore
everything to be said below, especially without direct experimental
confirmation, can be considered as a "pure fantasy". We will try to show
nevertheless that it is this particular variant of such a "fantasy" which is
based on relatively simple and clear physical principles (though not always
clearly formulated), which become especially attractive when taking into account
that all alternative attempts to achieve any progress at least in
qualitative understanding of microworld physics, up to now have been totally
unsatisfactory.

Mostly for historical reasons the fundamental theory at small distances of
the Planck scale
$\sqrt{\gamma_{\rm N}\over\hbar c}\sim 10^{-33}$~cm ($\hbar$ and $c$ denote
the Planck constant and vacuum speed of light
\footnote{In what follows, if not specially noticed, these constants are
formally put equal to unity, i.e. in relativistic physics of microworld
velocities will be measured in units of speed of light $c$, while actions in
units of the Planck constant $\hbar$.}, while
$\gamma_{\rm N}$ is the Newton gravitational constant), where it is
necessary to take into account effects of quantum gravity or, stated alternatively,
gravity becomes comparable with the other interactions, is called
{\em String Theory}
\footnote{There exists already vast literature on string theory
including few books
by people who "founded" the string theory (see \cite{Pol,GSW,Polch})
lots of review, including the reviews in Physics Uspekhi
\cite{Kni,Mor} etc.
However this branch of science is still developing so rapidly that
re-understanding even of the basic notions and concepts happens quite often.
Say, relatively new reviews \cite{SchwarzStrR}-\cite{LosevMR} are quite
useful my point of view, though this is certainly the
list far from being complete.}.
This name can be considered not ideal and other suggestions for
different names show up from time to time (say,
M-theory
etc). In what follows we will use
the "traditional" name, since though being not complete or exact term, it
"catches" in the best way one of the main principles of this theory -- a
natural "geometric" regularization of small distances by introducing
the extended objects of non-zero length (typically of the Planck scale).
The appearance of
strings in the role of such extended objects immediately leads to the
theory containing massless gauge bosons and gravitons (whose consistence though has
yet to be proven).

Let us point out separately that the widely used (especially
in popular literature) word "superstrings", seems to be much more
unacceptable because, first, it literally corresponds only to the narrow class
of string models and, second, it mixes two absolutely different and
mutually independent physical ideas. It couples the concept of strings
proper with
the very different idea of supersymmetry (or symmetry between bosons and fermions).
As we see below, the role of such symmetry is especially important in
quantum field theory, where supersymmetry allows even to extend the horizon
of applications of the Standard Model. In contrast to typically
field-theoretical role of supersymmetry aimed to cancellation of
the ultraviolet divergencies, for string theory of main importance is
that fundamental theory at small distances
is {\em not} a local quantum field theory, and this
is already encoded in its name.

Let us specially stop at this point. On one hand, string theory does not
contradict to the existence of quantum field theory as a reasonable
effective theory at energies much less than Planckian ($10^{19}$~GeV),
which naturally describes the physical processes at weak coupling. Within
its range of validity, quantum field theory automatically takes
into account the contribution of anti-particles and proposes the values for
the amplitudes and cross-sections which are in rather nice agreement with
experiment. Moreover (and this will be discussed below in detail), when
studying the processes where the contribution of gravity is inessential
or at energies much less than Planckian energy,
string theory often reduces to
quantum field theory -- to the theory of gauge vector fields. It is
exactly in this sense the field theory is often called an
{\em effective} theory for strings at large distances. Roughly speaking,
field theory arises in the low-energy limit of string theory, similar to how
non-relativistic limit of the field theory gives rise to quantum mechanics,
which in its turn as $\hbar\to 0$ reduces to classical mechanics.

On the other hand, one should immediately notice that historically the step
towards the string theory from quantum field theory is nothing else but
change of the {\em paradigm}, and within the frames of new paradigm quantum
field theory can no longer pretend to the role of fundamental physical
theory. Below we are going to discuss this point applying rather simple
physical principles, which lead to an understanding that any attempts to
construct theory of quantum gravity in the framework of quantum field theory
are almost absurd.

However, here one should definitely and honestly point out that the situation
within string theory itself is far from being perfect. Pretending to be the
fundamental theory of microworld and unifying theory of all interactions,
string theory has not only been formulated in closed form, but even
does not have any well-studied "sample example", demonstrating more or less
all its basic ingredients, like in the case of simple models of quantum
mechanics (a harmonic oscillator or an atom of hydrogen) or quantum field theory
(say, scalar field theories with $\phi^3$- or $\phi^4$-potentials,
or quantum electrodynamics). In fact, at present
only some "pieces" of
string theory, rather chaotically placed among other "pieces", are
available to be investigated and partially formalized. Nevertheless, during
recent years some definite progress has been detected (and is still taken
place!)
in the area of string theory, which certainly distinguishes it among other,
practically dead-end directions.

The main purpose of this review is to discuss basic {\em physical}
principles forming the base of string theory and try to demonstrate their
attractive features, reviewing some (in particular recent) achievements in
this sphere. Notice immediately, that these achievements are not at all
obvious to everybody and do not explain (yet?!) observable physical
phenomena. It seems nevertheless to be very important that only in the
framework of string theory at least the possibility to {\em raise} several
new questions of principal importance arose. One of the most well-known of
them is the problem of the space-time dimension, supposed to be solved
{\em dynamically} instead of usual fixing of the dimension "by hand".
This approach is totally new in comparison with traditional point of view
accepted in quantum field theory, where space-time belongs to a few initial
basic ingredients.

The dynamical nature of space-time is a direct consequence of
{\em definition} of string theory already at perturbative level by the
Polyakov path integral where the sum over all physically different
configurations is represented by the sum over all geometries
on two-dimensional string world-sheets. The arising "geometrization" of
string theory already at the perturbative level also plays an
essential role in the
attempts to go beyond the perturbation theory. Recent most striking
achievements are indeed related to the ideas to identify parameters of
physical theory (masses, condensates, coupling constants) with the
parameters or moduli of certain (complex) manifolds arising as a "compact
parts" of the full space-time, dynamically chosen by string theory.

To finish this introduction let us also point out that the specific
situation around string theory, quite untypical for physics, also leads to
a large amount of "social" problems, which are rather interesting in themselves
but their discussion goes beyond the scope of this review. For example,
string theory very often (and at least from my point of view
very unfair) is claimed to be "pure mathematics" in contrast to many other,
more traditional spheres of activity in theoretical physics considered
to be "physics by definition". In particular, many physicists got used to
the more traditional paradigm of quantum field theory and call
{\em all} problems of string theory "mathematics" only because they arose in this
particular context, while any technical problem of the formalism of quantum
field theory is considered as "physics".

It is certainly true that string
theory as any other interesting sphere in theoretical physics raised lots of
new mathematical problems and requires the application of branches of mathematics
previously not widely used in physics, moreover certain problems of string
theory are playing the role of "locomotive" for some directions of research
in mathematics. However, it seems to be completely wrong to stress this
particular aspect of the new theory and in what follows we will try to
discuss mostly simple and natural {\em physical} aspects of string theory.

Another social effect which is quite often (and again unfairy)
associated only with string theory is the widely spread invasion of "marketing"
principles into the modern science. Caused by purely social problems,
continuous advertisement of the string theory as a theory which has
{\em already} solved all possible problems of natural science (especially on
the background of absence of any strict arguments supporting this point of
view) does great harm to anybody willing to understand seriously this
interesting direction in modern physics. Together with the lack of relations
with experiment, existing for more traditional spheres of theoretical
physics, the wide advertising of string theory brought only negative
attention to this field of science especially among quite conservative
physicists. However, it is also necessary to stress that the development of
string theory in present conditions would be simply impossible without
bright and striking new ideas (see, for example, \cite{Pol}),
which only partially, and mostly many years after they had been
pronounced, were turned into the frames of more or less strict
formulations.
It is rather natural to get a lot of "garbage" along this way and one of
the main
difficulties is the opportunity to be killed by huge stream of various
literature which often does not contain any useful information. Without
pretending to objectivity, especially in such a delicate question, I
certainly understand that the reference list to this review contains only
very restricted fraction of existing literature, and the choice of these
particular references was mostly determined by (sometimes accidental)
my personal knowledge.

\bigskip\noindent
{\bf Content of the review}. We start in sect.~\ref{ss:gaugegrav}
with discussion of the Standard Model of gauge interactions
(electromagnetic, weak and strong) of elementary particles and (classical)
theory of gravity -- General Relativity. The main aim of this discussion is
to fix once more the status of quantum field theory as absolutely
satisfactory and verified experimentally model of observable interactions of
elementary particles, which however runs into serious difficulties in the
strong coupling regime and, mainly, which is absolutely useless as a
theory of quantum gravity.

In sect.~\ref{ss:string}, we will try to
formulate the main principles of string theory, coming mostly from geometric
formulation of string perturbation theory in terms of the Polyakov path
integral. The main message of this section is that it is two-dimensional
geometry -- the basic point of the Polyakov formulation -- which is
responsible for new string approach to the dynamical nature of
space-time and here is principle difference between string theory
and standard quantum field theory. We will also discuss supersymmetry as an
origin for appearance of the fermions and the Fradkin-Tseytlin effective
actions, being the most convenient "bridge" between string theory and
effective quantum field theories.

Sect.~\ref{ss:nonperturb} is devoted to recent attempts in string theory to
go beyond the perturbative regime. The main purpose of this section is to
explain the basic ideas of these attempts: the idea of duality between
the theories at strong and weak coupling and the classical extended
objects appearing necessarily in non-perturbative string theory. As an
illustration of the progress in studying the non-perturbative effects being
an outcome of applying new stringy methods, we will discuss the
Seiberg-Witten theory which allows, in particular, to make a new step in
understanding of the mechanism of confinement.

Sect.~\ref{ss:holography} is totally devoted to one of the most interesting
new problems in string theory -- an attempt of dual description of the
non-Abelian gauge theories at strong coupling in terms of gravity (or theory
of closed strings). Finally, in sect.~\ref{ss:new}, we review a few other
modern directions coming out of string theory, this section being written for
the most advanced reader (the same is true for the sect.~\ref{ss:is}).
Paragraphs of the text containing technical issues and therefore being more
difficult for understanding, are typed with a smaller font.

\section{Physics of Elementary Particles. Gauge Theories and Gravity
\label{ss:gaugegrav}}

There have been no essential changes in elementary particle theory during
the last decades. Still two main problems are at the center of interest, these
are confinement (or keeping of quarks locked inside the hadrons) and the quantum
theory of gravity
\footnote{More strictly these are the problems of elementary particle
physics "in a wide sense". From a more "narrow" point of view one may in
principle doubt in existence of the problem of quantum gravity.}, while
all the rest can be almost completely explained in the framework of
the Standard
Model. Mostly probable, the solution of these two problems is impossible
without progress in understanding of the properties of gauge theory and
general relativity at strong coupling, i.e. exactly where the standard
field-theoretical methods being the basic ones for the Weinberg-Salam model
of electroweak interactions and quantum chromodynamics (QCD) at high
energies become useless.

The Standard Model in its main features can be considered as a non-Abelian gauge
theory with the gauge group $SU(2)\times U(1)\times SU(3)$ (the last
factor corresponds to the "color" or strong interaction) and matter fields of
"three generations" \cite{Okun} (see also, e.g.  \cite{Andreev}). The
computations are performed using the technique of the gauge field theory
\cite{SlavFad} at small coupling constants -- i.e. by perturbation theory,
and the results of such computations are nicely consistent with experiment
(see footnote 1
\footnote{Apart from neutrino oscillations (see, for example,
\cite{NeOsc}).}).
From pure theoretical or kind of aesthetic point of view
the Standard Model is a little bit "ugly" due to presence of "external"
parameters -- such as the Weinberg angle, as well as due to absence of
completeness in some questions like spontaneous symmetry breaking or the
Higgs effect, which is responsible for masses of non-Abelian
$W$- and $Z$-bosons. Nevertheless, the Standard Model is an absolutely
consistent quantum field theory. It is a renormalizable quantum field theory,
which was already marked by the corresponding Nobel Prize in physics
\cite{tHoVeufn}.

If speaking about gravity, its "observable part" is still negligible in the
sense of the possible influence or this or that choice of the theory of
quantum gravity. At least to my knowledge by now there is no direct
experimental evidence of the existence of gravitons as well as any clear and
unambiguous data concerning the problems of dark matter and cosmological
constant (see, for example, \cite{Rubufn}). All experts agree only that dark
matter seems to exist and the cosmological constant looks like being nonvanishing.
Despite of growing precision of experimental methods in
astrophysics, the existing data are too scarce in order to put at least some
framework onto the set of existing theoretical models. Moreover, the very idea
of applicability of present physical theories to the model of Universe as a
whole seems to be rather "voluntaristic", while the attempts to formulate the
model of Universe in terms of microworld physics, i.e. in the language of
quantum mechanics or quantum field theory do not have any real physical
background and can be considered almost absurd. Thus, when discussing
the problems of quantum gravity one has to use only pure theoretical and
aesthetic criteria.

\subsection{Gauge Field Theories
\label{ss:gauge}}

Gauge theories or theories of massless vector fields describe all
interactions except for gravity. The theory of gauge fields or the
Yang-Mills fields can be formulated without even using stringy principles
and can be considered as a closed physical theory within some range of
energies. Nevertheless the viewpoint onto the theory of gauge fields
as being "derivative" from string theory leads to its much deeper understanding
and already brought us to new interesting results.

The progress achieved in gauge theories, especially in
their supersymetric versions has allowed many people to say that
gauge theory, or even any quantum field theory can be treated
beyond the level of perturbative expansion. However, and we want
to stress this point, when such words are pronounced it is usually
implied that something {\em extra} should be added to the standard
definition of the quantun field theory, base on particular well-known
manipulations with field-theoretical Lagrangian. In other words, this
implies some {\em new} definition of a quantum field theory (rather
different from a standard one) which is even more close to what we
call here by string theory. Despite this seems to be only a terminological
difference, using of the old expression "field theory" is not quite
adequate in this case, because the "new definition" of field theory in
practice leads to change of the paradigm, since the new effects cannot be
obtained on the level of formal manipulations with the Lagrangian.

In gauge theories matter interacts due to exchange by massless vector
fields. In case of electrodynamics or Abelian theory the gauge group is
$U(1)$, i.e. it contains the only vector field is
$A_{\mu}(x)$ (photons), if the theory is
non-Abelian (or equivalently the Yang-Mills gauge theory \cite{SlavFad}) the
fields can be conveniently represented by matrices from the Lie algebra of
the corresponding gauge group ${\bf A}_{\mu}(x)\equiv
\|A_{\mu}^{ij}\| $ (gluons), in the $SU(N)$ case, for example, by the
$N\times N$ (anti)Hermitean traceless matrices. The minimal interaction is
introduced by the "long" derivative
\be
\label{longder}
\d _{\mu} \rightarrow {\bf D}_{\mu} = \d _{\mu} +  {\bf A}_{\mu}
\ee
or $D_{\mu}^{ij} = \d _{\mu}\delta ^{ij} +  A_{\mu}^{ij}$, if the gauge
field interacts with matter from the representation of the gauge group whose
elements are labeled by index $i$. The gauge-invariant Lagrangian of the
Yang-Mills fields has the form
\be\label{langym}
{\cal L}_{\rm YM} = {1\over 2g^2}\Tr {\bf F}_{\mu\nu}^2
\ee
where
\be\label{curv}
{\bf F}_{\mu\nu}  = \d_{\mu}{\bf A}_{\nu} -
\d_{\nu}{\bf A}_{\mu} + [{\bf A}_{\mu},{\bf A}_{\nu}]
\ee
In the case of electrodynamics matrix-valued fields turn into numbers and
therefore the formula (\ref{curv}) does not contain commutators (leading to the
self-interaction in (\ref{langym})) and one may not write the trace $\Tr $
over the matrices.

For the Standard Model the gauge group is
$SU(3)\times SU(2)\times U(1)$ and one should add to Lagrangian (\ref{langym})
the Lagrangian of matter fields (electrons, quarks, etc) with the "long"
derivative (\ref{longder}). After that one can perform the standard
field-theoretical computations developing the perturbation theory in
coupling constant $g$. Such a theory will no more be fundamental
at the
level of field-theoretical perturbation theories, since it contains the
Abelian factor $U(1)$ with coupling constant growing at small distances,
while the theory with "controlled" behavior at small distances "should be"
non-Abelian. In what follows we will restrict ourselves to the compact (for
the integrality of charges!) non-Abelian $SU(N)$ groups, considering all
other gauge groups as "pure exotic".

\begin{figure}[tb]
\epsfysize=1.5cm
\centerline{\epsfbox{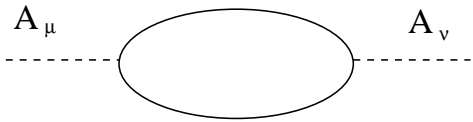}}
\caption{\sl One-loop diagram arising in the calculation of corrections
to the effective charge. In electrodynamics, as follows from Ward
identities, the computation may be restricted to only this diagram}
\label{fi:loop}
\end{figure}

The reason for the "non fundamental" nature of the Abelian theories is famous
"zero-charge" or "Moscow zero" in electrodynamics. In quantum field theory
parlance this means the growth of charge at small distances. The
physical origin of such behaivior comes from the screening of charge by
virtual electron-positron pairs, while the gauge $U(1)$ fields themselves
are not charged. Technically this means that one-loop corrections
(the simplest diagram for the computation of this effect,
say, in electrodynamics is depicted in fig.~\ref{fi:loop})
lead to the following dependence of effective charge on the energy scale
$\mu$
\be
\label{rg}
{d g\over d\log\mu} \equiv \beta(g) = b_0g^3 + \dots
\ee
where the coefficient
\be
\label{beta}
b_0 \propto N_F - N_V
\ee
is the {\em difference} of the contributions $N_F$ of matter fields
and $N_V$ of the gauge fields themselves,
propagating along the loop at the diagram in fig.~\ref{fi:loop}. In
electrodynamics the self-interaction of photons is absent, hence $N_V=0$,
and the coefficient in formula (\ref{beta}) is positive. This means the
growth of charge with $\mu$, or approaching small distances and as
a consequence electrodynamics at small distances is not well-defined,
i.e. cannot be a fundamental theory. Simultaneously electrodynamics
continues to be nice effective theory at {\em large} distances, where
$g_{QED}\equiv e$ is
small~\footnote{This is true in the elementary particle physics, but not in
condensed matter theory, where instead of
${e^2\over mc} \sim {1\over 137}$ the parameter of
perturbative expansion is ${e^2\over mv_{F}} \sim 1$.}.

\begin{figure}[tb]
\epsfysize=7.5cm
\centerline{\epsfbox{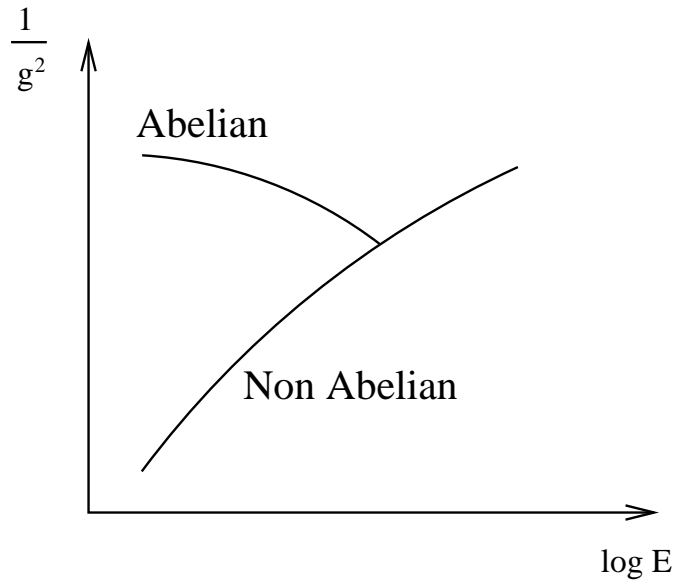}}
\caption{\sl Dependence of effective couplings upon energy in Abelian
and non-Abelian gauge theories. The top curve corresponds to
"zero charge" while the bottom curve corresponds to asymptotic freedom.}
\label{fi:renorm}
\end{figure}

The situation changes drastically for the case of
{\em non} Abelian gauge theories where extra {\em anti} screening of
charges by {\em charged} (in color) gauged fields exists so that
$N_V\neq 0$ due to self-interaction of gluons. This leads to the possibility
of "asymptotic freedom" \cite{GVP}, when interaction becomes weak at small
distances for $N_F<N_V$. The difference is demonstrated in
fig.~\ref{fi:renorm}, where the difference between zero-charge and
asymptotically free theories can be clearly seen.

A natural way out from such situation is to consider electrodynamics as a
"part" of some non-Abelian theory from which is "splits" at some scale where
non-Abelian symmetry is violated. In such a case the non-Abelian gauge theory
(especially in supersymmetric case) can be considered as "fundamental", at
least in some energy range where the effects of gravity have not yet
become necessarily taken into account. From this perspective
{\em renormalizability} of gauge theories has a quite simple meaning -- the
Lagrangian (\ref{langym}) is useful for the description of physics in rather
large energy range if for the coupling $g$ one would substitute its
corresponding effective value at given energy. With this substitution the general form
of the Lagrangian remains intact (and it does not require additional terms
when passing from one energy to another).

\subsection{Spontaneous Breaking of Gauge Symmetry}

Let us now discuss how at some energy scale the gauge group can (partially)
turn into Abelian. In the most natural way it happens if the theory contains
the scalar fields in the adjoint representation of the gauge group, for
example, as a consequence of supersymmetry. Suppose the scalar potential
has minima such that condensates or vacuum expectation values of scalars do
not vanish. For the field in adjoint representation of the gauge group
$SU(N)$ this means that vacuum values $\bphi$ may be chosen in diagonal form
\be\label{vacmatr}
\bphi = \left.\left(
\begin{array}{cccc}
\phi_1 &     &  &  \\
    & \phi_2 &  &  \\
    &     &\dots & \\
    &     &      & \phi_{N}
\end{array}\right)\right|_{\Tr\bphi = \sum \phi_j = 0}
\ee
using gauge invariance. For the convenient choice of gauge-invariant
quantities one may take parameters like
$\Tr\bphi^k$ or their "generating functions"
\be\label{polyn}
P_{N}(\lambda ) =\det (\lambda - \bphi) =
\prod_{i=1}^N(\lambda - \phi_i)
\ee
The total number of algebraically independent parameters
$\{\phi_i\}$ is equal to the rank of the group, in the mostly
well-known case this is $\rank [SU(N)] = N -1$. It is customary
to say that these
parameters are co-ordinates in the parameter space or
moduli space of gauge theory.
Due to the Higgs effect the off-diagonal part of the matrix of gauge field
${\bf A}_{\mu}$ for $\bphi\neq 0$ becomes massive, since the interaction
\be
[\bphi , {\bf A}_{\mu}]_{ij} = (\phi_i-\phi_j){\bf A}_{\mu}^{ij}
\ee
literally turns into the mass terms
\be\label{Wmass}
\sum (\phi_i-\phi_j)^2\left({\bf A}_{\mu}^{ij}\right)^2 =
\sum (m_{\rm W}^{ij})^2\left({\bf A}_{\mu}^{ij}\right)^2
\ee
in the Lagrangian. At the same time the diagonal part, as follows from
(\ref{Wmass}), remains massless, i.e. the gauge group $G = SU(N)$ is
broken by Higgs mechanism to $U(1)^{\rank G} = U(1) ^{N -1}$
\footnote{In the situation of "general position", i.e. when
$\phi_i\neq\phi_j$ for $i\neq j$. If the eigenvalues (\ref{vacmatr})
partially coincide, the broken group still contains non-Abelian factor
$SU(K)$ with $K<N$.}.

Thus, in the generic situation at the scale $\bphi$
(the scalar field has a dimension of mass) non-Abelian gauge group is broken
down to Abelian which in the simplest $SU(2)$ case is exactly that of
electrodynamics. In what follows, even in general
$U(1)^{N -1}$ case we would call such an Abelian theory (generalized)
electrodynamics and refer to the corresponding charges as electric
charges.

\subsection{Nonperturbative Effects: Instantons and Monopoles
\label{ss:instmon}}

In contrast to electrodynamics the non-Abelian gauge theories are
essentially nonlinear since the Lagrangian (\ref{langym}) contains cubic and
quartic terms in the Yang-Mills fields. It means that equations of motion
are nonlinear even without the matter fields. Nonlinear equations typically
do have lots of nontrivial solutions, related in the case of non-Abelian
gauge theories to nontrivial topological properties of the gauge groups.

Do these solutions affect elementary particle physics? The exact answer
to this question is still only hypothetical, but from general arguments it
is clear that the influence can be essential in the strong
coupling regime. Indeed, from general properties of quantum theory we know that the
main contribution of a classical trajectory to quantum amplitude (the
Feynman path integral) is nothing but $\exp (-S/\hbar)$, where $S$ is
the classical action on given configuration. For the theory of non-Abelian
gauge fields the corresponding action, or
Lagrangian (\ref{langym}),  integrated over space-time,
will give rise to the contributions of the form
$\exp \left(- {{\rm const}\over g^2}\right)$, which are exponentially
suppressed at weak coupling. However, by the same logic it is quite
possible, that the same contribution would be much more essential at strong
coupling, i.e. exactly there, where the main and unclear yet phenomena are
"hidden". Hence, the classical solutions look like being very important for
studying the strong-coupled phase.

At present among all classical solutions in non-Abelian gauge theories the
most essential role belongs to {\em instantons} or pseudoparticles
\cite{Pol75,BPST,VZNSUFN}. By instanton one usually means the
configuration of fields "localized" in
four-dimensional Euclidean space, which satisfies
the (anti) self-duality equations
\be
\label{selfdual}
{\bf F} = \pm^*{\bf F} \ \ \ \ \ \ \ \ \ {\rm или}
\nn \\
{\bf F}_{\mu\nu} = \pm\2
\epsilon_{\mu\nu\lambda\rho}{\bf F}_{\lambda\rho}
\equiv \pm\widetilde{\bf F}_{\mu\nu}
\ee
($\mu,\nu=1,\dots,4$).
Any solution to the self-duality equations (\ref{selfdual}) is automatically
a solution to the Yang-Mills equations of motion
${\bf D}_{\mu}{\bf F}_{\mu\nu} = 0$ (the opposite is incorrect!) due to the
Bianchi {\em identities}
${\bf D}_{\mu}\widetilde{\bf F}_{\mu\nu} =
\epsilon_{\mu\nu\lambda\rho}{\bf D}_{\nu}{\bf F}_{\lambda\rho} \equiv 0$
(i.e. relations, true for any fields). For the instantons
\be
\label{instact}
S = {1\over 2g^2}\int_{d^4x}\Tr {\bf F}_{\mu\nu}^2 =
{1\over 2g^2}\int_{d^4x}\Tr {\bf F}_{\mu\nu}\widetilde{\bf F}_{\mu\nu} =
{8\pi^2 n\over g^2}
\ee
where $n$ is the topological charge, counting how many times the
three-dimensional sphere of large radius in four-dimensional space-time
"winds" around the compact gauge group (in fact around its $SU(2)$ subgroup). We
will see below that in certain important examples the nonperturbative
configurations in some sense are "exhausted" by instanton configurations.

The simplest one-instanton solution \cite{BPST} to the self-duality
equations (\ref{selfdual}) has the "bell-shaped" form
\be
\label{bpst}
{\bf A}_\mu \propto \bfeta_{\mu\nu}{x_\nu\over x^2 + \rho^2}
\\
{\bf F}_{\mu\nu} \propto \bfeta_{\mu\nu}{\rho^2\over (x^2 + \rho^2)^2}
\ee
in four-dimensional space-time with the center, chosen in
(\ref{bpst}), to be at the point $x_0=0$. In eq.~(\ref{bpst}) we have
introduced $\bfeta_{\mu\nu}$ -- the 't~Hooft ${\bf C}$-number matrices
(see, for example \cite{VZNSUFN}). Solution (\ref{bpst}) corresponds to the
topological charge of the instanton $n=1$.

Another important nonperturbative effect is the monopole or a particle with
magnetic charge. In the Abelian theory monopoles can arise only as
external sources, but in the framework of non-Abelian theory they can be
identified with certain configurations of extra (scalar or Higgs)
fields \cite{tHoPo}. The simplest monopole configuration arises as a result
of reduction of the self-duality equation (\ref{selfdual}), when fields do
not depend on time and ${\bf A}_0=\bPhi$ is considered as an extra scalar.
Under such reduction the self-duality equations (\ref{selfdual}) turn
into the Bogomolny equations
\be
\label{bpseq}
{\bf D}_i\bPhi = \2\ \epsilon_{ijk}{\bf F}_{jk}
\ee
($i,j=1,\dots,3$). As in the instanton case the topological configuration of
monopoles is nontrivial -- they cannot be obtained by continuous deformation
of configurations with trivial (vanishing) fields. The obstacle is
topological charge. The monopole masses are similar to the actions of
instanton configurations. For the so called
BPS-monopoles \cite{BPS}, being exactly the solutions to
equations (\ref{bpseq}), the masses are equal to
\be
\label{mbpsm}
m_{\rm mon}^{ij} = {4\pi\over g^2}m_{\rm W}^{ij} = {4\pi\over g^2}
\left(\phi_i - \phi_j\right)
\ee
It follows from this formula that at weak coupling the monopoles are very
{\em heavy} particles. However, the situation can again change
after passing to the strong coupling area, though the formula
(\ref{mbpsm}) is literally incorrect. However, at strong coupling the
monopoles might become even more {\em light} than ordinary, i.e.
electrically charged particles. In such circumstances the condensation of
light monopoles can bring us to confinement of electric charges similar to
the Meissner effect in superconductivity.

Thus, the nonperturbative effects related to nontrivial classical
configurations may play an important role when describing the theory at strong
coupling. On of the attendant technical problem is that these effects are
usually "screened" by the perturbative corrections. In order to get a
clearer picture of the nonperturbative effects one should pass to
supersymmetric theories
\cite{SUSYGL,SUSYVA,SUSYWZ} (see also the papers \cite{NS,R,GeSa},
the books \cite{SUSYWest,GSW} and reviews \cite{Scherk,SUSYOM,SUSYL,VyNe}).

\subsection{Supersymmetric Gauge Theories
\label{ss:susyga}}

The main distinguishing feature of the supersymmetric theories is that they
contain an equal number of bosonic and fermionic excitations. Therefore, due to
the different signs of the bosonic and fermionic contributions into loop
diagrams one gets essential cancellation of divergences. This effect is
easily seen, say, directly in the formula
(\ref{beta}), if one puts $N_F$ to be the contribution of fermionic loops,
while $N_V$ -- the contribution of bosonic loops.

Adding to the corresponding Lagrangians the superpartners of the vector and
matter fields one may consider non-Abelian gauge theories as quite
satisfactory for the description of all interactions (except for gravity) in
some vast range of energies. Renormalizability still means that theory is
described by (supersymmetric) Lagrangian of the Yang-Mills fields with
matter terms added in some range of scales and the {\em only} thing to be
added to such Lagrangian is prescription how the coupling
$g = g(\mu)$ depends on the scale $\mu$. This is governed by the
renormalization group equation (\ref{rg}), which looks much simpler in
supersymmetric theories due to cancellation of loop corrections in
perturbation theory.

One of the main "phenomenological" problems of supersymmetric gauge theories
\footnote{The phenomenology of supersymmetric quantum field theories
goes beyond
the scope of this review (see, for example recent review in Physics Uspekhi
\cite{VyNe}). This is a quite interesting and fashionable
topic, whose only weak point is the
absence of experimental confirmation of supersymmetric particles. From our
point of view it is much more important that supersymmetric theories play
the role of a nice "theoretical laboratory" for studying nonperturbative
effects in realistic gauge theories.}
is the presence of scalar fields in their spectra. The scalar fields are
necessary superpartners for the matter fermions and even for the Yang-Mills
fields in the case of extended supersymmetry, i.e. when each field has more
than a single superpartner. Due to supersymmetry the excitations of the
scalar fields should have the same masses as the excitations of fermions (and
vector fields) which totally contradicts to the observable spectrum in
nature. It means that in our world supersymmetry is broken at least at some
scale and the dynamical derivation of such a scale is one of the main problems of
the theory. However, if we believe that this problems will be solved, beyond
this scale (at small distances) the supersymmetric theory is a good object
for study since it is not so "polluted" by loop corrections.

In contrast to nonvanishing vacuum expectation values of
the other fields the scalar
condensates $\langle\bphi_A\rangle\neq 0$ do not violate the space-time
symmetry. Then in low-energy effective theory all parameters of the
effective Lagrangian (masses, couplings) become in general nontrivial
functions of these condensates. As we already mentioned such functions are
usually called functions on the moduli space of supersymmetric
gauge theories. In gauge theories with extended supersymmetry (when number
of supersymmetry generators in terms of the Majorana spinors is \N2 and higher)
one cannot write down potential energy for Abelian fields not violating
supersymmetry. In non-Abelian theories the only choice for such a potential
term, not violating extended supersymmetry, is to take the sum of
commutators of the matrix-valued fields $\{\bphi_A\}$ of the form
$\sum_{A<B}\Tr [\bphi_A,\bphi_B]^2$. In theories with such potential
energy only the light Abelian fields "survive" at large distances, i.e. one
gets electrodynamics (see (\ref{Wmass})) together with massless scalars or
moduli -- the fields whose vacuum values can be arbitrary. Hence, in
gauge theories with extended supersymmetry there exists an infinite number
(parametric family!) of vacua and the problem of the theory is to find the
spectrum and effective couplings of the low-energy effective theory as
functions of the vacuum condensates. An important circumstance is that
supersymmetry imposes extra requirements on the space of condensates, in
particular this space should be complex (and sometimes moreover K\"ahler,
special K\"ahler or hyper-K\"ahler) so that the class of available functions
is essentially restricted. All these general arguments are applicable
only in the case when supersymmetry (or any other symmetry) is the exact
symmetry of quantum theory, i.e. is not violated by quantization.

In the theories with "minimal" \1N supersymmetry the Abelian superpotential
is generated and moduli, in general, become massive and acquire fixed vacuum
expectation values. In complex co-ordinates on moduli space the
superpotential is a holomorphic function $W(\phi_A)$, and vacua are
defined by the equation $dW=0$, since potential
$V(\phi,\bar\phi) \propto \sum_A \left|{\d W\over\d\phi_A}\right|^2$. The
geometrical meaning of the appearance of the complex manifolds in field
theory is absolutely unclear, but, as we see below, it is rather
natural to consider this phenomenon as an "artefact" of string theory. It is
very nontrivial that complex geometry sometimes allows one to predict the
{\em exact} form of the low-energy effective Lagrangians which already
account for the nonperturbative effects (see sect.~\ref{ss:sw} below).

\subsection{General Relativity as Effective Theory}

The discovery of instantons and other nonperturbative solutions essentially
extended the behavior of the theory of strong interactions. It has been
demonstrated that the elementary particle physics does not reduce to
perturbation theory, whose frames in QCD are determined by high energies
(the asymptotic freedom regime), where the standard formulation of the
gauge field theory based on perturbation theory works quite well
\cite{SlavFad}. Nevertheless, the instantonic computations appeared to be
only the next approximation in QCD far not enough to describe confinement
and other effects of strong coupling. As for quantization of gravity, even
supersymmetry \cite{SUSYWZ,SUSYWest} as a mechanism for cancellation of
divergencies does not allow any dream about the possibility of a
consistent theory
of quantum gravity in the framework of quantum field theory
(see, for example, \cite{Scherk}). Despite many attempts to construct a
theory
of quantum gravity in the framework of quantum field theory, say, as a field
theory with infinite-dimensional group of gauge symmetry, such an approach
seems to be based on nothing for a few quite simple reasons. We will
try to discuss these reasons in this section and will come back to them many
times below when speaking about string theory.

Let us first notice that by quantum field theory, if nothing opposite is
stated directly, we will understand the {\em local} quantum field theory,
satisfying the renormalizability criterium. The local quantum field theory
(with Lagrangian depending upon not higher than second derivatives)
guarantees a well-defined procedure of quantization of a {\em free} field
-- an infinite system of particles and anti-particles, corresponding to
the quadratic in fields part of the Lagrangian. The interaction in such
a theory
is introduced by terms of higher degree in the fields and in weak
coupling approximation the relativistic quantum field theory nicely describes
the scattering of particles. It automatically takes into account the
contribution of
antiparticles into the physical processes, which can be considered at
present as its main achievement.

A much more delicate aspect is renormalizability -- the dependence of coupling
constants upon the energy scale. In a renormalizable quantum field theory
an interaction can be described by a {\em finite} set of couplings
(often even a single coupling, as in gauge theories, see sect.~\ref{ss:gauge}), whose
dependence of scale is rather {\em weak}. In reality this "weak dependence"
means logarithmic dependence of the dimensionless coupling constants, like
in gauge theories or $\lambda\phi^4$-theory  in four dimensions.
Renormalizability means that in some wide range of energies the theory is
described by a single Lagrangian -- new interaction vertices should not
be added and the corresponding couplings weakly depend on the scale.

In the theories with dimensional coupling constants and/or an
infinite set of
interaction vertices these features lose any sense. The dimension of
coupling constant, more exactly "negative mass" dimension like the
dimension of the Newton gravitational constant $\gamma_{\rm N}\sim 1/M^2$
in four dimensions (in $D$-dimensional space-time
$\gamma_{\rm N}^{(D)}\sim M^{2-D}$) leads to unbounded growth of the
perturbative corrections of the form
\be
\label{gravcor}
1 + \gamma_{\rm N}\Lambda^{D-2}+\dots
\ee
when one removes the cutoff $\Lambda\to\infty$. This means that the
theory at any
finite scale depends on what happens at small distances. This completely
contradicts the idea of renormalizability, i.e. the idea that after
introducing scale-dependence of the couplings one may completely forget
about small distances.

Such a concept appears to be totally acceptable for
renormalizable (supersymmetric) gauge theories, but is absolutely useless
for the theory of gravity. Gravity (with dimensional coupling and
infinitely many interaction vertices of gravitons $h_{\mu\nu}(x) =
G_{\mu\nu}(x) - \delta_{\mu\nu}$) "remembers" small distances and
is {\em not} renormalizable field theory. The same conclusion follows from
the study of "lattice" or discretized gravity (except for two-dimensional
case \cite{ds,dsBK,dsDS,dsGM}, directly related to string theory),
where the continuum limit is
not well-defined, in contrast to, say, lattice gauge theories.

The difference between gravity and quantum field theory is in fact far
deeper. Quantum field theory computes only the "relative" but {\em not}
"absolute"
value of a physical quantity, i.e. only the difference between the value of
some quantity at given scale $\mu$ and its value at some "normalizing
point" -- at some fixed scale $\mu_0$. Of course, in renormalizable quantum
field theories (for example in gauge theories) it is enough to fix
only  a finite (and usually small) set of quantities at the
"normalizing point", then
the theory is capable to predict {\em any} cross-sections. However, this
circumstance does not abolish this principle feature of quantum field
theory, especially transparent in condenced matter physics, where
a natural "cutoff" exists (say the scale of elementary atomic lattice)
and it is
possible to distinguish between the "macroscopic" quantities, which do not
depend upon this scale and the "microscopic" ones. Moreover, in the condenced
matter physics usually only the {\em relations} between the
microscopic quantities
but not the quantities themselves do not depend on the lattice cutoff,
and this situation is very similar to what one gets in quantum field theory.

The simplest example is energy of any state, which is defined already in
free field theory not as an absolute quantity, but compared to, say, "vacuum
energy". Naively the "vacuum energy" gets a contribution from the infinitely
many vacuum energies of harmonic oscillators
\be
\label{evac}
E_{vac} \propto {\hbar\over 2}\int d{\bf p}\omega({\bf p}) =
{\hbar\over 2}\int d{\bf p}\sqrt{{\bf p}^2 + m^2}
\ee
In field theory without gravity this quantity is {\em not} observable and
can be considered as a reference point, i.e. one may put, say
$E_{vac}=0$. When including gravity, according to the principle of
equivalence the vacuum energy is a source for gravitational field. The
field theoretical expression (\ref{evac}) gives a value absolutely
uncomparable to the value of the cosmological constant with any cutoff (or,
better, with any scale of supersymmetry breaking). The fundamental theory
containing gravity must know how to compute "absolute" values, and this means
that such theory in principle cannot be quantum field theory. The problem of
vacuum energy or the cosmological constant is one of the principle unsolved
problems of modern physics and we will come back to the questions not once
below.

From the structure of corrections (\ref{gravcor}) it is clear that at small
distances $l^{-1} \sim M_{\rm Pl} = \gamma_{\rm N}^{1\over 2-D}$ gravity,
generally speaking, becomes strong. The problems of strong gravitation
interaction and related issues of strong gravitational fields, say, in black
holes, are even less studied that the problems of strongly coupled gauge
theories. One of the well-known effects from the theory of black holes is
the linear relation $S = {{\rm Area}\over 4\gamma_{\rm N}}$ between the
number of states or entropy $S$ and {\em area} of the horizon
(${\rm Area}$) of a black hole \cite{BeHaw}. This statement cardinally
contradicts the expectations of quantum field theory, where the number of
states is always proportional to the {\rm volume} (but not to the area).
This is a kind of indirect argument in favor of the point of view that in
strong gravitational fields one may find some fundamental one-dimensional
structures; for a detailed discussion of this issue see \cite{BiSuTAHo}. Of
course, not being decisive, this is one of the indirect arguments in favor of
string theory.

\setcounter{equation}0
\section{Main Principles of String Theory
\label{ss:string}}

In order to get a consistent theory of quantum gravity one should crucially
change the theory at Planckian scales and replace the pointlike objects by one
dimensional extended objects -- strings. String theory by definition
possesses a dimensional constant, which for historical reasons (see
formula (\ref{regge}) below) is denoted as $\alpha'$. This constant has
dimension of the {\em square} of length. In "fundamental" string theory,
pretending to be the theory of quantum gravity, this parameter can be
nothing else but the Planck length, i.e. $\sqrt{\alpha'}\sim 10^{-33}$~cm.
However, more generally, its value may be chosen depending on problem
under consideration. For example, in string theory applied to the theory of
strong interaction at large distances this parameter should be of the order
of the hadron size $10^{-13}$~cm.

Let us point out that $\alpha'$ is the {\em only}
constant, put "by hand" into string theory. It has a clear sense of
the {\em scale} where stringy effects become essential. There are no other
constants in string theory, even the dimensionless string coupling
$g_{\rm str}$, as we see below, is not really a parameter, but is rather
related to the vacuum condensate of a background field -- the so called
dilaton. In other words, this constant is a dynamical parameter of the
theory.

String theory drastically differs from quantum field theory. We will be
coming back to the discussion of this issue many times, so let us now briefly
formulate the main points. In string and field theory:

\begin{itemize}
\item there is a different "counting in loops", i.e. in field theory and
string theory the intermediate state propagating along the loops are counted
with different weight factors;
\item there is an essential difference in how dimensional reduction
looks like,
moreover, these theories are especially different in space-times with
compact directions;
\item space-time shows up in field theory and string theory in totally
different ways; string theory is characterized by a "dynamical" nature of
space-time. In particular there exist, say, "mirror pairs", i.e. the
manifolds which are {\em not distinct} by string theory;
\item locality and causality also appear in different ways.
\end{itemize}

As was first noticed by Scherk and Schwarz \cite{ShSch},
string theory naturally leads to unification of gauge fields and gravity
into one single theory, since in the spectrum of string one automatically
gets {\it massless} vector fields together with massless fields of spin two.

\subsection{Gauge Fields and Gravitons
\label{ss:ym-f-str}}

Let us start the discussion of foundations of string theory from an
old observation that the theory of one-dimensional extended objects
naturally contains vector
fields and gravitons. The simplest (though not the most strict) way to see
this is to consider a string field or a functional of string contour
$\Phi [X_{\mu}(\sigma )]$ and its expansion in string harmonics (with the
Fourier coefficients $\alpha ^{\mu}_n$)
\be\label{string}
X_{\mu}(\sigma ) = x_{\mu} +
\sum _{n\neq 0}{\alpha ^{\mu}_{-n}\over n}\ \exp ({in\sigma})
\ee
This expansion obviously has the following form
\be\label{str-field}
\Phi [X_{\mu }(\sigma )] = \phi (x) + A_{\mu }(x)\alpha ^{\mu}_{-1} + \dots
\ee
After quantization $[\alpha ^{\mu}_n,\alpha
^{\nu}_m]=n\delta_{n+m,0}\delta^{\mu\nu}$ the Fourier coefficients turn into
the creation and annihilation operators of string excitations. Then
formula (\ref{str-field}) can be better thought of as the action of the
{\em operator} $\Phi [X_{\mu }(\sigma )]$ on the Fock vacuum $|0\rangle$ in
the space of states of an open string. The first term means that vacuum
corresponds to the wave function of a scalar field $\phi(x)$, the next
neighbor state $\alpha ^{\mu}_{-1}|0\rangle$ is related to the vector field
$A_{\mu }(x)$. In expansion (\ref{str-field}) one may take into account
only the string harmonics (the coefficients of decomposition in (\ref{string}))
$\alpha^{\mu}_n$ with $n<0$ (creation operators), since
$\alpha^{\mu}_n|0\rangle = 0$ when $n>0$.
\begin{figure}[tp]
\epsfysize=4cm
\centerline{\epsfbox{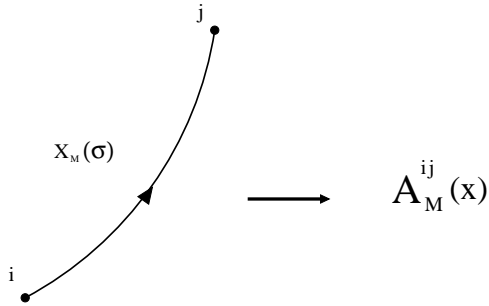}}
\caption{\sl Massless non-Abelian vector field as a string with quarks at
the ends}
\label{fi:string-gauge}
\end{figure}

String quantum mechanics and requirement of invariance
under reparameterizations of "internal" co-or\-di\-nates on the world-sheet
immediately leads to the condition that the vector field $A_\mu(x)$ should
be massless. The simplest explanation of this fact is that
reparameterizations of co-ordinates on world-sheet have "eaten up" two
degrees of freedom, so that physical degrees of freedom are only the
{\em transverse} excitations, say $\alpha_{-1}^i$, $i=1,\dots,D-2$, if
speaking about the vector field. Hence, the vector has only $D-2$ physical
components, where $D$ is the space-time dimension. This automatically means
that the vector field is massless or gauge field, since a massive vector
must have
$D-1$ physical components. More strictly it can be demonstrated considering
the operator of string mass or energy of string excitations
\be
\label{masstr}
M^2 = {1\over\alpha'}
\left(\sum_{n=1}^\infty \alpha^i_n\alpha^i_{-n} - 1\right)
\ee
which shows that the string spectrum contains the massless gauge field. However,
this spectrum starts from the tachyon $\phi(x)$, resulting in additional
problems; one of the most effective tools to overcome this problem is
supersymmetry.

In order to make vector field $A_{\mu}(x)$
non-Abelian one should assign the extra indices to the ends of string
\cite{ChaPa} (for example, of the quark- or antiquark- fundamental
representations). Then the vector field becomes matrix
$\| {\bf A}^{ij}\| $ transforming under {\em adjoint} representation
of the corresponding gauge group (see fig.~\ref{fi:string-gauge}).
For quite a long period of time this procedure was performed "by hand"
(amplitudes were simply assigned by the Chan-Paton factors), until it
finally has become clear that a non-Abelian theory naturally arises if one
allows existence of so called D-branes (see sect.~\ref{ss:dbr}).
Since it is massless vector field which appears
in string spectrum, one gets exactly gauge quantum field theories
in the field theory limit $\alpha '\rightarrow 0$, when
masses (\ref{masstr}) of all other string harmonics
$M^2 \sim {N\over\alpha'}$ (with $N$ being the eigenvalue of the operator
$\sum \alpha^i_n\alpha^i_{-n}$ of the "number of particles" -- string
harmonics in formula (\ref{masstr})) become very large and their excitations
in low-energy effective theory, i.e. at the distances much larger than
$\sqrt{\alpha'}$ can be neglected.

\begin{footnotesize}
In supersymmetric string theory the massless sector contains vector
supermultiplets, where the rest of the states are constructed by
supersymmetry. In the low-energy limit this leads to a supersymmetric theory
of the Yang-Mills fields as an effective theory of massless modes over
the possible
vacua of string theory. According to modern general philosophy
quantum field theories (in particular, supersymmetric gauge
theories got in this way) can be considered as an effective description of physics near
different vacua of string theory. These vacua can be related to each
other by duality transformations -- some {\em discrete}
transformations, exchanging different vacua of string theory and, therefore,
different quantum field theories.
\end{footnotesize}

\begin{figure}[tp]
\epsfysize=6cm
\centerline{\epsfbox{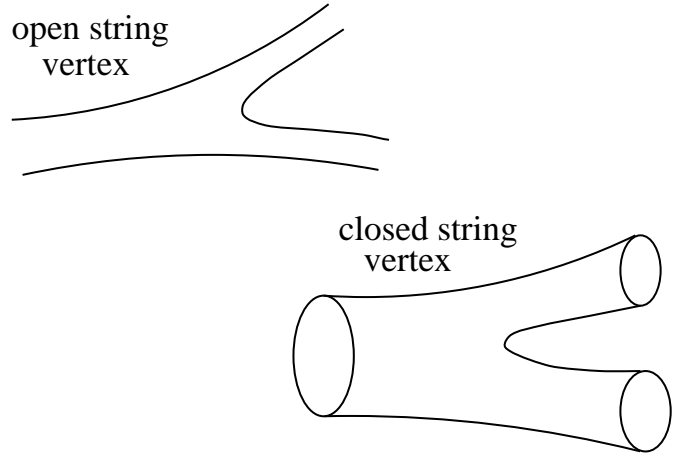}}
\caption{\sl Interaction vertices of open (at the top) and closed (in
the bottom) strings}
\label{fi:interact}
\end{figure}

The expansion over modes of a closed string is similar to formula
(\ref{str-field}) but since the interaction (say with the background fields)
in the closed sector takes place over the whole world-sheet, one should
consider two sets of string harmonics corresponding to left and right
waves independently
propagating over the string world sheet. These waves
are solutions to the equations of motion of free string:
$\alpha^\mu_n e^{in(\tau+\sigma)}$ и $\tilde\alpha^\mu_n
e^{in(\tau-\sigma)}$. The spectrum again starts from the tachyon (the
different one with the modulus of mass squared twice that of the
tachyon of the open string spectrum). Massless fields correspond to the
states $\alpha_{-1}^\mu \tilde\alpha_{-1}^\nu|0\rangle$, or, more exactly to
their linear combination. Dividing the second rank tensor into irreducible
representations of the Lorentz group, it is easy to see that the
corresponding fields consist of, first
\be
\label{B}
\left(\alpha_{-1}^\mu\tilde\alpha_{-1}^\nu -
(\mu\leftrightarrow\nu)\right)|0\rangle\cdot B_{\mu\nu}(x)
\ee
or the antisymmetric tensor field $B_{\mu\nu}$, second
\be
\label{dil}
\delta_{\mu\nu}\alpha_{-1}^\mu\tilde\alpha_{-1}^\nu|0\rangle\cdot \varphi(x)
\ee
i.e. the massless scalar usually called a dilaton. It is the vacuum value
of the dilaton which gives the value of string coupling constant. Finally,
the rest of the components form the massless and traceless symmetric tensor
\be
\label{grav}
\left(\alpha_{-1}^\mu\tilde\alpha_{-1}^\nu - {1\over D}
\delta_{\mu\nu}\alpha_{-1}^\lambda\tilde\alpha_{-1}^\lambda\right)|0\rangle
\cdot G_{\mu\nu}(x)
\ee
or the {\em graviton}.

\begin{figure}[tp]
\epsfysize=5cm
\centerline{\epsfbox{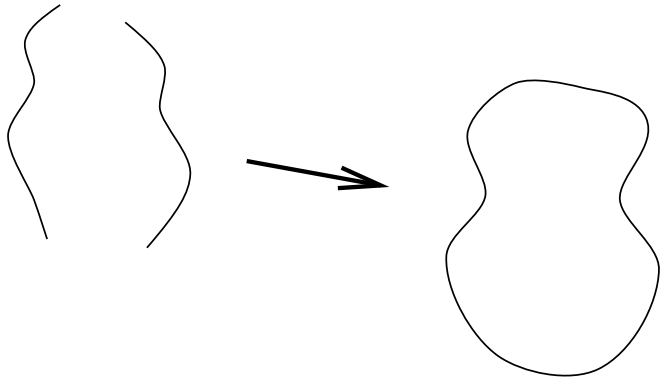}}
\caption{\sl Interaction of open strings leads to closed strings}
\label{fi:opclo}
\end{figure}

All considerations of this section are based by now on the simplest quantum
mechanics of the free string. Switching on the interaction
(see fig.~\ref{fi:interact}), one may easily verify the two following
important properties of the theory:
\begin{itemize}
\item Tree amplitudes of scattering of massless states of open strings in
the limit $\alpha'\to 0$ turn into the scattering amplitudes of vector
gauge bosons, and similar the scattering amplitudes of the states
(\ref{grav}) of the closed sector turn into the amplitudes of graviton scattering
\cite{ShSch}.
\item Interaction of two open strings leads to appearance of the closed
strings (see fig.~\ref{fi:opclo}). Together with the previous remark this
means that gauge field theories, constructed in the framework of string
theory, necessarily lead to the appearance of gravity.
\end{itemize}

\subsection{Massive Fields and Ultraviolet Cutoff}

Let us turn now to massive fields of string spectrum, Their masses
$M$ (see (\ref{masstr})) are measured in units of the (inverse) string
length or the Planck mass $\sqrt{n/\alpha '}$, where $n$ is the number of
corresponding string harmonics or excitation level. It is easy to understand
that this number is linearly related to the (maximal possible) spin of the
excitation $J$. The exact relation can be written in the form of so called
Regge trajectory -- the linear function
\footnote{Let us stress once again that dimensional parameter $\alpha'$
characterizes the scale when string effects become to be essential.
Therefore the exact value of this quantity is different for strings, arising
as effective description of strong interactions at large distances and
"fundamental" strings, corresponding to quantum gravity. Using the notation
originally introduced in the context of hadron physics, we will consider however,
if the opposite is not stated directly, this parameter to be equal to square
of the Planck length.}
\be
\label{regge}
J = \alpha(M^2) \equiv \alpha_0 + \alpha'M^2
\ee
and from (\ref{masstr}) it immediately follows that $\alpha_0=1$
for an open string. The relation between spin and mass (\ref{regge}) was known long
ago in the theory of strong interactions, which after the works of Veneziano
\cite{Ven}, Nambu and Goto \cite{NG} became a "parent" of string theory.
Notice immediately that all excitations with higher spins in string theory
do have masses of the order of the Planck mass. Therefore their absence in
visible spectrum does not contradict to their presence in the theory, unlike
of the non-removable well-known defect of the quantum field theories with
higher spins.

In the limit $\alpha'\to 0$ string theory reproduces the theory of pointlike
particles. From the whole "tower" of fields only the massless fields
survive (under assumption that the tachyon problem is solved;
this problem will be discussed in detail
in sect.~\ref{ss:susy} and \ref{ss:tachpot}). The size of a string can be
estimated, for example, computing the correlator
\be
\label{size}
\langle 0|\int_{d\sigma}\left(X(\sigma)-x\right)^2|0\rangle =
\alpha'\sum_{n>0}{1\over n^2}\langle 0|\alpha_n\alpha_{-n}|0\rangle \propto
\alpha'\sum_{n>0}{1\over n}\propto
\\
 \propto \alpha'\log n_{max} = \alpha'\log
(\sqrt{\alpha'}E_{max})
\ee
where $n_{max}$ and $E_{max}$ are "number" and energy of maximal excited
string harmonic. This formula shows that the size of string is of the order
of $\sqrt{\alpha'}$ (it grows very slowly with energy), which
justifies the interpretation of the only dimensional parameter of string
theory $\alpha'$ as a square of string length.

\begin{figure}[tp]
\epsfysize=7cm
\centerline{\epsfbox{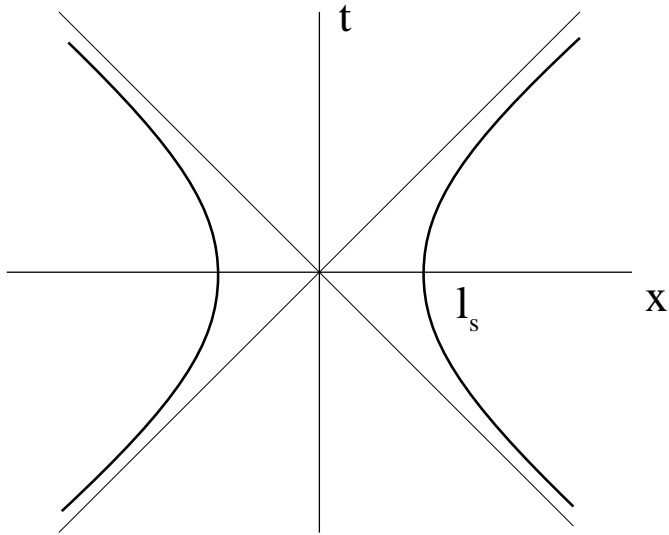}}
\caption{\sl The singularities of propagator in string theory. In contrast
to local quantum field theory they are located inside the hyperboloid,
getting into the distances of $l_s=2\pi\sqrt{\alpha'}$ into the space-like
domain.}
\label{fi:hyperb}
\end{figure}

Notice finally, that the number of quantum states in the
string spectrum grows
rapidly with the energy of excitations. At large energies the spectral
density behaves as
\footnote{The numeric coefficient in front of $\sqrt{\alpha'}M$ in the formula
(\ref{rho}), generally speaking, depends on the particular string model.
Literally in (\ref{rho}) it is written as in the theory of closed strings,
where it is maximally universal. One of the simplest methods
to derive this coefficient for any string model is to consider the
singularities of string propagators \cite{Mprop}.}
\be
\label{rho}
\rho (M) \propto \exp ({2\pi\sqrt{\alpha'}M})
\ee
This behavior leads to absolutely unusual (and different from quantum field
theory) properties of string theory at small distances or large energies --
i.e. at the Planck scale.
\begin{itemize}

\item One of the ways to see this already in the theory of non-interacting
strings is to consider the thermodynamics of string states. Neglecting
interaction free energy has the form
\be
F(\beta) \sim \int dE \rho (E) \exp({-\beta E})
\ee
and for the density of string states (\ref{rho}) this integral converges
only at $\beta > \beta_H = 2\pi\sqrt{\alpha'}$ or at the temperatures less
than the Hagedorn temperature $T_H = {1\over\beta_H} =
{1\over 2\pi\sqrt{\alpha'}}$
\footnote{The Hagedorn temperature coincides with the Hawking temperature of
the black hole whose gravitational radius is equal to string length
$M\gamma_{\rm N} \sim \sqrt{\alpha'}$.}. It means that at the Hagedorn
temperature the phase transition is possible \cite{AtWi}. Simple
calculations show that at hight temperatures the number of (gauge-invariant)
states in string theory is much less than in quantum field theory. For the
"normalized" free energy in string theory independently of space-time
dimension $D$ one has
${F\over VT} \stackreb{T\to\infty}{\propto} T$ instead of
${F\over VT} \stackreb{T\to\infty}{\propto} T^{D-1}$ in field theory.
Not being yet finally understood, this property demonstrates the qualitative
agreement between the high-energy properties of string theories with
corresponding (hypothetical) properties of gravity.

\item Another manifestation of the same effect is violation of
"microlocality" in string theory, related to the growth of spectral density
according to (\ref{rho}). Computing the Green function or propagator of
string between "pointlike" initial and final states (see fig.~\ref{fi:prop}),
\begin{figure}[tp]
\epsfysize=4cm
\centerline{\epsfbox{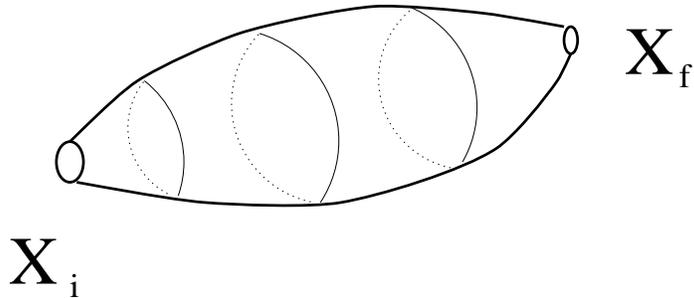}}
\caption{\sl Propagator of closed string with fixed boundary contours.
Choosing these contours as points the propagator becomes a function of two
variables $G(X_f,X_i)$, and can be compared to a
similar object in quantum field theory.}
\label{fi:prop}
\end{figure}
and studying its singularities it is easy to see that they  look like
singularities of {\em non localizable} theory, i.e. lie within some
hyperboloid getting into the distance of the order $\sqrt{\alpha'}$ into the
space-like region \cite{FM2} (see fig.~\ref{fi:hyperb}).

\item Scattering amplitudes in string theory at large energies crucially
differ from the corresponding amplitudes in quantum field theory by
softer behaivior, this can be seen already at the level of the Veneziano
amplitude (see, for example, \cite{GSW}). Due to summing over infinitely
many states in the intermediate channels, the amplitudes of string theory
contain the "cutting" factor at high energies.

\end{itemize}

Notice finally, that the opposite limit to field theory $\alpha'\to\infty$
(the so called "nill-strings") is very singular. Being a complicated
technical problem, this limit is most likely senseless from the
physical point of
view. It corresponds to the theory at the energies much more than Planckian,
i.e. in the region where neither field theory nor even string theory are
literally applicable and taking such limit is similar to an attempt to use
field theory beyond the scale of ultraviolet cutoff.

\subsection{String Perturbation Theory -- Sum over Two-dimensional
Geometries
\label{ss:strpol}}

\begin{figure}[tp]
\epsfysize=6cm
\centerline{\epsfbox{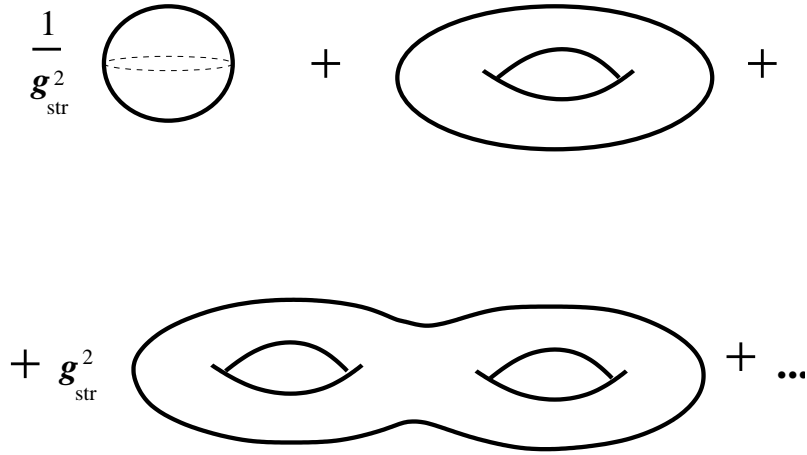}}
\caption{\sl String "Feynman diagrams" corresponding to the
first three terms of
the perturbative expansion (\ref{ppi0}) for closed strings. The tree-level
contribution (of the order of ${1\over g_{\rm str}^2}$ in "string
normalization") (\ref{ppi0})) corresponds to the sphere, the
one loop contribution is given by the torus, the two-loop by the
figure of eight, etc.}
\label{fi:perturb}
\end{figure}

The perturbative structure of string theory can be defined by the
"loop expansion", see fig.~\ref{fi:perturb},
\be
\label{ppi0}
{\cal F} = \sum_{g=0}^\infty g_{\rm str}^{2g-2}F_g
\ee
or by expansion over topologies or {\em genera} of the world sheets being
two-dimensional Riemann surfaces. The role of parameter of this expansion is
played by$g_{\rm str}$ -- the string coupling constant. Notice immediately that
expansion (\ref{ppi0}) is written for the free energy or the {\em logarithm} of
the partition function (in contrast to quantum field theory) since it
includes summation only over "connected diagrams". Literally the loop
expansion on fig.~\ref{fi:perturb} is valid for the theories with only
closed strings. These theories include the closed bosonic strings as well as
so called superstrings of type II
\footnote{When D-branes are absent, see sect.~\ref{ss:reduct} and
\ref{ss:dbr}.}, on which we will mostly concentrate below. If the theory
contains open strings together with closed one should also add the
world-sheets with boundaries.

Let us also note that the normalization in
(\ref{ppi0}) as well as in fig.~\ref{fi:perturb} is chosen in such way that
the contribution of any genus is proportional to the
particular power of string
coupling $g_{\rm str}$, which is equal, up to a sign, to the
{\em Euler characteristic} of the corresponding world-sheet. Due to this
normalization the expansion starts with $g_{\rm str}^{-2}$ and includes for
closed strings only {\em even} powers of string coupling. In the theory of
open strings for the world-sheets with boundaries one would also get the odd
degrees of the coupling constant. This means that the string coupling in
closed sector is in fact proportional to the square of the open string
coupling and this fact will be important below when discussing the
nonperturbative theory.

The contribution of each genus is computed by the Polyakov path integral
\cite{Pol81} over the string co-ordinates and two-dimensional geometries or
metrics on world-sheet
\footnote{Since string theory {\em by definition} contains an integral over
two-dimensional metrics it is often identified with two-dimensional quantum
gravity. Indeed the parallels between string theory and quantum gravity in
two dimensions are very useful for studying both theories. However, one
should remember the principle difference in space-time interpretation,
which for string theory is multidimensional and the observables in string
theory are defined in multidimensional space-time.}
\be\label{ppi}
F_g = \int Dh_{ab}D{\bf X} \exp\left({- \int _{\Sigma _g}\partial{\bf X}
\bar\partial{\bf X}}\right)
\ee
where ${\bf X}$ are co-ordinates of string, being at the same time from the
point of view of two-dimensional world sheet theory the fields of a
free field theory, and $h_{ab}$ denote metrics on Riemann surface
$\Sigma _g$ of genus $g$. The summation over two-dimensional geometries
in (\ref{ppi}) was originally formulated by Polyakov as integration over
metrics. If, however, one takes into account the invariance under
reparameterizations on world sheets, the sum is really taken over the
"equivalence classes" of metrics (with respect to changes of co-ordinates or
reparameterizations) and it is these equivalence classes which correspond to
physically different configurations.

On the first glance the action
$\int _{\Sigma _g}\partial {\bf X}\bar\partial {\bf X}$ in formula
(\ref{ppi}) does not at all depend on two-dimensional metric
$h_{ab}$. Two out of three its components may be immediately "killed" by two
reparameterizations of the world-sheet co-ordinates, say, the metric can be
brought by reparameterizations to the conformal form
$h_{ab}=\exp({\varphi})\delta_{ab}$, when it is determined by a single
function $\varphi$ on string world-sheet. It is easy to check that free action
(\ref{ppi}) does not at all depend upon the "conformal factor"
$\varphi$ and the integral over metrics in (\ref{ppi}) looks like being
trivial. However, this is not true. The reason is that two-dimensional
theory (\ref{ppi}) is a simple quantum mechanics but with
{\em infinitely} many degrees of freedom and therefore the integral in
(\ref{ppi}) should be regularized.

If we require that regularized theory should be independent of the choice of
co-ordinates on string world-sheet (and such arequirement is absolutely
necessary from physical point of view -- the sensible physical theory must
not depend on co-ordinates on unobservable world-sheet of the Planckian
size) the regularization (for example, cutoff) should be introduced {\em
covariantly}. This means that {\em quantum} theory (\ref{ppi}) in general
does depend on metric $h_{ab}$ or at least on its conformal class. Such
phenomenon is called as anomaly (see, for example, the review
\cite{MorAn} and references therein), and in our case we deal with
two-dimensional conformal or gravitational anomaly. Calculation of this
anomaly in
\cite{Pol81} has demonstrated that two-dimensional geometry
essentially restricts the properties of space-time which is a target-space
for string theory. The origin of these restrictions is that contribution to
anomaly of "physical degrees of freedom" should be compensated by the
contribution of two-dimensional geometry (supergeometry) itself.
And it is this constraint which leads to well-known
critical dimensions $D=26$ (or $D=10$) "fixed by God". Such
restrictions are not as strong as one had thought at the beginning of
string era, but nevertheless string theory in some sense chooses the
space-time "itself". The space-time in string theory should be essentially
multidimensional, though partially these dimensions can be "small" -- i.e.
responsible for the "internal" degrees of freedom in spirit of the
Kaluza-Klein models \cite{KaKle}.

The computation of anomaly \cite{Pol81} shows that in quantum theory the
conformal factor $\varphi$ "alives" and acquires the meaning of extra (singled
out) co-ordinate of the space-time. Anomaly adds the kinetic term for the
field $\varphi$ to the action (\ref{ppi}), so that (in flat space-time) the
total action acquires the form
\be
\label{freeact}
\int_{\Sigma}\left(\d{\bf X}\bar\d{\bf X}+
\d\varphi\bar\d\varphi + \dots\right)
\ee
where in some natural normalization the field $\varphi$ should be regarded
as
imaginary. In other words formula (\ref{freeact}) is naturally interpreted
as a free action in Minkowski space. The interpretation of time as "scale
factor" arising in the framework of string theory is a bit similar to
analogous interpretation in general context of gravity and cosmology.

Let us return to the properties of the path integral (\ref{ppi}) over
two-dimensional geometries. In the case of pointlike particles this integral
is reduced to the finite-dimensional integral over the Feynman parameters,
which have the meaning of invariant
lengths of the trajectories of particles. In such away the Feynman diagrams
(say, in the $\phi^3$-theory) arise directly at the first-quantized level.
The main physical problem coming out of the integrals over Feynman
parameters (and hence from the integral over one-dimensional geometries) is
the appearance of ultraviolet divergencies due to contributions of
trajectories of infinitely small lengths. In string theory these
singularities are naturally regularized when one passes from world-lines
intersecting at some points to smooth world sheets (this immediately
leads to the fact that only cubic interaction is possible in string theory).

A more
delicate effect is that two-dimensional geometry {\em regularizes} the
contribution of small distances since this contribution is geometrically
equivalent to the contribution of trajectories of large lengths. According
to the main principle of quantum physics the summation should be taken only
over the independent configurations. One should immediately conclude that in
order to avoid "double counting" the contribution of the trajectories with
small lengths should {\em not} be counted at all, if all equivalent
"infrared" configurations are already taken into account. As a result
of this logic we get a striking consequence that in string theory
если учтены все эквивалентные им конфигурации в
{\em by definition} the ultraviolet problems of the quantum field theory are
absent, more strictly there are no ultraviolet divergencies if there are no
infrared
\footnote{This is not the case for many string models due to presence of
tachyons.}.
This statement follows from the analysis of finite-dimensional part of the
integral over two-dimensional geometries given by the integral over moduli
spaces of complex structures of Riemann surfaces
(this issue is in the center of discussion of the main part of review
\cite{Kni}).

\begin{footnotesize}
According to the Belavin-Knizhnik theorem \cite{BKni} the integral over
metrics (\ref{ppi}) is reduced to the integral over the moduli space of
complex structures of the Riemann surfaces
\be\label{tbk}
F_g = \int _{{\cal M}_g}d\mu (y)|f(y)|^2
\ee
where ${\cal M}_g$ is the (finite-dimensional) moduli space of complex
structures of the Riemann surface $\Sigma _g$. The concrete choice of the
integration measure depends on particular choice of a string model, for
the bosonic string this is the Mumford measure \cite{BKni}. It is the
modular invariance of the integrand in (\ref{tbk}) leading to the fact
that contributions of the trajectories of small lengths and the
trajectories of large lengths are physically equivalent. The formulation
(\ref{ppi}), (\ref{tbk}) allows one in principle to use the symmetry
properties in order to get some nonperturbative information, though by its
own definition this is just a perturbative expansion around some vacuum
and the integral (\ref{ppi}) computes only the  $g$-\-loop correction of
the expansion of string perturbation theory.
\end{footnotesize}

\subsection{Dynamical Nature of Space-Time and Two-dimensional Conformal
Theories
\label{ss:conf}}

Let us come back to the fact that the contribution of the new co-ordinate
coming from two-dimensional metric allows to cancel the conformal anomaly.
This condition is not empty (in the sense that it does not take place
everywhere) and leads to {\em dynamical restrictions} on the properties of
physical space-time. The basic restrictions look as follows:

\begin{itemize}
\item In flat space-times string theory exists only in some
{\em distinguished} or critical dimensions. The simplest bosonic
string (\ref{ppi}), (\ref{freeact}) demands the total number of dimensions
to be $D=26$ (including time), and the theory of fermionic or
supersymmetric strings (two-dimensional supergravity) fixes the critical
dimension to be $D=10$.
\item In nontrivial background fields, say, when metric is not flat, the
background fields should satisfy the classical equations of motion, in
particular the Einstein equations
\be
\label{einstein}
R_{MN}(G) - \2 G_{MN}R(G) - T_{MN} = {\cal O}(\alpha')
\ee
up to the string corrections. In eq.~(\ref{einstein}) $G_{MN} = G_{MN}(X)$
is the space-time
metric, $R_{MN}(G)$ is its Ricci tensor and $T_{MN}$ is the
stress-energy tensor of the other background fields.
Moreover, in presence of nontrivial background fields the
anomaly cancellation condition is changed. In such case the critical
dimension ($D=26$ or $D=10$) "moves", i.e. changes due to contribution
of corrections in $\alpha'$ to the anomaly  -- the terms, starting with
$\alpha'R(G)$.

\end{itemize}

Generally speaking, the space-time should not be necessarily Minkowski
space or the Euclidean flat space
\footnote{The problems of signature of space-time are still beyond the
framework of string theory and we will not discuss it here. Let us only
point out that we imply everywhere a possibility of smooth analytic
continuation of the theory in Minkowski space to the Euclidean space and
we will not distinguish between these two formulations below.},
say ${\bf R}^4$, it may have a nontrivial metric (satisfying the Einstein
equations due to the two-dimensional symmetries \cite{FraTse}). It can be
even a nontrivial compact manifold (or, more exactly has a compact part),
corresponding, as already mentioned above, to the internal (gauge) degrees
of freedom in spirit of the Kaluza-Klein models. The Polyakov path
integral (\ref{ppi}) should be then understood in "generalized" sense when
instead of free infinite-dimensional quantum mechanics (or two-dimensional
field theory (\ref{ppi}) with the fields $X$, to be interpreted as
space-time co-ordinates) one should deal with some generic sigma-model
\be
\label{sigma}
\int_\Sigma \left(G_{MN}(X)\d X^M\bar\d X^N + {\cal R}^{(2)}\Phi(X) +
\dots\right)
\ee
where ${\cal R}^{(2)}={\cal R}^{(2)}(h)$ is the curvature of the
two-dimensional
metric, while $G_{MN}(X)$ and $\Phi(X)$ are nontrivial background fields
for the space-time metric and dilaton. A principal new moment in string
theory is that the theory "adjusts" to itself the space-time where it exists.
More strictly, it imposes essential constraints on the characteristics of
the target space-time and forces the background fields to be solutions to
the equations of motion. Let us also point out that comparing eqs.~(\ref{sigma})
and (\ref{ppi0}) and using the Gauss-Bonnet theorem
$\int_\Sigma {\cal R}^{(2)}(h) = 2-2g$, (where $g=g(\Sigma)$ is genus of
the Riemann surface $\Sigma$) one gets the relation between the "zero
mode" $\Phi_0$ of the dilaton field $\Phi(X)$
(more exactly of its vacuum expectation value) and the string coupling
constant $g_{\rm str} = \langle \exp(\Phi_0)\rangle$.

Considering string theory in the external background fields, including
nontrivial metric of the space-time (such theories for historical reasons
are usually called as two-dimensional sigma-models), it is necessary all
the time to look after the condition of conformal invariance, which is
reminiscent of the reparameterization invariance after the metric
$h_{ab}$ has been chosen in conformal form, see (\ref{ppi}), (\ref{freeact})
and (\ref{sigma})). In other words, nontrivial background fields should
necessarily correspond to the two-dimensional conformal
sigma-models, or, more directly to the two-dimensional conformal
theories \cite{BPZ}. The difference between these two notions is only in
the fact that majority of known two-dimensional conformal theories have only
an approximate description in sigma-model terms. Usually, an
explicitly known
nontrivial sigma-model can correspond only to "bare" values of the
background fields, while the exact background fields, which hypothetically
describe the exact conformal theory are not really known. In such a case the
conformal field theory can nevertheless be formulated axiomatically \cite{BPZ}
or, in terms of free field theories \cite{bosonizaFF,bosonizaDF,bosonizaGMMOS},
corresponding to the simplest dilaton background
\footnote{A nice exception consists of two-dimensional sigma-models on
group manifolds and conformal theories corresponding to them \cite{WZW}.
However, even in this case it is simpler and more natural to construct the
conformal theory just {\em requiring} that conformal symmetry is an
exact quantum symmetry consistent with the current algebra, always
existing on group manifold \cite{KZ}.}.

\begin{footnotesize}
Two-dimensional conformal field theories \cite{BPZ} are the theories with
invariance under the action of the {\em infinite-dimensional}
(only in two-dimensions!) group of conformal symmetry. This group if
formed by holomorphic reparameterizations on world sheets, keeping metric
in conformal form $h_{ab}=\exp({\varphi})\delta_{ab}$. The generators of
such transformations form the Virasoro algebra
\be
\label{vir}
[{\cal L}_n,{\cal L}_m]=(n-m){\cal L}_{n+m} + {c\over 12}\delta_{n+m,0}
\ee
and in the "classical" case (at $c=0$) may be represented as
${\cal L}_n = - z^{n+1}{d\over dz}$, i.e. form the basis of holomorphic
vector fields on the world-sheet $\Sigma$, parameterized by complex
co-ordinates $(z,\bar z)$. Implying that conformal symmetry is an exact
symmetry of quantum theory (and this is again a natural requirement of
independence of physics of the choice of co-ordinates on the world-sheet
of the Planckian size), one gets immediately an infinite number of
constraints (the Ward identities) on the correlation functions in
two-dimensional theory \cite{BPZ}. This allows in principle to calculate
any two-dimensional correlator, being the "building blocks" for string
amplitudes. It turns out that the same statement can be formulated
alternatively: despite all conformal theories corresponding to nontrivial
manifolds in space-time being not free theories (\ref{freeact}) in the
literal
sense, for any conformal theory there exists a representation in terms of
free fields or so called {\em bosonization}
\cite{bosonizaFF,bosonizaDF,bosonizaGMMOS}. This means that in any
nontrivial space-time, consistent with two-dimensional conformal
invariance, string theory is {\em in principal defined} perturbatively and
the integrals (\ref{ppi}) and (\ref{tbk}) can be calculated. Bosonization
effectively reduces the computations in nontrivial conformal theories to
the calculation (of quite complicated correlation functions) in the
theories with quadratic action
\be
\label{dilback}
S_{CFT}(\varphi) = \int_\Sigma \left(\partial\varphi{\bar\partial}\varphi +
\alpha _0{\cal R}^{(2)}\varphi\right)
\ee
where the constant $\alpha _0$ (or constant vector in case of many fields)
is related to the central charge
$c_{CFT} = 1 - 12\alpha _0^{\ 2}$. This is the way how non-integer central
charges of nontrivial theories arise from the free theories with central
charges just equal to the number of fields, $c=D$. It is also useful, as
follows from comparison with (\ref{sigma}), to interpret action
(\ref{dilback}) as the action of a string in the external
{\em linear} dilaton background $\Phi(\varphi) = \alpha_0\varphi$. We will
see below that such a background is also singled out
in string theory also from other points of view.

Besides, for generic conformal theories one should specially notice that a
single conformal theory may correspond in general to strings on {\em
different} manifolds ${\cal X}_1$ and ${\cal X}_2$. Such manifolds are
called mirror manifolds \cite{mirrorVafa,mirrorWitten}. The simplest
example is a free theory of a field, taking values on a circle
-- the theories on circles ${\cal X}_1 = S_R$ of radius $R$ and
${\cal X}_2 = S_{\alpha'/ R}$ with the radius $\alpha'/R$ are equivalent,
see sect.~\ref{ss:r1r}).
\end{footnotesize}

Let us recall once more that the amplitudes in string theory are built
from the correlation functions of two-dimensional conformal field theory.
More exactly, the scattering amplitudes of, say, massless excitations
above some vacuum do correspond to the particular correlators in
two-dimensional conformal field theory corresponding to this vacuum. These
operators are fixed by the set of corresponding quantum numbers and by
condition of conformal invariance -- the consequence of reparameterization
invariance on the world-sheet. It is remarkable that conformal invariance
immediately leads to all physical requirements on the operators of
physical particles. Let us demonstrate this on the example of the operator
of emission or absorbtion of a photon (in a flat space-time)
\be
\label{photon}
\epsilon\cdot\d X \exp({ip\cdot X})
\ee
with momentum $p$ and polarization vector $\epsilon$. First, conformal
invariance says that a "physical operator" must have unit dimension, then
and only then the result of integration over the boundary of the
world-sheet (in case of open strings, or over the whole world-sheet in
case of closed strings) will not depend on the choice of co-ordinates. For
the operator of photon (\ref{photon}) it means (due to unit dimension of
pre-exponent) that $p^2=0$, or, alternatively, that the
({\em anomalous} in the sense of two-dimensional conformal field theory)
dimension of the exponent in (\ref{photon}) vanishes. Thus, from the
condition of {\em two-dimensional conformal invariance} one immediately
obtains that the photon is massless. In fact this derivation is just a
little bit more strict variant of the argumentation from the beginning
of sect.~\ref{ss:ym-f-str}.

Slightly more detailed analysis of the conformal invariance leads
rapidly to the transversality of physical photon $\epsilon\cdot p=0$,
or to {\em gauge} invariance. Indeed, decomposing the polarization vector
into the transversal and longitude parts $\epsilon_M = \epsilon_M^\bot
+ \epsilon_M^\|$, so that $\epsilon^\bot\cdot p=0$ и $\epsilon_M^\|
\propto p_M$, one easily finds that
\be
\label{prodol}
\epsilon^\|\cdot\d X \exp({ip\cdot X}) \propto p\cdot\d X \exp({ip\cdot X})
\propto \d\left(\exp({ip\cdot X})\right) = {\cal L}_{-1}\cdot \exp({ip\cdot
X})
\ee
i.e. the contribution of the longitudinal part is the total derivative and
disappears after the integration over the boundary of the world sheet. In
other words, using the last equality in (\ref{prodol}), one may say that
the operators or states corresponding to physical particles are defined
in the language of two-dimensional conformal theories up to the "gauge"
states of the form ${\cal L}_{-1}|\Psi\rangle$ and with the vanishing
norm. Thus, the "ghost-free" requirement of two-dimensional theory leads
to the gauge invariance in physical string spectrum.

\subsection{Supersymmetry and Fermions
\label{ss:susy}}

Let us now briefly discuss the extra world-sheet fields and related internal
degrees of freedom. One of the important properties of string theory is that
by
introducing supersymmetry on world-sheet one immediately obtains the
space-time fermions
\footnote{Here one should make a few extra comments. This property in fact can
be detected already at the level of pointlike particles. Moreover, in some
sense (without using the notion of supersymmetry) it was known long before
the string theory appeared. Nevertheless, it seems to be extremely important
that only in string theory or on two-dimensional world-sheets, this
property arises naturally and without "pathologies" of the one-dimensional
case.}.

Already in the degenerate example of string -- the relativistic particle --
it is enough to introduce the world-line supersymmetry
\cite{BDHDZ}, to get the space-time fermions. The world-line action can be
defined requiring the invariance under the (one-dimensional!) supersymmetry
with Grassmann parameter $\epsilon$
\be
\delta X = \epsilon\Psi
\\
\delta\Psi = -\epsilon\left({\dot X} + \2\chi\Psi\right)e^{-1}
\\
\delta\chi = - 2{\dot\epsilon}
\\
\delta e = - \epsilon\chi
\label{susy}
\ee
The corresponding invariant action
\be
\label{dirpart}
\2\int_{dt}\left({\dot{X}^2\over e} + \Psi{\dot\Psi} +
{\chi\over e}\Psi{\dot X} + m^2(e + {1\over 4}\chi d_t^{-1}\chi)\right)
\ee
includes in addition to co-ordinates $X_M$ and one-dimensional "metric"
$e$ the Grassmann "gravitino" $\chi$ and fermionic variables $\Psi_M$ with
the first-order kinetic term, such that these variables coincide with their
own momenta $\Psi_M = {\delta S\over\delta\dot\Psi_M}$. After quantization
one gets the relations $[\Psi_M,\Psi_N]_+ = \delta_{MN}$,
i.e. the Grassmann variables $\Psi_M$ turn into the Dirac gamma-matrices and
the wave function carries now also the {\em space-time} spinor index, since
it becomes a vector of certain representation of the Clifford algebra. The
corresponding representation in terms of the (one-dimensional analog) of
the
Polyakov path integral with the action (\ref{dirpart}) allows to
compute Green functions in the theory of Dirac fermion.

Notice, that the world-line supersymmetry (\ref{susy}) (as well as its
direct generalization -- the supersymmetry on the string world-sheet) is
practically identical to the well-known supersymmetry in quantum mechanics.
The simplest example of supersymmetry in quantum mechanics is a particle in
magnetic field, which can be considered as a quantum mechanical system with
the Hamiltonian $H = \left({\bsigma}\cdot {\bf P}\right)^2$ (with the Pauli
matrices $\bsigma$ being the simplest representatives of the Dirac
matrices). The role of supergenerator is played by the Dirac operator
${\bsigma}\cdot {\bf P}$, and this exactly corresponds to the interpretation
of supersymmetry transformations as "square roots" of the energy-momentum
operators. The essential feature of supersymmetry in quantum mechanics (in
particular that of (\ref{susy})) is that the related "fermionic number"
is not really "fermionic" from the point of view of
space-time.

Indeed, when the role of Hamiltonian is played by the square of the
Dirac operator, the "fermionic number" is nothing but direction of spin.
Therefore from the perspective of physical space-time the supersymmetric
"bosons" and "fermions" just correspond to different directions of spin of a
"real space-time" fermion, whose wave function satisfies the Dirac equation.
As we see below the world-sheet supersymmetry in string theory reminds one
a lot
the supersymmetry in quantum mechanics apart from details with the boundary
conditions due to a extra co-ordinate on the world-sheet. It is quite
nontrivial that this "auxiliary" supersymmetry of a quantum-mechanical type
leads to the "real" space-time supersymmetry in string spectrum.

Hence, things are much more interesting for the fermionic string -- the
first-quantized theory with the world-sheet action
\be
\label{fstring}
{1\over 2\pi\alpha'}
\int_\Sigma\left(\partial X\bar{\partial} X+
  \Psi\bar{\partial}\Psi + \bar{\Psi}\partial\bar{\Psi} +
\chi\Psi\bar{\partial}X + \bar{\chi}\bar{\Psi}\partial X +
\2\bar{\chi}\chi\bar{\Psi}\Psi\right)
\ee
invariant under the transformations of two-dimensional supergravity
\cite{BDH&DZ}. First three terms in the expression (\ref{fstring})
(at $\chi=\bar\chi=0$) correspond to the action, invariant under the
{\em global} two-dimesional supersymmetry transformations on world-sheet
\cite{GeSa}. Depending on the boundary conditions (periodicity or
antiperiodicity or their analogs in the open string case) the fermionic
fields $\Psi$ either do not or do contain the "zero mode" -- the constant
component $\Psi^{(0)}_M$, which in complete analogy with the example of
fermionic particle may turn into the set of Dirac matrices after
quantization $[\Psi^{(0)}_M,\Psi^{(0)}_N]_+ = \delta_{MN}$.

Thus, depending on the choice of boundary conditions, there are
two sectors in fermionic string. The wave functions of one sector possess an
index of a representation of the Clifford algebra and correspond to the
space-time fermions, while the wave functions of another sector do not have
such indices and correspond to the space-time bosons. The corresponding
two-dimensional conformal theory \cite{bosonizaFMS,bosonizaVV} allows to
compute the correlation functions, corresponding to arbitrary scattering
amplitudes in the fermionic string.

After all that it is natural to ask how the states of the fermionic string
spectrum corresponding to space-time bosons and space-time fermions are
related to each other. At first glance these two sectors -- bosonic and
fermionic -- differ too much from each other, for example, the bosonic sector
(or the Neveu-Schwarz sector \cite{NS}) contains tachyon, while the
fermionic sector (or the Ramond sector \cite{R}) is tachyon free.
Nevertheless, there exists a natural "GSO-projection"
(i.e. procedure leaving only half of the states in
the spectrum) \cite{GSO}, which results in leaving in the spectrum the equal
number of states from both sectors in such a way that the full spectrum
(after projection) becomes space-time supersymmetric!

Moreover, at the level
of the one-loop partition function this projection arises naturally after
summing over all possible boundary conditions of fermionic fields
\cite{SWss}. All this means that supersymmetry on the world-sheets of
fermionic strings leads to the supersymmetry in (ten-dimensional)
space-time. The resulting theory -- the "reduced" fermionic string with
ten-dimensional supersymmetry, after John Schwarz is often called
superstring.

In the open string sector the GSO-projection leaves in the Neveu-Schwarz
sector the subsector with {\em odd} "fermionic number" (in the sense of
world-sheet fermions), for example the massless vector
$\Psi^\mu_{-1}|0\rangle_{NS}$ is left in the spectrum of open superstring
while the naive "vacuum" or the Neveu-Schwarz tachyon $|0\rangle_{NS}$ is
"killed" by the GSO-projection. In the Ramond sector the GSO projection
leaves only the space-time fermions with fixed chirality (the eigenvalue of
the operator $\2(1\pm "\Gamma_5")$, $"\Gamma_5"\propto\prod_{M=1}^{10}\Gamma_M$,
acting on the ten-dimensional Majorana
spinors), the number of such
fermionic states (at each mass level) is exactly equal to the number of
states in the Neveu-Schwarz sector with the odd "fermionic number".
Hence, in the
theory of closed strings one may have two different superstring
theories. One would contain the fermions of different chiralities while the
other -- the fermions of the same chirality: the first is called a type IIA
theory while the second -- a theory of type IIB.

It turns out that superstrings can be reformulated without two-dimesional
world-sheet Neveu-Schwarz-Ramond type fermions. There exists
anaternative Green-Schwarz formulation \cite{GSsst}, using the extra
Grassmann fields $\theta_\alpha(\sigma,\tau)$
(spinors in ten-dimensional space-time in contrast to the ten-dimensional
vectors $\Psi_\mu(\sigma,\tau)$) explicitly invariant under the
ten-dimensional supersymmetry transformations. However, the variables
$\theta_\alpha(\sigma,\tau)$ behave as scalars with respect to
two-dimensional reparameterizations of co-ordinates and two-dimensional
supersymmetry is not a symmetry of the Green-Schwarz superstrings.

The investigation of anomalies, started in \cite{GS84}, has brought us to the
following list of anomaly-free superstring models: type IIA and type IIB
theories (closed string non-chiral and chiral theories with
\N2 in ten dimensions), type I theory (which includes open strings) and
theories of heterotic strings \cite{heter} (the string models where, say,
left or holomorphic part corresponds to the twenty-six-dimensional bosonic
string with extra compactification while the right or antiholomorphic part
-- to the ten-dimensional superstring) with the gauge groups
$SO(32)$ and $E_8\times E_8$.

\begin{footnotesize}
Unfortunately the ten-dimensional superstring pretending to be the most
successful among existing string models is strictly defined, in general,
only at tree and one-loop levels. Starting from the two-loop corrections
(the last diagram depicted at fig.~\ref{fi:perturb}) to the scattering
amplitudes all expressions in the perturbative superstring theory are
really {\em not} defined. The reason for that comes from the well-known
problems with supergeometry or integration over the "superpartners" of  the
moduli of complex structures.

In contrast to the bosonic case (\ref{tbk}), where the integration measure is
fixed by the Belavin-Knizhnik theorem, the definition of the integration
measure over supermoduli (or, more strictly, the odd moduli of super-complex
structures) is still an unsolved problem \cite{loopsst,loopsstR}. The moduli
spaces of the complex structures of Riemann surfaces are non compact, and
the integration over such spaces requires special care and additional
definitions. In the bosonic case, when the integrals over moduli spaces
diverge, the result of integration in (\ref{tbk}) is defined only up to
certain "boundary terms" -- the contributions of degenerate Riemann surfaces
or the surfaces of lower genera (with less "handles", see
fig.~\ref{fi:perturb}). In the superstring case one runs into more serious
problems since the very notion of the "boundary of moduli space" is
{\em not defined}. Indeed the integral over the Grassmann odd variables does
not "know" what is the boundary term. This is the fundamental reason why the
integration measure in fermionic string is not well-defined and depends on
the "gauge choice" or the particular choice for the "zero modes"
$\chi$ in the action (\ref{fstring}). For two-loop contributions this
problem can be solved "empirically" (see \cite{loopsst,loopsstR}), but in
the general setup the superstring perturbation theory is not mathematically
well-defined. Moreover, these are not problems of the formalism:
the same obstacles arise in less geometrical approach of Green and Schwarz
\cite{KaMo}.
\end{footnotesize}

\subsection{Effective Actions for Background Fields
\label{ss:fratse}}

By analogy with the generating functionals for particles in external fields
one may introduce the interaction of strings with background fields. The
integration over the string degrees of freedom will give rise to certain
effective functionals, depending already only upon the local fields in
space-time. Such functionals are called the Fradkin-Tseytlin effective
actions \cite{FraTse}, and can be considered as the most efficient way for
getting effective field theories from string theory.

Such an approach looks very transparent and clear from an ideological point
of view. Indeed, at observable energies massive string modes are not excited
and only the massless local fields "fly out" into our low-energy world. The
interaction of string with local fields can be easily written down from
certain symmetry requirements, say adding an exponential of the interaction
term with the gauge field
\footnote{Notice that the operator (\ref{photon}) literally corresponds to
the first term in formula (\ref{gauge}), if one takes for the role of gauge
field the solution to the equations of motion in the form of plane wave
$A_M (X) \propto \epsilon_M \exp({ip\cdot X})$.}
\be
\label{gauge}
\int_{\d\Sigma} dt\left({\dot X}_M(t) A_M(X(t)) +
{e(t)\over 2}F_{MN}(X(t))\Psi_M(t)\Psi_N(t)\right)
\ee
(the ordered $P$-exponent in the non-Abelian case). The procedure here is the
same as for relativistic particle, one should only remember that an integration
in (\ref{gauge}) is taken over the boundary of the world-sheet $\d\Sigma$,
while in the case of a particle the integration was taken along the whole
world-line. This means that only the open strings interact with the vector
fields. In the closed string sector the situation is similar, and the action
is defined by the terms like (\ref{sigma}), where the interaction (and thus
the integration) is performed over the whole surface of the world-sheet.

In quadratic approximation the effective string actions {\em must} coincide
with quadratic terms in the Lagrangians of the corresponding field theories
for the background fields. The direct derivation of this correspondence is
impossible due to vanishing of the two-point correlators on the world-sheets
of the simplest topology (this is again a direct consequence of
two-dimensional geometry). An indirect argument in favor of such a
coincidence is self-consistency of the theory. Indeed, two-dimensional
conformal invariance requires that background fields satisfy equations of
motion, which in their turn would require the appropriate kinetic terms in
the effective Lagrangians. The higher terms in background fields and derivatives
in the effective actions follow straightforwardly from the calculation
of string amplitudes.

One of the most interesting (and one of the few computable) examples of the
non-local effective actions, arising for strings in the external gauge
fields is the Dirac-Born-Infeld action (in any even-dimesnional space-time)
\be
\label{dbi}
S_{DBI} = \int_{d^Dx} \sqrt{\det_{MN}(G_{MN} + 2\pi\alpha'F_{MN})}
\ee
It comes out directly from the calculation of the effective string action
for external electro-magnetic field, interacting with the string
world-sheets of the open strings having the simplest possible topology of a
disk \cite{FTDBI}.

This is a rather nontrivial fact -- all the corrections in $\alpha'$, or
loop corrections from the point of view of two-dimensional field theory (let
us recall here that from the point of view of string theory any computation
on disk counts only the "tree-level" contributions) sum up to the compact
formula (\ref{dbi}). This formula is really valid at large fields
$F_{MN}\sim \alpha'^{-1}$ of the order of string tension. The action
(\ref{dbi}) has supersymmetric and even non-Abelian analogs which are rather
interesting for the investigation of effective actions in nonperturbative
string theory.

In the closed string sector one gets an effective action for the Einstein
gravity
\be
\label{effgr}
\int_{d^Dx} \sqrt{G}e^{-2\Phi}\left(R(G) +{1\over 2}(\nabla\Phi)^2 + \dots
\right)
\ee
where $G\equiv\det_{MN} G_{MN}$, with the only difference that the scale or
normalization of "string" metric differs from the "scale" or normalization of
the Einstein metric by (exponent of the ) vacuum value of the
dilaton field $\Phi$. It leads in particular to the fact that the
"real" Newton constant or the Planck mass in ten-dimensional theory is
connected to the string tension via
\be
\label{10rel}
\gamma_{\rm N}^{(10)} = \left(M_{\rm pl}^{(10)}\right)^{-8} = g_{\rm str}^2\alpha'^4
\ee
where $g_{\rm str} = \langle\exp(\Phi)\rangle$. This relation will be
essentially used below in discussion of the nonperturbative string theory.

\setcounter{equation}0
\section{Strings without Strings. Non-perturbative Theory
\label{ss:nonperturb}}

\subsection{M-theory
\label{ss:Mtheory}}

Let us turn now to some achievements in string theory of the last ten years,
related mostly with the attempts to go beyond the perturbation theory. As we
already discussed in the context of quantum field theory, one immediately
looses any "solid background" since this is the field where there is no
reliable formalism. All possible statements can be based on a few
"semi-qualitative" considerations
\footnote{An exception can be found, if any, in the framework of so called
"discretized"  versions of quantum field theory, for example, in the so
called "lattice" theories which are beyond the scope of this review. Note
also that the progress in understanding of non-perturbative effects in
lattice gauge theories is seriously "screened" by additional problems of the
correspondence between the lattice theory and its continuum limit.}.
Nevertheless, these attempts can have some success and there still exists a
hope that they will be mostly successful in the framework of string theory.
This hope is based on the existence of certain deeply "hidden" symmetries which
may manifest themselves at nonperturbative level.

Note here that in contrast
to the widely spread opinion about the pure mathematical character of the
problems of string theory (which is not too far from being true if we
restrict ourselves to the string perturbation theory), the problems of nonperturbative
string theory have more fundamental and physical character. Let us repeat
that the main problem is that nonperturbative string theory (as well as
nonperturbative quantum field theory) does not exist in adequate physical
form, i.e. does not exist in the form of any reasonable formalism. What is
called at the moment nonperturbative string theory or M-theory
is just a set
of purely "philological" postulates reminding one, say, the "Butlerov theory",
well-known from the high-school course of organic chemistry .

The main hypothesis formulated at present in this or that way implies
existence of some unique nonperturbative string theory or M-theory
\cite{MtheoryT,MtheoryWi} (see also the reviews
\cite{TownsendMR}-\cite{LosevMR}) which has a large set of vacua understood
in the sense of perturbative string theory. In other words, the
perturbation theory around these vacua corresponds to
(different!) two-dimensional conformal field theories considered above,
interacting via
anomaly with two-dimensional gravity. The fact that different perturbative
expansions describe different phases of the same theory is encoded in the so
called duality -- not very well-defined and often only intuitively
understood similarity of certain objects from the different phases of
M-theory.

In the limiting case this means that there exist duality transformations,
relating different quantities in quantum field theories. These relations can
be established even between the quantities in absolutely different regimes,
for example the particle-like states in one theory may be related to the
soliton-like states in the dual one and vice versa. This is the reason why
such duality cannot in practice be verified by standard methods of quantum
field theory (except maybe in the two-dimensional theories, where,
for example the
well-known duality between the sine-Gordon and Thirring models exists). On the
other hand it allows to consider the well-known problems from an
absolutely new
perspective and sometimes leads to surprising new results.

The hypothetical properties of M-theory make it a little bit similar to the
field theory which contains together with "particle-like" states the
collective nontrivial excitations like solitons, monopoles etc. However, in
contrast to conventional quantum field theory, depending on the values of
parameters or moduli of M-theory (for example the vacuum condensates of the
scalar fields) the same observable objects (say electrically and/or
magnetically charged particles) may be described equally as elementary
and/or soliton-like particles with {\em different} field-theoretical
Lagrangians.

Speaking about M-theory we will still use the term "string theory"
despite the fact that in nonperturbative theory the very concept of
fundamental one-dimensional extended objects acquires much more "hidden"
form. In various considerations of M-theory a huge amount of hypersurfaces
of arbitrary dimension (or, better to say, of arbitrary co-dimension) take
part. From the naive point of view the one-dimensional extended objects are not
at all singled out among other, and strings are just particular case of so
called $p$-branes (number $p$ measures the dimension of brane). For example,
particle corresponds to $p=0$, string -- to $p=1$, the membrane from
which is derived the word {\em brane}), -- to $p=2$ and so forth.

However, the special role of strings is still caused by the fact that only
strings can pretend to be the fundamental objects. We cannot really add
anything here to the arguments of sect.~\ref{ss:ym-f-str}, with
the only difference being that now one should discuss separately the particular
domains of moduli space of nonperturbative theory. In different domains there
can exist (and do exist) different theories of fundamental strings. In such
situation the fundamental string of one of perturbative theories can be,
generally, the heavy "composite object" in another perturbative theory.
Moreover, only strings are naturally charged with respect to vector fields
which leads, on one hand, to the non-Abelian theories, and
on the other hand the gauge invariance of the theories of vector fields (and
gravity) allows opportunity for existence of {\em light} strings (more
strictly light excitations of strings) while light
membranes etc are absent.

The notion of duality, at least in the sense to be used below
has mostly stringy origin and is related to the properties of complex
manifolds often arising already in perturbative string theory. In
perturbative string theory these properties belong to "unobservable"
geometry of world-sheets, but, quite unexpectedly, analogous properties
arise in the context of complex manifolds, being the "auxiliary" nontrivial
part of the multidimensional space-time. The dualism between the structures
on world-sheets and in target-space is rather surprising and not yet
well-studied phenomenon in string theory, a manifestation of this intrinsic
connection -- the relation between world-sheet and space-time
supersymmetries was already discussed in sect.~\ref{ss:susy}. The simplest
example of duality between anomaly free string models -- the so called
T-duality relating IIA and IIB superstring theories -- is a direct
consequence of the famous $R\leftrightarrow {\alpha'\over R}$ duality, to be
considered in detail in sect.~\ref{ss:r1r}. Other duality transformations
typically relate to each other two theories with at least one of them being
in strong coupling phase. Thus, their verification is an absolutely nontrivial
problem.

Let us now try to list the main {\em postulates} of M-theory:
\begin{itemize}
\item {\bf M-theory and eleven-dimensional supergravity}. The low-energy
limit of M-theory is supergravity in a space-time of $D=11$ dimensions
\footnote{Dimension $D=11$ is singled out (by a slightly strained arguments)
already directly from geometric interpretation of the Standard Model with
the gauge group $U(1)\times SU(2)\times SU(3)$
(see, for example \cite{Pol}, page 275). If we consider that the group of
Standard Model naturally acts on some manifold of compactification, then the
natural dimension of such manifold can be determined as a sum of unity for the
$U(1)$-factor, two ($\dim ({\bf S}^2)=\dim ({\bf CP}^1)=2$) for the
$SU(2)$-factor and four ($\dim ({\bf CP}^2)=4$, if it is implied that group
acts on complex manifold) for the $SU(3)$-factor. Together with four "visible"
dimensions this gives $D=1+2+4+4=11$.} \cite{11SUGRA}. This is the maximal
possible supergravity and, thus, maybe the only distinguished and nice
theory from all supergravity models. Its bosonic sector contains only the
metric $G_{MN}$ and antisymmetric tensor field (the 3-form) $C_{MNK}$. The
only (dimensional) parameter in this theory -- the eleven-dimensional Planck
mass $M_{\rm pl}\equiv M_{\rm pl}^{(11)}$. Under dimensional reduction of
eleven-dimensional supergravity one gets the ten-dimensional supergravity of
the type IIA -- the field theory limit of IIA string theory. This leads to
the relation between the square of string length (or inverse string tension)
$\alpha'$, radius of the compact dimension $R$ and eleven-dimensional Planck
mass, which reads
\be
\label{mpl11}
\alpha' R M_{\rm pl}^3 = 1
\ee
The relation between ten-dimensional and eleven-dimensional Planck masses
\be
\label{relM}
M_{\rm pl}^9R = \left(M_{\rm pl}^{(10)}\right)^8 =
{1\over g_{\rm str}^2\alpha'^4}
\ee
can be obtained directly from the reduction of the Einstein action
of supergravity (the first equality in formula (\ref{relM})). The connection
between ten-dimensional Planck mass and string coupling constant (the second
equality in (\ref{relM})) is a consequence of the difference between the
string and gravitational "definitions" of metric, differing by
$\langle \exp({-2\Phi})\rangle= 1/g_{\rm str}^2$, where $\Phi$ is the
dilaton field (see~(\ref{effgr}), (\ref{10rel})). Altogether this leads to
the equality
\be
\label{gR}
R = g_{\rm str}\sqrt{\alpha'}
\ee
demonstrating that with the growth of the string coupling constant
$g_{\rm str}$ the radius of a hidden compact dimension $R$ {\em blows up}.
This leads to a possible interpretation of M-theory as a string theory in a
strong coupling regime.

\item {\bf M-theory as type IIA string theory at strong coupling}. M-theory
is {\em not} a theory of fundamental strings in the sense of
sect.~\ref{ss:strpol}, already because there are no anomaly free
perturbative string theories with the space-time dimension $D=11$.
Nevertheless, the arguments presented above allow one to consider M-theory
as a
type IIA string theory at strong coupling, where the extra compact dimension
shows up and the size of this dimension is related to the strong coupling
constant by eq.~(\ref{gR}).

\item {\bf Strings and extended objects in M-theory}. The analysis of
extended objects being solutions to the equations of motion in M-theory (in
reality -- the equations of motion of eleven-dimensional supergravity) and
their dimensional reduction to $D=10$ leads to rather natural parallels
between branes in M-theory and branes in string theory. Say, the
hypothetical membrane of M-theory winding along the compact dimension
becomes a string. One more similar relation will be discussed below in
sect.~\ref{ss:dbr} when we are going to discuss the exact nonperturbative
results in supersymmetric gauge theories. It turns out that it is M-theory's
$5$-brane which plays the main role in geometric formulation of these results.

\end{itemize}

As well as ten-dimensional perturbative string theory, the
eleven-dimensional M-theory may manifest itself in a four-dimensional world
only after some "compactification". One of the differences between
perturbative and non-perturbative theories in this context is that presence
of the extended objects leads after compactification to some new nontrivial
effects. A remarkable property of supersymmetry is its relation to the
complex geometry of (especially nontrivial part of) space-time. It is
reflected in the fact that the nontrivial complex manifolds of string
compactification correspond to effective supersymmetric quantum field
theories. Parameters of such theories (coupling constants, vacuum
condensates, masses etc) are parameters or moduli of the complex
manifolds of the corresponding string compactification, for example of the
Calabi-Yau manifolds \cite{GSW}. The duality transformations in this case
can be identified with action of the corresponding modular group.

\begin{footnotesize}
In order to get the macroscopic four-dimensional gauge theory, one should find
some four-dimensional reduction. There is a standard way in string theory
coming back to the old Kaluza-Klein idea: the full space time can be
presented as a direct product of four-dimensional Euclidean space and some
complex manifold $K$. The "internal" space $K$ determines the "color"
properties of the theory, the number of four-dimensional supercharges etc.
Supersymmetry requires the compact manifold $K$ to be the three-dimensional
complex manifold in the ten-dimensional picture (or, say, to be the product of
the three-dimensional complex manifold with a circle from the
eleven-dimensional point of view).

Moreover, it turns out that sometimes the nontrivial part of this
three-dimensional complex (or six-dimensional real) manifold can be
presented by a one-dimensional complex curve (or two-dimensional
Riemann surface $\Sigma $). Starting from eleven-dimensional M-theory
one should choose some particular compactification scheme down to four
dimensions, such that the resulting theory would get an appropriate
four-dimensional supersymmetry, the required gauge group (in majority of real
situations $SU(N)$) and an appropriate set of matter multiplets. According to
\cite{WittM97}, there exists a compactification scenario when the complex
geometry can be formulated in terms of Riemann surfaces and this scenario
leads exactly to the Seiberg-Witten effective theories \cite{SW}.

It is the (complex) analytic structure which distinguishes a class of
theories where the exact nonperturbative results can be formulated. These
results are formulated using the technique of holomorphic (meromorphic)
functions. The idea to use holomorphic objects goes back to the application
of complex analysis to the theory of instantons and the Belavin-Knizhnik
theorem \cite{BKni,Kni} of perturbative string theory, see
sect.~\ref{ss:strpol}. In the simplest class of problems under
discussion the moduli of physical theories may be identified exactly with
the moduli of {\em one-dimensional} complex manifolds -- the (space-time!)
complex curves or Riemann surfaces $\Sigma $, which {\em a priori} have no
relation to the world-sheets of string theory. However, to study these
objects one may successfully use the same technical tools which were used
when studying the perturbative string theory (see sect.~\ref{ss:string}).
An analogous picture may be expected for the theories where physical moduli
spaces are identified with the moduli spaces of higher-dimensional complex
manifolds (two-dimensional complex manifolds $K3$, Calabi-Yau three-folds
etc, see details, e.g. in \cite{GSW}). Moreover, there exists a unifying picture
of string compactification which implies that complex curves can be
considered as degenerate cases of more general compactification manifolds,
for example when the Calabi-Yau manifold effectively degenerates into
one-dimensional complex curve $\Sigma $ \cite{VafaC}. A nontrivial
topological structure of the curve $\Sigma $ is essentially nonperturbative
information, since in the perturbation theory this curve arises only "locally" as
a scale parameter. This means, in particular, that the string effects play
an essential role in the structure of the exact nonperturbative solutions of
gauge theories and the topological degrees of freedom, playing a
decisive role
for the construction of an effective theory, are directly related to
"windings" of strings around nontrivial cycles in the manifolds of string
compactifications.
\end{footnotesize}

\subsection{Strings in Compact Dimensions
\label{ss:r1r}}

In the brightest form the difference between string theory and quantum
field theory appears in the case of topologically non-trivial space-time,
and the simplest example of such space-time is the space-time with compact
dimensions or just a "box" with periodic boundary conditions at the ends.
The structure of such "compactified" string theories implies the
existence of a very nontrivial symmetry (duality) relating different string
models. In particular, these models can be related in such
a way that the perturbative regime in one of the models allows to propose
some reasonable hypothesis about nonperturbative effects in another. In
other words, duality transformations allow to consider string models as
perturbative expansions (\ref{ppi}) as expansions around different vacua of
the same theory. The only weak point (at present) of this concept is the
absence of any reliable or strict statements in the mathematical sense.

The main example of duality is symmetry in the theory of {\em closed}
strings in a space-time with compact dimensions (in the simplest case --
with the only co-ordinate taking values on some "circle"
$\phi \sim \phi + 2\pi Rn$, with $n\in \bf Z$ being any integer). The
spectrum of such theory and one-loop partition function are invariant under
discrete transformation $R \leftrightarrow {\alpha '\over R}$
\cite{gauss}. This invariance follows from the fact that in addition to
the standard discrete spectrum of particle on a circle with the quantized
momentum $p \propto {n\over R}$, $n\in{\bf Z}$
(existing certainly as well in ordinary quantum field theory with compact
dimensions) there exists also another type of string excitations: a
string can wind around a circle and the energy of such winding mode is
${mR\over\alpha '}$, also with $m\in{\bf Z}$.

In the "decompactification" limit $R\rightarrow\infty $, the first part of
spectrum will become continuous (again, as in ordinary quantum field
theory), while the string winding excitations would become infinitely heavy
and their contribution to the partition function can be neglected. However,
the full spectrum
\be\label{egauss}
M^2_{n,m} = \left({n\over R}\right)^2 + \left({mR\over\alpha '}\right)^2
\ \ \ \ \ \ \forall\ n,m
\ee
is obviously {\em invariant} under the change
$R \leftrightarrow {\alpha '\over R}$.

The presence of the second term, or the spectrum of string winding modes in
eq.~(\ref{egauss}) is sometimes interpreted as stringy modification of the
uncertainty principle. Indeed, expression (\ref{egauss}) allows to think
that the uncertainty principle $\Delta X \sim {1\over E}$ is valid literally
up to scales of the order of $\sqrt{\alpha'}$, while beyond this scale
the formula should rather be replaced by something like
$\Delta X \sim {1\over E} + \alpha'E$.

It is relatively easy to see that the duality transformation
$R \rightarrow {\alpha '\over R}$ leaves invariant the holomorphic
quantities, say the current $\d\phi_L(z) \rightarrow \d\phi_L(z)$, but
changes the sign of the anti-holomorpic ones:
$\bar\d\phi_R(\bar z) \rightarrow - \bar\d\phi_R(\bar z)$.
It means, for example, that the operators of emission and absorbtion of
"particles" of the form
\be
V_p \propto \exp \left(ip\phi(z,\bar z)\right) =
\exp \left(ip\phi_L(z) + ip\phi_R(\bar z)\right)
\ee
become {\em non-local} (from the point of view of the field
$\phi(z,\bar z) = \phi_L(z)+\phi_R(\bar z)$) operators of the world-sheet
"vortices"
\be
\exp \left(ip\phi_L(z) - ip\phi_R(\bar z)\right)
\ee
and vice versa.

The same is true for the action of the duality transformations
$R \rightarrow {\alpha '\over R}$ on the holomorphic and/or
anti-holomorphic (on the equations of motion) world-sheet fermions:
$\Psi_L(z) \rightarrow \Psi_L(z)$, but, at the same time
$\Psi_R(\bar z) \rightarrow - \Psi_R(\bar z)$. This immediately leads to
nontrivial consequences for the type II superstrings in ten-dimensional
space-time ${\bf R}^9\times {\bf S}^1$ with one compact dimension. One can
forget for a moment about the nine non-compact co-ordinates and consider what
happens in such a theory under the transformation
$R \rightarrow {\alpha '\over R}$.

In the bosonic sector the winding modes still replace the Kaluza-Klein modes
and vice versa, but the components of the two-dimensional fermionic fields
$\Psi$ along the compact direction
corresponding to the left- and right- movers behave differently: one
preserves the sign while the other one changes it. It
follows then that the $"\Gamma_5"$-matrix, and therefore the operator
of chirality projection changes sign only in one of the sectors. Hence,
the non-chiral IIA theory under the transformation
$R \rightarrow {\alpha '\over R}$ turns into the chiral IIB theory and vice
versa. The transformation $R \rightarrow {\alpha '\over R}$ in
multidimensional space-time with a single compact direction, exchanging the
type IIA and type IIB theories is usually called T-duality.
This is the only
duality of string theory which can really be verified, since it relates
the theories, which can be both considered at weak
coupling. In a similar way T-duality relates
the heterotic
string models with the gauge groups $SO(32)$ and $E_8\times E_8$.

Now, if we consider an effective action for string theory, say, in
$D+1$ dimensions and reduce it to $D$ dimensions, the size of the
compact dimension arises as factor in front of the ($D$-dimensional) action,
and can be further interpreted as a coupling constant. It allows one to turn
$R \leftrightarrow {\alpha '\over R}$-duality into relation between the
effective theories such that one of these theories is at strong coupling
while the other is weakly coupled. As a result of such reasoning one gets
a {\em hypothesis} that some quantum field theory on a given manifold and at
weak coupling is equivalent
\footnote{In the above sense. Such equivalence usually implies (partial)
coincidence of spectra and {\em certain} correlation functions in dual
theories.} to a different theory, generally on a different manifold and in
the strong-coupling regime. It is quite surprising that applying this sort
of arguments to particular supersymmetric gauge theories, it is possible
sometimes to make explicit predictions about the exact spectra and exact form
of low-energy effective actions.

\begin{footnotesize}
To finish this section let us stop once more at so called
"mirror symmetry" in string theory
\cite{mirrorVafa,mirrorWitten,mirrorAspinwall,mirrorMorrison}. We are not
going to discuss the mathematical issues of this problem, related to the
fact that string theory allows one hypothetically to establish certain relations
between the complex and K\"ahler structures of some manifolds. For us it is
more important that string theory in principle possesses the possibility of
"non-distinction" of the space time, in the sense that for given string
model the space-time may not be determined uniquely. The simplest example of
such phenomenon is discussed above -- string models on the circles
with the radii $R$ and $\alpha'/R$ coincide at least at the
level of spectrum
\footnote{Let us recall once more that identifying different string models
by duality transformations one should strictly fix what is exactly
identified and in what sense. Typically only the spectrum and {\em some}
correlation functions are borne in mind.}. Passing from circles to
tori it is easy to see that the same symmetry is preserved. Under such
process the type A theory on two-dimensional torus
${\bf T}={\bf S}^1_{R_1}\times {\bf S}^1_{R_2}$ would become equivalent to
the type B theory on the torus $\tilde{\bf T}=
{\bf S}^1_{R_1}\times {\bf S}^1_{1/R_2}$ and vice versa.
Notice now that the area of the torus
${\rm Area}({\bf T})=R_1R_2$ and the modulus of complex structure
$\tau({\bf T})=iR_1/R_2$ are up to imaginary unity, in different order,
correspondingly the
modulus $\tau (\tilde{\bf T})$ and area ${\rm Area}(\tilde{\bf T})$ of the
"mirror torus" $\tilde{\bf T}$. Thus, we come to the
statement of "mirror symmetry" about the equivalence of the A and B theories on
mirror manifolds -- the manifolds for which the moduli of complex and
K\"ahler structures replace each other.

The physical nature of the mirror symmetry is rather transparent, though it
contains a paradox at first glance. Replacement of momentum by the energy of
the winding mode roughly speaking corresponds to the replacement of momentum by
co-ordinate, and therefore the mirror symmetry is in some sense the symmetry
between co-ordinates and momenta. It is clear that our world does not have
such a symmetry, since we can always single out the space of co-ordinates or
configuration space and the phase space is its cotangent bundle.

Hence, what should we do with mirror symmetry in string theory? The resolution of
this puzzle is in the simple fact that such symmetry is possible only at the
scales of order of $\sqrt{\alpha'}$, for example from dimensional
requirement $p\leftrightarrow x/\alpha'$. Therefore, the mirror manifolds
identified by string theory are in principle unobservable in the
"macroworld"!
Moreover, at such scales the phase space may not necessarily be a cotangent
bundle. Say, the quantum mechanics of spin, formulated in adequate terms
(see, for example, \cite{NiRo,AFS}), corresponds to the phase space, having
configuration of sphere, which is not at all a cotangent bundle. Another,
maybe even more simple example from quantum mechanics is a particle in
magnetic field. In this example there is a "natural replacement" of the
configuration plane transversal to the direction of magnetic field by the
"phase plane" on the distances of the order of magnetic length
$l \sim \sqrt{\hbar c/ eB}$.
\end{footnotesize}

\subsection{Dimensional Reduction in String Theory and D-branes
\label{ss:reduct}}

Formula (\ref{egauss}) leads to rather nontrivial conclusions about
dimensional reduction in string theory. In field theory or the theory of
pointlike particles the second term on the right hand size of
(\ref{egauss}), proportional to $(\alpha')^{-2}$, can be omitted and we
obtain
the conventional Kaluza-Klein spectrum. For the compactified quantum field
theory it means that reducing the field theory from $D$ to $(D-1)$
dimensions via the compactification of one dimension onto the
circle of radius $R$ with further limit $R\to 0$,
the $D$-dimensional field can be conveniently written in terms of the
Fourier series (not the Fourier integral) with respect to compact co-ordinate
$x_0$
\be
\phi (x,x_0) = \sum_n \exp\left({i\pi n{x_0\over R}}\right)\phi_n(x)
\ee
After substitution of this expansion into the action
\be
\int_{d^{D-1}x} dx_0 \sum_{M=1}^D(\d_M\phi)^2 = \int_{d^{D-1}x} \sum_n
\left(\sum_{\mu=1}^{D-1}\d_\mu\phi_n\d_\mu\phi_{-n} + {n^2\over R^2}
\phi_n\phi_{-n}\right)
\ee
one gets the sum over $(D-1)$-dimensional fields $\phi_n(x)$ with the
masses, exactly corresponding to the first term in (\ref{egauss}).
At $R\to 0$ all fields with $n\neq 0$ become infinitely heavy and at
distances much more than $R$ one may forget about them. Thus, after
compactification and dimensional reduction we obtain from
$D$-dimensional field theory the field theory in $(D-1)$ dimensions.

This rather natural conclusion remains intact even in the case of the open
string theory, where the nontrivial winding modes corresponding to the
second term in eq.~(\ref{egauss}) are absent. However, for the theory of
a closed string one comes to a different conclusion. In the limit
$R\to 0$ the
Kaluza-Klein modes with the masses ${n/ R}$ still would become
infinitely heavy, i.e. inessential for the limiting spectrum, but, in
contrast to them, masses of all states corresponding to windings vanish!
This
means that at such reduction from $D$ dimensions to $(D-1)$ dimensions,
the Kaluza-Klein "tower" corresponding to an extra dimension disappears as
in field theory ... but in the same procedure the equivalent "tower of
fields" reappears due to the light at $R\to 0$ modes of the closed
string winding around the compact direction. Thus, the extra tower of fields remains
in the spectrum of closed string, i.e. no reduction
to $(D-1)$ dimensions really happened and the theory remains
$D$-dimensional!

Now, consider the same procedure in the theory with both closed and open
strings. The conclusion is a bit of paradox: as $R\to 0$ closed strings
would be still propagating in $D$-dimensional space-time, while the theory
of open strings will be $(D-1)$-dimensional. Alternatively, if we require
consistency and "smooth" behavior of string theory under the change of
parameter or moduli $R$ -- the size of a compactified dimension, one has to
allow the existence of absolutely new nontrivial vacua, containing certain
distinguished hypersurfaces (the example considered above contained a
hypersurface of unit codimension, however, it is easy to see that
compactifying several dimensions the codimension can be made arbitrary).
These hypersurfaces are characterized by the fact that only there the open
strings can keep their ends. In modern terminology such hypersurfaces are
called the Dirichlet or D-branes, and the volume between branes
is called the bulk.

Let us now list the main properties of D-branes, essential for the study of
nonperturbative string theory:

\begin{itemize}

\item Since vector fields arise in the open string sector (see
sect.~\ref{ss:ym-f-str}), in the theory (or, better to say in the vacuum)
with D-branes the vector fields are {\em localized} on the D-brane's
hypersurfaces. Hence, D-branes proposed a new, purely string mechanism of the
localization of vector fields, which is absent in quantum field theory.
Notice also, that the theory with open strings in all $D$-dimensional
space-time can be interpreted as a vacuum with the Dirichlet brane (or
several Dirichlet branes in the case of nontrivial Chan-Paton factors) of
dimension $p=D-1$, see fig.~\ref{fi:dbrane} and sect.~\ref{ss:dbr}).

\item In the theories with space-time supersymmetry D-branes are the BPS
states, invariant under the action of half of the supersymmetry generators.
This is due the fact that in the open string sector there are twice fewer
supersymmetry generators than in the closed sector, since the fields on the
boundary of the world-sheet are constrained by the boundary conditions. The
BPS nature of D-branes is also related directly to the fact that they are
charged with respect to antisymmetric tensor fields of the Ramond-Ramond
sector. Namely, the D$p$-brane is charged with respect to the $(p+1)$-form,
which can be integrated over the world-volume of the D$p$-brane as
$\int C^{(p+1)}$, and the corresponding charge arises as a central extension
of the supersymmetry algebra. This central extension breaks, however,
the $D$-dimensional Lorentz-invariance as well as the very existence of the
hypersurface of D-brane.

\item The D-brane tension is proportional to the {\em first} power of
the string coupling constant. One of the arguments supporting this relation is
interaction of D-brane with the {\em open} strings, whose perturbation
theory contains the expansion in $g_{\rm str}$, and not in
$g_{\rm str}^2$, see sect.~\ref{ss:strpol}. This distinguishes D-branes from
so called solitonic branes, interacting only with closed strings. The
corresponding effective action of the background fields
(see sect.~\ref{ss:fratse}) can be roughly written as
\be
\label{Deffa}
\int_{d^Dx}\sqrt{G}\left( e^{-2\Phi}(R(G) - H^2) - (dC)^2\right)
\ee
where $\Phi$ is the dilaton, $\langle \exp({-2\Phi})\rangle = g_{\rm str}^{-2}$;
$R(G)$ is the curvature of $D$-dimensional metric,
$G\equiv\det_{MN} G_{MN}$, $H=dB$ is the field-strength of antisymmetric tensor
field, related to the solitonic branes while $dC$ is the field-strength
of the Ramond-Ramond $(p+1)$-forms. It is the different dependence on
dilaton of the terms $(dB)^2$ and $(dC)^2$ in eq.~(\ref{Deffa}) that leads
to the fact that the "thickness" of the solitonic brane does {\em not}
depend on $g_{\rm str}$
(for constant dilaton equations obtained from variation of the terms
$\sqrt{G}(R(G) - H^2)$ in formula (\ref{Deffa}) and their solutions do not
depend on $g_{\rm str}$), and its mass or tension is proportional to
$g_{\rm str}^{-2}$, while the "thickness" of the D-brane (solution to the
equations following from variation of the terms
$\sqrt{G}\left(e^{-2\Phi}R(G) - (dC)^2\right)$ in (\ref{Deffa}))
is proportional to $g_{\rm str}$, and its tension is proportional to
$g_{\rm str}^{-1}$. This means that at weak coupling D-brane can indeed be
considered as a very thin hypersurface "glued" to the ends of the
open strings.

\end{itemize}

Note, that due to the absence of "normal" nonperturbative theory these
properties are established only with the help of certain mostly qualitative
arguments (see, for example, \cite{Polch,PolchDbr}). In what follows we will
restrict ourselves to a "minimal use" of these properties, i.e. we will use
them only where the D-brane picture leads to more or less clear physical
consequences.

\subsection{D-branes and non-Abelian Gauge Fields
\label{ss:dbr}}

Let us now discuss in detail how the (four-dimensional) supersymmetric gauge
theories arise in the context of string theory.
One should start with any supersymmetric string theory without anomalies.
There exist several examples of such
theories (defined originally as {\em perturbative} expansions in terms
of the path integrals (\ref{ppi})) and their basic feature is that they live
in $D=10$ and have at least \1N ten-dimensional space-time supersymmetry
(see the end of sect.~\ref{ss:susy}).

\begin{figure}[tp]
\epsfysize=8cm
\centerline{\epsfbox{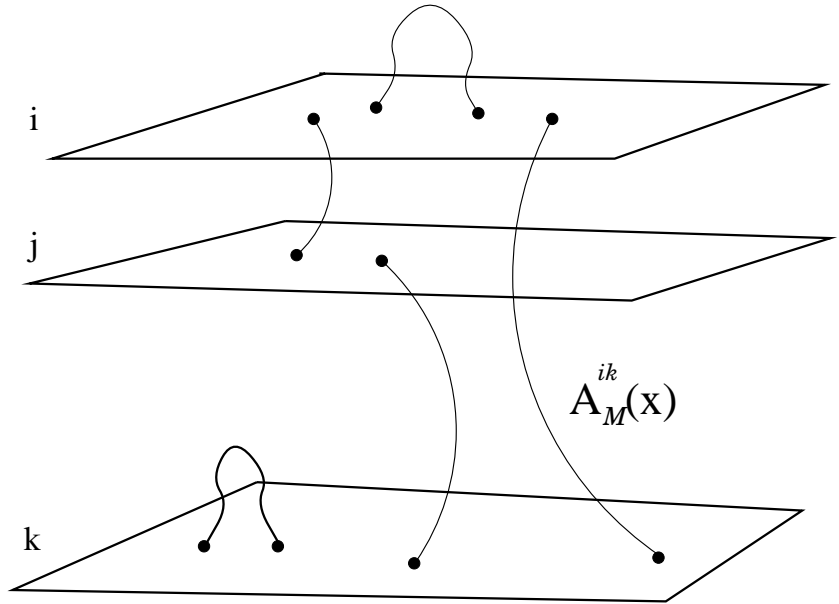}}
\caption{\sl
D-branes. The interaction is carried by strings attached by
their ends to different D-branes or parts of the same D-brane. In the
background of several D-branes one naturally gets non-Abelian vector
fields in the spectrum of strings since the fields become labeled by
the numbers of D-branes they are attached to.}
\label{fi:dbrane}
\end{figure}

One of the main ingredients of the relation between strings and
gauge theories are the above mentioned appearence of the
D-brane configurations in non-perturbative string theory
\cite{duadbr,WittD}. D-branes are classical ("heavy") objects which
can be thought of as certain hypersurfaces in a target space and whose basic
feature is the possibility of interaction via emission and absorption of open
strings (see fig.~\ref{fi:dbrane}) -- even in the theories with no bulk
open string interactions (for simplicity we will restrict ourselves only to
such theories, called as type II theories, see sect.~\ref{ss:susy}).
As we already discussed in sect.~\ref{ss:reduct}, such hypersurfaces
naturally arise in compactified string theory, implying that it behaves
"smoothly" under the change of parameters of the compact manifold.

It is easy to see that the configuration of $N$ parallel D-branes on
fig.~\ref{fi:dbrane} leads naturally to the $SU(N)$ gauge group
(more strictly to the group $U(N)=SU(N)\otimes U(1)$ with
inessential for the fields in the adjoint
representation $U(1)$ factor),
broken down generally to $U(1)^{N-1}$.
Indeed, consider $N$ parallel D-branes, then the (oriented) open
string stretched between the $i$-th and $j$-th brane
($i,j=1,\dots,N$) (see fig.~\ref{fi:dbrane}) contains a vector field
${\bf A}^{ij}$ in its spectrum. The mass of this vector field is
proportional to the length of the string (since the
energy or mass of a string is
proportional to its length), i.e. to the distance between the
$i$-th and $j$-th branes.

Thus, the $U(1)^{N-1}$ massless gauge fields will come out of the
strings with both ends glued to the same D-brane, while
the fields ${\bf A}^{ij}$ with $i\neq j$
will acquire the "Higgs" masses, proportional to the vacuum condensates of
scalar fields (more strictly to the differences of these condensates for the
corresponding components). These vacuum values are determined by
the "transverse" co-ordinates of the D-brane
$\phi \sim{\sqrt{{\vec x}_{\bot}^2}\over\alpha '}$. Thus if the open strings
themselves naturally lead to the appearance of massless vector gauge fields,
the open strings in D-brane vacua rather naturally correspond to the
theories with (in general broken) non-Abelian gauge symmetry
\footnote{Let us recall that before this fact was understood, non-Abelian
gauge theories were constructed "by hand", "gluing" quarks to the ends of
open strings (see fig.~\ref{fi:string-gauge}), or introducing the
non-Abelian Chan-Paton factors \cite{ChaPa} directly into string
amplitudes.}.

The next step is -- again looking at fig.~\ref{fi:dbrane} -- to see how from
ten-dimensional string theory one gets for such a configuration a theory in
a much fewer number of dimensions (an ideal result would be to get
four-dimensional theory). Indeed, it is easy to understand that the gauge
theory "localizes" to the D-brane world-volume, i.e. the real number of vector
indices is equal to the dimension of this world-volume. The D-brane
hypersurface breaks full ten-dimensional Lorentz-invariance, therefore only
the components corresponding to the directions "along" the world-volume form
a real vector. The rest of the components, from the point of view of
unbroken space-time theory on the D-brane world volume look like set of
scalars, what is in complete analogy with the dimensional reduction of the
theory of a vector field (see, for example \cite{Scherk}).

\begin{figure}
\epsfysize=8cm
\centerline{\epsfbox{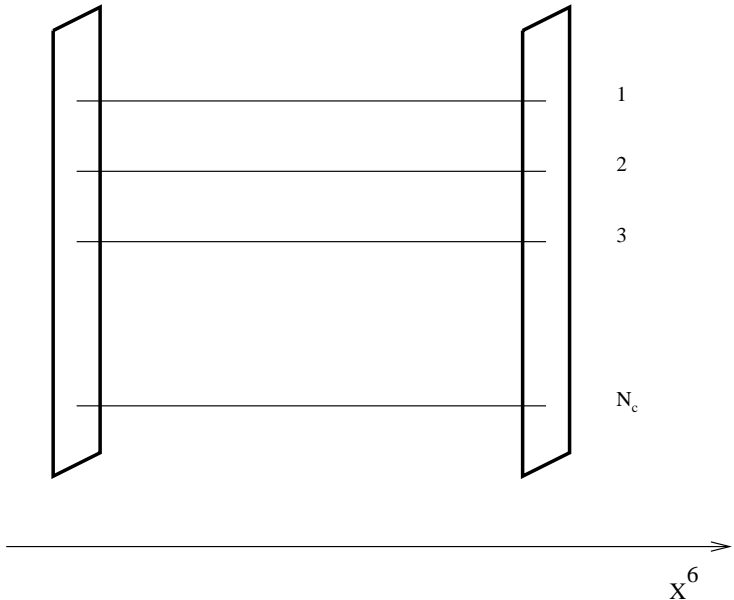}}
\caption{$4$-branes restricted by $5$-branes to a finite volume
(in the horizontal $x^6$-direction) give rise to macroscopically 4-dimensional
theory.}
\label{fi:walls}
\end{figure}

The Dirichlet $p$-brane world-volume
\footnote{To avoid misunderstanding let us again point out the
accepted terminology. D-brane is short for "Dirichlet brane"
and has no relation with the
dimension of this hypersurface, which is conventionally noted by the letter
$p$. Sometimes even the notation D$p$-brane is used, i.e. the
$p$-dimensional Dirichlet brane with the world-volume of dimension $(p+1)$.
Let us repeat once more that $p=2$ corresponds to a membrane (the origin of
the word "brane"), one would often meet in the literature, D1-branes, or
D-strings, D0-branes or Dirichlet particles or even D(-1)-branes or
D-instantons, as well as branes of dimensions $2<p\leq D-1$, where in the
last inequality $D$ means already the dimension of space-time and does not
come from the word Dirichlet.}
has dimension $p+1$ (including time!), i.e. naively in order to get
four-dimensional gauge theory one should consider parallel
D3-branes. This scenario is quite possible but gives rise to \4N supersymmetry
in four dimensions; in order
to get less trivial \N2 (or even \1N) theory it is better to use
another option, the Diaconescu-Hanany-Witten "ladder" brane configuration
\cite{DHW,WittM97} with $N$ parallel
D4-branes stretched between two vertical walls (see fig.~\ref{fi:walls}),
so that naive
five-dimensional D$4$ world-volume theory becomes macroscopically (in the light
sector) four-dimensional by the famous  Kaluza-Klein argument for a system
compactified on a circle or put into a box. Certainly there are many other
constructions based on discrete symmetries, orientifolds etc, however the
"brane zoology" is beyond the scope of this review (see, for example
\cite{branezoo}) and we will discuss only the
simplest "ladder" example, especially since it is this example that
corresponds to one of the strongest statements about non-perturbative
supersymmetric gauge theories.

The role of vertical walls should be, best of all, played by $5$-branes
\cite{WittM97}, then dimensional arguments lead to the logarithmic
behavior of the macroscopic coupling constant ${1\over g^2} \sim\log\mu$
(cf. with formula (\ref{rg})). In the leading approximation this comes up
since the corresponding "compact" co-ordinate, which turns into a coefficient
in front of the action (\ref{langym}), has logarithmic
behavior as a function of "transverse" directions, i.e. satisfies the
{\em two}-dimensional Laplace equation, where the effective two dimensions
are formed by the ends of D$4$-branes in $5$-branes. More generally the fact
that the logarithm (of the complex argument) is the Green function of the
two-dimensional Laplace operator is one of the "foundations" for the D-brane
constructions of supersymmetric gauge theories.

\begin{figure}
\epsfysize=10cm
\centerline{\epsfbox{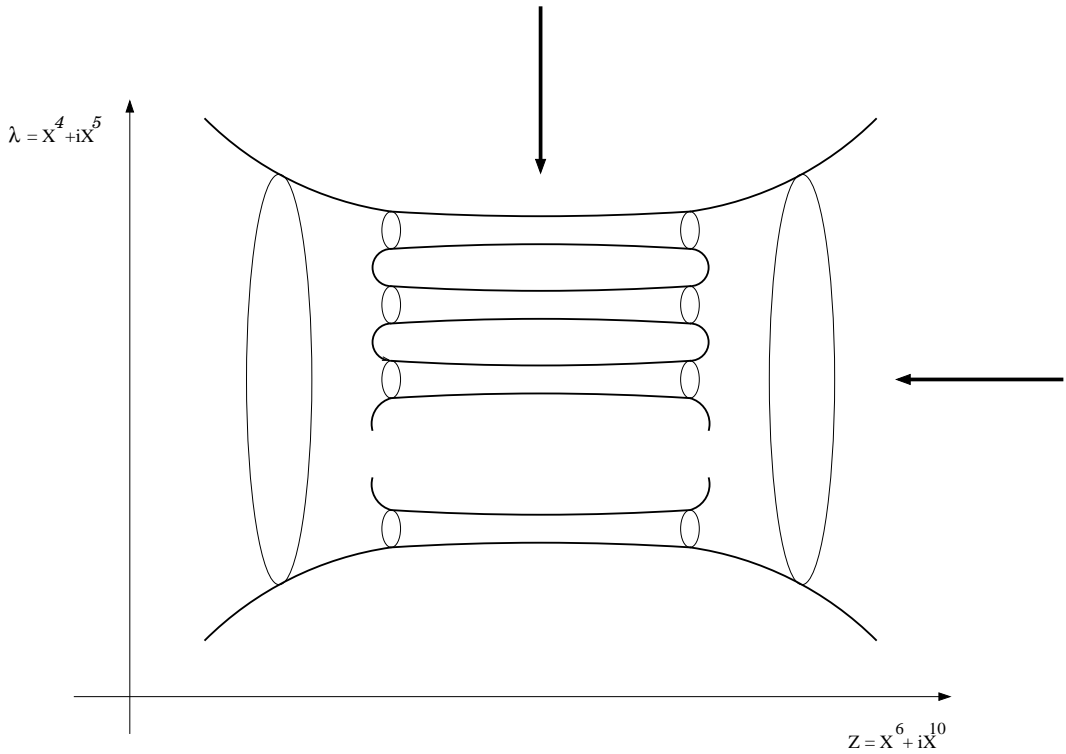}}
\caption{Brane configuration, represented as a result of
"resolution" the previous picture --
the "thin" ladder turns into a "swedish ladder" -- the hyperelliptic
Riemann surface being at the same time
$N$-fold covering of the horizontal cylinder.}
\label{fi:todacu}
\end{figure}

This picture of $4$- and $5$- branes in ten dimensions is certainly
very rough and true only in a (quasi)\-classical approximation. In particular it
is naively singular at the points where 4-branes meet 5-branes. These
singularities were resolved in a nice way in \cite{WittM97} where
it was proposed to
"raise" the whole picture into an eleven-dimensional target space of M-theory
and to consider D$4$-branes as $M$-theory $5$-branes compactified onto
eleventh dimension with $x^{10}$ being the corresponding extra compact
co-ordinate.
Then the picture in fig.~\ref{fi:walls} turns into the surface of
a "swedish ladder" and apart from macroscopic directions
$x^0,\dots,x^3$ looks like a (non-compact) Riemann surface with rather
special properties (see Fig.~\ref{fi:todacu}).

In other words, as a result of "resolution of singularities
one gets a unique smooth $5$-brane, which
leaving aside four flat dimensions ($x^0,x^1,x^2,x^3$)
looks like $N$ cylinders $R\times S^1$ embedded
into the target space along, say, $(x^6,x^{10})$ dimensions
(and which can be parameterized by complex co-ordinate $z = x^6+ix^{10}$).
The cylinders are separated in the "orthogonal" space
$V^{\bot} = (x^4,x^5,x^7,x^8)$, but they are all glued
together (see fig.~\ref{fi:todacu}) by vertical walls,
and the "effective" two-dimensional subspace of $V^{\bot}$
can be described by the complex coordinate $\lambda = x^4 +ix^5$.

\begin{footnotesize}
Let us try to establish the relation between the brane configurations and
complex manifolds. The simplest way to describe a nontrivial complex manifold
is analytic, i.e. by certain (polynomial) equations in multidimensional
complex space ${\bf C}^n$. Let us demonstrate now how the pictures in
fig.~\ref{fi:dbrane} and fig.~\ref{fi:todacu} can be rewritten in terms of
algebraic equations on complex variables.

Introducing the co-ordinate $w = \exp({z})$ to describe a cylinder,
we see that the system of non-interacting branes (fig.~\ref{fi:dbrane}) is
given by the $z$-independent equation
\be
P_{N}(\lambda) = \prod_{i = 1}^{N}
(\lambda - \phi_i) = 0,
\label{nibr}
\ee
while their bound state (fig.~\ref{fi:todacu}) is described by a complex
curve $\Sigma $ (a single equation on two complex variables)
\be
w + \frac{\Lambda^{2N}}{w} = P_{N}(\lambda)
\label{todacur}
\ee
In the weak-coupling limit  $\Lambda \rightarrow 0$
(i.e. $1/g^2 \sim  \log\left(1/\Lambda\right) \rightarrow \infty$)
one comes back to the set of disjoint branes (\ref{nibr}).
Eq.~(\ref{todacur}) presents an
analytic formulation of fig.~\ref{fi:todacu} --
$5$-brane of topology $R^3\times\Sigma$ embedded {\em holomorphically}
into a subspace $R^5\times S^1$ (say, spanned by $x^1,...,x^6,x^{10}$)
of the full space-time.

A somewhat more transparent way to get the same equations is related to
the theory of integrable systems \cite{MMaM} and uses the fact that
in vacuum state the scalar fields satisfy
the BPS-like condition -- the first-order equation (cf. with (\ref{bpseq}))
\be\label{bps-cond}
D_M\Phi \equiv \partial _M\Phi + [A_M,\Phi ] = 0
\ \ \ \ \ F_{MN} = 0
\ee
It acquires exactly the form of eq.~(\ref{bps-cond}) when only one
of the fields
$\Phi^{(4)},\ldots,\Phi^{(8)}$ is nonvanishing -- otherwise it would also
contain the scalar interaction terms. This is essentially the case of
the configuration depicted in fig.~\ref{fi:todacu},
which implies that some scalar field,
say $\Phi\equiv \Phi^{(4)}+i\Phi^{(5)}$, develops a nonvanishing
$z$-dependent vacuum expectation value.
In order to explain or "derive" fig.~\ref{fi:todacu}, it is necessary to
demonstrate that eq.~(\ref{bps-cond}) has a
{\em non-trivial} solution $\Phi (z)\neq const$ and the reason for this is
that non-trivial boundary conditions are imposed on $\Phi$ at
$z\rightarrow\pm\infty$. This procedure is considered in detail in
\cite{MMaM} and results in the so called {\em Lax representations} for the
algebraic equations of nontrivial complex manifolds -- in this case for the
complex curves \cite{LaxCo}. Under such procedure eq.~(\ref{bps-cond}) turns, for
example, into
\be\label{hi-sol}
\bar\partial\Phi ^{ij} + (q_i-q_j)\Phi ^{ij} =
m(1-\delta ^{ij})\delta (z - z_0)
\ee
with the solution
\be
\Phi ^{ij}(z) = p_i\delta ^{ij}+
m(1-\delta ^{ij})
{\theta_{\ast}(z - z_0 + {\Im\tau\over\pi}(q_i-q_j))\theta_{\ast}'(0)
\over\theta_{\ast}(z - z_0)\theta_{\ast}({\Im\tau\over\pi}(q_i-q_j))}
\exp\left({(q_i-q_j)(z-{\bar z})}\right)
\ee
where $\theta_{\ast}(z)$ is the odd Jacobi theta-function. Equation
$\det (\lambda - \Phi(z))=0$ (literally corresponding to the theory with
broken \4N supersymmetry) in the limit $m\to\infty$ and $\tau\to +i\infty$
with $m^N\exp({i\pi\tau})=\Lambda^N$ turns exactly into eq.~(\ref{todacur}),
the details and references can be found in \cite{Mbook,Mtmf}.

In this way one can derive the analytic representation of the complex curve
(\ref{todacur}) "from first principles". The next step is to derive the
effective action of the low-energy four-dimensional theory. According to
\cite{WittM97}, this problem can be solved starting from the effective
action on the $5$-brane world-volume or the theory of self-dual two-form
$C = \{ C_{MN}\}$, $dC =^{\ast} dC$. Roughly speaking it means that instead
of open strings, as in fig.~\ref{fi:dbrane}, the interaction is effectively
performed by "open membranes".
The theory of two-forms is essentially Abelian. Even if one introduces
the matrices $C_{MN}^{ij}$ in the adjoint representation of $SU(N)$ associated
with the membranes attached between $i$-th and $j$-th cylinders,
the non-Abelian interacting theory cannot arise since such
interaction is inconsistent with the gauge invariance. Such a theory may
contain only a non-linear interaction of non-minimal type
-- like $\Tr(dC)^4$, i.e. depending upon the tension of $C$. These
terms, however, contain higher derivatives (powers of momentum)
and they are irrelevant in the low-energy effective actions.
The "Abelian" nature of the theory of two-forms makes the
description of the Lax operator
(vacuum expectation value of the scalars
of the supermultiplet which describe the transverse
fluctuations of the $5$-brane), and thus the derivation
of the shape of the curve
$\Sigma$ in the type IIA picture, a nontrivial problem.
Instead, exactly due to the fact that the action on
(flat) world-volume is essentially quadratic
\be\label{act}
\int _{d^6x}|dC|^2 + \hbox{supersymmetric\ \ terms}
\ee
there are no corrections to the form of the
effective four-dimensional action, once the curve $\Sigma$ is given.
It is enough to
consider the dimensional reduction of (\ref{act}) from six down to four
dimensions \cite{WittM97}, implying that the two-form $C$ can be
decomposed as
\be\label{ver}
C_{\mu z} = \sum _{i=1}^{N-1}\left(A_{\mu}^i(x)d\omega _i(z) +
{\tilde A}_{\mu}^i(x)d{\bar \omega} _i(\bar z)\right)
\ee
where $d\omega _i$ are canonical holomorphic one-differentials
on $\Sigma$, $d{\bar \omega} _i$ -- their complex conjugate, and
the fields $A_{\mu}^i$, ${\bar A}_{\mu}^i$ depend only on the four
co-ordinates $x=\{ x^0,x^1,x^2,x^3\} $.

Choosing the metric on $\Sigma $ to be such that
$\ast d\omega _i = - d\omega _i$, $\ast d{\bar \omega} _i =
+ d{\bar \omega} _i$, the
self-duality of $C$ implies that the one-forms $A$
and $\bar A$ in
(\ref{ver}) correspond to the anti-selfdual and selfdual
components of the four-dimensional gauge field with the curvature
(tension) $F = \{ F_{\mu\nu}\}$:
\be
dA^i = F^i -^{ \ast} F^i
\nn \\
d{\tilde A}^i = F^i +^{ \ast} F^i
\ee
It remains to substitute this into (\ref{act})
to get $T_{ij}$ -- the period matrix of $\Sigma$
(which depends on the vacuum expectation values of the transverse
scalar fields once the shape of the
curve $\Sigma $ or its embedding into the
$(x^4,x^5,x^6,x^{10})$-space is already fixed).
The result for the four-dimensional effective action reads
\be
\label{effact}
\int _{d^4x}\Im T_{ij}F_{\mu\nu}^iF_{\mu\nu}^j +
\hbox{supersymmetric\ \ terms}
\ee
where effective couplings are expressed through (the imaginary part of) the
period matrix $\Im T_{ij} = \int_{\Sigma}d\omega_i\wedge d\bar\omega_j$
of the auxiliary Riemann surface (\ref{todacur}). The action
(\ref{effact}) coincides with the result of the Seiberg-Witten theory
\cite{SW}, up to the topological $\theta$-term, which can be restored by
slightly more delicate operating with the action of a self-dual two-form.
\end{footnotesize}

\subsection{Seiberg-Witten Theory
\label{ss:sw}}

The construction of the
exact nonperturbative effective actions for the low-energy
\N2 supersymmetric gauge theories is called Seiberg-Witten theory \cite{SW}.
The exact nonperturbative formulas
\cite{SW} contain the information about the spectrum of the BPS excitations
("W-bosons" and monopoles, see sect.~\ref{ss:instmon}) and the Wilsonian
effective action of the light fields (see, for example, \cite{Wils,ShiVa}).

As we already pointed out in sect.~\ref{ss:susyga},
supersymmetry leads to strong requirements on the form of the effective
action. In the case of \1N supersymmetry in four dimensions
the "classical" form of the superpotential $W$ is
{\em not renormalized} (and this allows to study vacua of the
theory -- the critical points of the superpotential $dW=0$)
while the kinetic terms
are governed by the K{\"a}hler metric or the K{\"a}hler potential.
For the extended supersymmetry the situation is even more restrictive --
there are no Abelian potential terms (and it means that instead of distinct
vacuum "points" one gets the continuous set of vacua described by
parametric families or {\em moduli}) and the effective action, say
for the vector multiplets,
can be written in terms of a single holomorphic function of several
complex variables \cite{Seib94,SW} -- a {\em prepotential}. In other
words the geometry of moduli space is {\em special} K\"ahler.

Let us turn now to the Seiberg-Witten theory for the
\N2 supersymmetric Yang-Mills theory without matter
\footnote{In supersymmetric gauge theories one usually means by matter only
the multiplets of the fermionic and scalar fields in the fundamental
representations of the gauge group -- the analogs of quarks in usual
(non-supersymmetric) QCD. In the theories with extended supersymmetry there
are also the fermions and scalars in the adjoint representation -- the
superpartners of the gauge fields.}.
The scalar potential in \N2 supersymmetric
gauge theory is essentially "non-Abelian" and has the form $V(\bphi ) =
\Tr [\bphi, \bphi ^{\dagger}]^2$. Its minima after factorization over the
gauge group correspond to the diagonal ($[\bphi, \bphi ^{\dagger}] = 0$),
and in the theory with the $SU(N)$ gauge group -- to the traceless matrices
(\ref{vacmatr}). Due to spontaneous breaking of the gauge group
this results (in general position) in the effective \N2 Abelian gauge
theory with the effective Lagrangian ${\cal L}_{\rm eff}(\Phi_i)$,
which can be written, say, in terms of the superfields
$\Phi_i$, whose vacuum values $\langle\varphi_i\rangle = \phi_i$ coincide
with the diagonal elements of (\ref{vacmatr}). Therefore the function of
complex variables ${\cal F}(a)
= \left.{\cal F}(\bphi)\right|_{\sum \phi_i = 0}$ (where the independent
variables $a_i$ -- in
perturbation theory -- can be chosen,
for example, as $a_i = \phi_i-\phi_N$, $i=1,\dots,N-1$) indeed determines
the Wilsonian effective action for the massless fields by means of the
following substitution
\be\label{subst}
{\cal L}_{\rm eff} \propto \Im\int d^4\vartheta
{\cal F}(\phi_i \rightarrow \Phi _i) =
\Im\left({\partial ^2{\cal F}\over\partial a_i\partial a_j}\right)F_{\mu\nu}^i
F_{\mu\nu}^j + \hbox{supersymmetric\ \ terms}
\ee
Notice immediately that the effective action (\ref{subst}) exactly coincides
with (\ref{effact}), after identifying the matrix elements of the period
matrix $T_{ij} = {\d^2{\cal F}\over\d a_i\d a_j}$ with the second
derivatives of prepotential.

In \N2 perturbation theory formula (\ref{subst} can be checked by explicit
computation of quantum corrections, which in conventional \N2
supersymmetric gauge theory
are reduced to the one-loop diagram (see fig.~\ref{fi:loop}).
Integrating over momenta
propagating along the loop one comes to the result
\be\label{effcharge}
T_{\rm 1-loop} \propto \sum_{\rm masses}
\log {({\rm mass})^2\over\Lambda ^2}
\ee
where $\Lambda\equiv\Lambda_{QCD}$ is a scale parameter of the theory and
the sum in (\ref{effcharge}) is taken over the masses of propagating fields
in the loop. In the easiest form this result can be written in terms of the
"Coleman-Weinberg" formula for the prepotential
\be\label{mass}
{\cal F}_{\rm 1-loop} =
\frac{1}{4}\sum_{\rm masses}({\rm mass})^2\log{({\rm mass})^2\over\Lambda ^2}
\ee
In pure supersymmetric Yang-Mills theory all masses in (\ref{mass})
are generated by
the Higgs effect (\ref{Wmass}), so finally the perturbative result
(\ref{mass}) acquires the form
\be\label{coleman}
{\cal F}_{\rm pert} =  {\cal F}_{\rm 1-loop} =
{1\over 4}\Tr\left( \bphi ^2\log{\bphi ^2\over\Lambda
^2}\right)
\ee
The same computation can be performed in the general case: one should
take the sum of the terms
like (\ref{coleman}) corresponding to the contribution of each multiplet;
the trace for each term
$\Tr\equiv\Tr_R$ should be taken in the corresponding representation and
the sign of each contribution depends of the type of the
multiplet (it is "$+$" for the vector and "$-$" for the hypermultiplet).
As for the massive excitations, it turns out that at least the
BPS massive spectrum
\be\label{mass-centr}
M \propto |{\bf n}\cdot{\bf a} + {\bf m}\cdot{\bf a}_D|
\ee
is related to the prepotential ${\cal F}$ by the formulas \cite{SW}
\be\label{aad}
{\bf a}_D = {\partial {\cal F}\over\partial {\bf a}}
\ee
The integer-valued vectors ${\bf n}$ and ${\bf m}$ in
eq.~(\ref{mass-centr}) correspond respectively to "electric" and "magnetic"
charges of "surviving" $U(1)^{N-1}$ gauge group.

With the instantonic contributions things are not so simple. The well-known
part contains the generic structure of the effective action which
implies that prepotential has an asymptotic expansion for large
values of the condensates $\langle\Phi\rangle\gg\Lambda$
\be\label{expansion}
{\cal F} = {\cal F}_{\rm pert} + {\cal F}_{\rm inst} =
{1\over 4}\sum _{\{ I\} }a_{\{ I\} }^2\log {a_{\{ I\} }^2\over\Lambda ^2} +
\sum _{\{ I\} }a_{\{ I\} }^2\sum _{k=1}^{\infty }{\cal F}_{\{ I\} ,k}
\left({\Lambda\over a_{\{ I\} }}\right)^{2Nk}
\ee
with some unknown coefficients ${\cal F}_{\{ I\} ,k}$, where the
multiindex $I$
corresponds to different components of the vector ${\bf a}$.
The terms with fixed $k$ in the r.h.s. of (\ref{expansion}) corresponds
to the sector with fixed instantonic number $k$ in the $SU(N)$ Yang-Mills
theory. For example, in the $SU(2)$ case the integral over the
size of each instanton has the form $\int {d\rho\over\rho ^5}$ giving rise
to the $\Lambda ^{4k}$ scale dependence for $k$ instantons.
However, {\em all} coefficients ${\cal F}_{\{ I\} ,k}$ in principle cannot
be computed by standard field-theoretical methods. Each of them can be
written in the form of some integral over the (each time different) moduli
space of an instanton configuration, therefore their "relative
normalization" simply cannot be defined. On the other hand, such
normalization can be fixed in some "natural way", and all performed
instantonic calculations (mostly with the $SU(2)$ gauge group) confirm the
Seiberg-Witten hypothesis.

\begin{figure}[tb]
\epsfysize=9cm
\centerline{\epsfbox{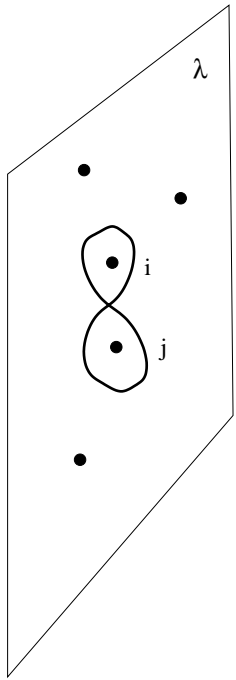}}
\caption{\sl "Figure-of-eight contour drawn around the points
$\lambda=\phi_i$ and $\lambda=\phi_j$ in $\lambda$-plane, which is an analog
of the $A$-cycle on the Seiberg-Witten curve.}
\label{fi:eight}
\end{figure}

According to the Seiberg-Witten hypothesis the BPS masses
${\bf a}$ and ${\bf a}_D$ can be expressed
through the periods of a meromorphic differential $dS$ on auxiliary
{\em Riemann surface} $\Sigma $
\begin{figure}[tp]
\epsfysize=3.2cm
\centerline{\epsfbox{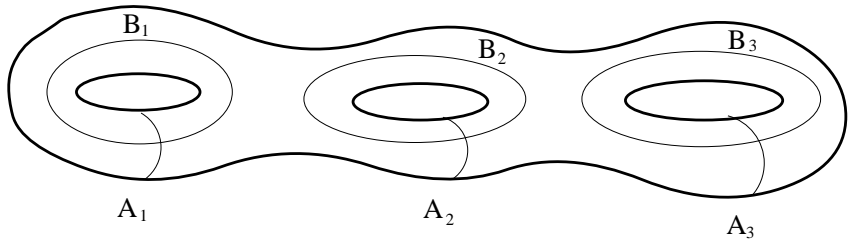}}
\caption{\sl Compact two-dimensional Riemann
surface of genus $g=3$. The canonical basis of
${\bf A}$ and ${\bf B}$-cycles has the intersection form
$A_i\circ B_j = \delta _{ij}$. An analogous picture arises in
fig.~\ref{fi:todacu} if one adds "by hand" both "infinity points"
$\lambda=\infty$.}
\label{fi:riemann}
\end{figure}
and depend on the vacuum expectation values
of scalar fields, as upon certain co-ordinates on the moduli
space of complex structures of $\Sigma $. In particular, in these specific
co-ordinates the matrix of effective charges
$T_{ij}({\bf a})= {\d^2{\cal F}\over\d a_i\d a_j}$ plays the role of the
{\em period matrix} of Riemann surface $\Sigma $.
For example, in the case of pure gauge
theory with the $SU(N)$ gauge group the auxiliary Riemann surface
has exactly the form (\ref{todacur}) \cite{sun}, where the coefficients of
the polynomial $P_N(\lambda)$ are expressed through the vacuum values of the
scalar fields (\ref{polyn}). The exact quantum values of the BPS masses are
related to the vacuum condensates through the {\em periods} over the so
called ${\bf A}$-cycles (see fig.~\ref{fi:riemann})
\be
\label{Wm}
{\bf a} = \oint_{\bf A}dS
\ee
for the $W$-bosons, and the ${\bf B}$-cycles for the monopoles
\be
\label{Mm}
{\bf a}^D = \oint_{\bf B}dS
\ee
of the meromorphic differential
\be
\label{dS}
dS = \lambda{dw\over w}
\ee
whose properties ensure (see the details, say, in
\cite{Mbook,Mtmf}), that the period matrix of the Riemann surface (\ref{todacur})
can be expressed in terms of the derivatives
\be
T_{ij} = {\d a^D_i\over\d a_j} = {\d^2{\cal F}\over\d a _i\d a_j}
\ee
Eq.~(\ref{todacur}) can be explained (but not derived!) in the
following way. Perturbatively, the masses of ``particles'' -- the
$W$-bosons and their superpartners are proportional to the differences of
$\phi_i$'s or the {\em roots} of the ``generating''
polynomial (\ref{polyn}). Thus they can be
``extracted'' from the polynomial (\ref{polyn}) via the residue formula
\be
\label{pertsample}
m_{ij} \propto \oint_{C_{ij}} \lambda d\log P_{N}(\lambda )
\ee
which for a particular contour $C_{ij}$ -- a "figure-of-eight",
drawn around the points
$\lambda=\phi_i$ and $\lambda=\phi_j$ (see fig.~\ref{fi:eight}) --
gives rise directly to (\ref{Wmass}).
The contour integral (\ref{pertsample}) can be viewed as defined on {\em
degenerate} Riemann surface -- a ("double") $\lambda $-plane with
$N$ removed points: in the roots of the polynomial (\ref{polyn}).
Then the formula (\ref{todacur}) can be interpreted in the following way.
The only non-perturbative effect in terms of this
Riemann surface is blowing up its singularities by the simplest possible
procedure -- replacing the marked points at
$\lambda = \phi_i$ by the ``handles'':
$w + {\Lambda^{2N}\over w} \sim \lambda - \phi_i$, and passing in this way
from the $\lambda$-plane with marked points to a smooth Riemann surface
(fig.~\ref{fi:todacu}).
\begin{figure}[tb]
\epsfysize=9cm
\centerline{\epsfbox{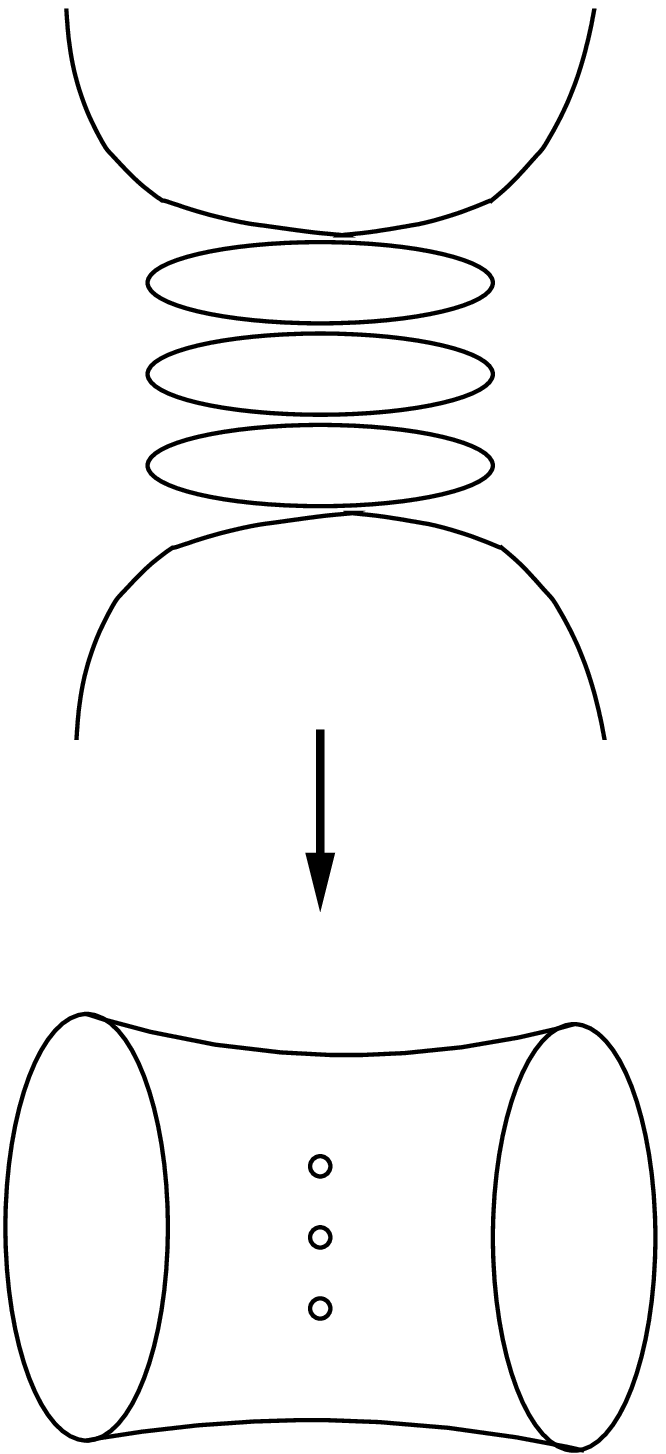}}
\caption{\sl Two degenerate limits of the smooth curve from
fig.~\ref{fi:todacu}.}
\label{fi:persol}
\end{figure}

A degenerate Riemann surface --
"two copies" of the $\lambda $-plane with $N$ marked points is depicted at
the top of fig.~\ref{fi:persol}. This degenerate limit, was already
mentioned before, corresponds to weak coupling in \N2 supersymmetric gauge
theory and, therefore, can be computed straightforwardly using one-loop
perturbation theory. The opposite degenerate limit is much more
interesting and corresponds to the degenerate Riemann surface in the
bottom of fig.~\ref{fi:persol}. This limit is stable when the extended
supersymmetry is broken down to \1N (the corresponding values of moduli of
this degenerate curve are exactly in the minima of \1N potential). It is
this limit, when the periods (\ref{Mm}) vanish
(the ${\bf B}$-cycles correspond to small circles on fig.~\ref{fi:persol} while
the differential (\ref{dS}) does not have any singularities at corresponding
points) and it means that the corresponding masses of magnetic monopoles
also vanish in this limit. The effective \1N superpotential acquires the
form
\be
{\cal W} = {\tilde Q}a^D(u)Q + \mu u
\ee
where $u = \langle\Tr\bphi^2\rangle$, $Q$ and ${\tilde Q}$ are the vacuum
values of the monopole supermultiplets and $\mu$ is the scale of violation
of \N2 down to \1N. The function $a^D(u)$ is defined by the integral
(\ref{Mm}). It follows from here that in the minimum
$\langle{\tilde Q}Q\rangle \sim \mu$, or the monopoles in
\1N theory condense at this leads to the (dual to well-known in
superconductivity) effect when the electric field is "forced out",
i.e. to (Abelian) confinement. Thus, the supersymmetric Seiberg-Witten
theory becomes a nice "exactly solvable" laboratory for studying the
properties of real QCD \cite{Stra,Yung}.

\subsection{Exact Nonperturbative Results and Integrable Systems
\label{ss:is}}

The fact that string theory possesses an extremely high symmetry allows
one in practice for the first time to raise a question about
the computation
of the exact correlation functions in absolutely nontrivial theories,
moreover not belonging formally to the class of quantum integrable models
at least in canonical sense. The main idea of getting exact answers from
symmetry considerations is based on deriving the relations, which
correlation functions should obey. If the symmetry is high enough these
relations may lead to the exact solution. It was in the framework of string
theory (more strictly in the framework of its simplest models) that
such program
was completely carried out and it turned to be possible to get exact (in
particular nonperturbative) information about the correlation functions.

\begin{footnotesize}
First, some progress was achieved in the theories "without matter" or in the
theories of two-dimensional gravity interacting with "minimal"
($c \leq 1$) matter
(let us recall, that the central charge $c$ counts the number of degrees of
freedom). It turned out that such theories can be effectively described in
terms of the matrix models of two-dimensional gravity \cite{mamo}, i.e.
in terms of the {\em finite-dimensional} matrix integrals of the form
\be\label{mm}
Z = \int DM \exp({- V(M)})
\ee
where $DM \propto \prod _{i,j}dM_{ij}$ denotes the simplest integration
measure over the finite-dimensional matrices. The loop expansion or the
expansion over topologies of the matrix graphs
\cite{thooft} of the integral (\ref{mm}) reproduces the (discretized
version) of the loop expansion (\ref{ppi0}) of
$c \leq 1$ string models. The double-scaling limit of the formula (\ref{mm})
\cite{ds} allows to identify ${\cal F} \propto \log Z$ directly with the
full generating function of the string theory correlators
\be\label{genfun}
\langle {\cal O}_{i_1}\dots{\cal O}_{i_n}\rangle =
{\d ^n{\cal F}\over\d T_{i_1}\dots\d T_{i_n}}
\ee
and/or with the effective action. The information about the function $\F$
can be encoded in the set of nonlinear integrable equations.

The generating function depends on variables of two types. The first
type of variables is the set of sources for physical operators
\be
\label{sources}
{\cal F}(g_{\rm str},{\bf T}) =
\sum_{g=0}^\infty g_{\rm str}^{2g-2}\F_g({\bf T}) = \\ =
\sum_{g=0}^\infty g_{\rm str}^{2g-2}\int Dh_{ab}\ D{\bf X}\
\exp\left({- S_{CFT}({\bf X},h_{ab}) + \sum T_k{\cal O}_k}\right)
\ee
and the derivatives of (\ref{genfun}) over these sources determine
the correlation functions in the theory. Expression
(\ref{sources}) does depend upon the choice of basis of the operators
${\cal O}_k$ or parameters $T_k$, and only in some fixed basis (not
necessarily convenient from the point of view of the world-sheet theory) it
can be elegantly described in terms of non-linear partial differential
equations or unitarity-like relations for the correlators. In general, such
relations are well-known in traditional quantum field theory (the Ward
identities, the Schwinger-Dyson equations etc) but the situation in string
theory is singled out by the fact that there equations can be written in the
form of {\em closed} system of {\em integrable} equations completely fixing
the generating function (\ref{sources}). As a function of parameters
${\bf T}$, the generating function (\ref{genfun}), (\ref{sources}) can be
defined only in the sense of formal series, whose coefficients are
identified with the correlation functions, but the series itself has
vanishing radius of convergency. This fact reflects the well-known
properties of the perturbative expansions in string theory and quantum field
theory and moreover it is consistent with the existing explicit formulas for
the exact nonperturbative solutions. If they exist these formulas are usually
known in the form of integral representations and may sometimes be written
in terms of the matrix integrals (\ref{mm}). However, the particular terms
of the series for (\ref{sources}), for example
$\F({\bf T})\equiv \F_0({\bf T})$ can be found and written in terms of
well-defined functions.

Another set of parameters, which the partition functions or generating
functions depend on, are the physical or space-time moduli of the theory.
The space of these parameters is usually finite-dimensional, in the
considered cases it is also often complex and may be interpreted as moduli
space of complex manifolds. I repeat that complex
curves or Riemann surfaces arising in this context have the
"space-time origin" (say come out of the
string compactification) and are not related to the world-sheets of string
theory!

As a function of moduli the generating function is a normal (say,
meromorphic) function of many complex variables and can often be computed
more or less effectively. The moduli parameters can be interpreted as the
low-energy values of the background fields (the Higgs scalar condensates,
moduli of physical metric -- the K\"ahler and complex structures etc) and as
a function of moduli the function ${\cal F}$ has usually the sense of an
effective action. The existing relation between the geometry of complex
manifolds and integrable systems allows one to identify the functions $\F$ with
solutions to nonlinear integrable equations.

In general the dependence upon the generating parameters and moduli is
rather different
\footnote{In topological two-dimensional gravity and in some topological
string models (of the $A_p$-\-type), the dependences on the moduli $t$
and the sources $T$ almost coincide (the $(t+T)$-\-formula \cite{LGGKM}).}
and both functions are independently interesting problems. For example, in the
Seiberg-Witten theory now there exists a reasonable answer only to the
first question
\footnote{Something about dependence on generating parameters and an analog
of the $t+T$-\-formula in Seiberg-Witten theory can be found in
\cite{RGWhi}.}, and it is very important that the Wilsonian effective action
in the massless sector can be expressed
(\ref{subst}) via a {\em function} of several complex variables. Thus, it is
the knowledge of the function ${\cal F}$ as function of moduli and all its
derivatives, say the expansion over the sources ${\bf T}$ which
gives the most complete information about the theory.

The effective theory can be formulated in terms of (a classical) integrable
system. This formulation is universal in the sense that it does not depend
on many properties of the "bare" theory. For example, it does not really
depend even on the dimension of a bare theory: two-dimensional,
four-dimensional, and even five-dimensional theories look absolutely
similar from this point of view. Moreover, so obtained effective theories
remind a lot the {\em topological} field theories. They possess many
properties of two-dimensional topological field theories, though the "bare"
theories are essentially multidimensional and, what is especially important,
contain massless propagating particles.

Let us now list the main types of differential equations arising in
nonperturbative string theory.

\begin{itemize}
\item {\bf The "Virasoro" constraints} (more strictly -- the Virasoro-like
constraints) \cite{vir,virmamo,virGKM}. This is one more manifestation of
the not
yet clear duality between the world-sheet and space-0time structures. The
"Virasoro constraints" arising in matrix models of two-dimensional gravity
and topological theories have the general form
\be
\label{lntau}
{\cal L}_n \exp({\cal F}) = 0
\ee
where ${\cal L}_n$ are the differential operators in parameters $\{ T_n\}$,
forming the Virasoro algebra (\ref{vir}). Note that equations of this
type already arise in some effective space-time formulations of string
theory. In contrast to Virasoro generators of the world-sheet
reparameterizations, the operators ${\cal L}_n$ in this context have a purely
space-time interpretation.

Solution to the constraints (\ref{lntau}), can usually be expressed through
the tau-functions of the hierarchies of integrable equations. Sometimes
these tau-functions can be written in terms of the matrix integrals (about
appearance of the Virasoro constraints in matrix models and the relation
between the matrix models and integrable systems see, for example,
\cite{Mormat}). For the generating functions, written in the form of
matrix integrals (\ref{mm}), the Virasoro constraints follow from the
loop equations or Ward identities $\langle\delta V\rangle = 0$
(the average is understood in the sense of partition function
(\ref{mm})), which are basically the simplest analogs of the Ward identities
in gauge field theory.

\item {\bf The associativity equations} \cite{WDVV}. A nontrivial
over-determined system of differential equations to the generating function
${\cal F}$, containing its third derivatives. Collecting the third
derivatives into the matrices
$\|{\sf F}_{i}\|_{jk}= {\d^3\F\over\d T_i\,\d T_j\,\d T_k}$, the
associativity equations can be written in compact form \cite{MMM}
\be
\label{WDVV}
{\sf F}_i {\sf F}^{-1}_{j}{\sf F}_{k}=
{\sf F}_k {\sf F}^{-1}_{j}{\sf F}_{i}
\;\;\;\;\;\;\forall\;\; i,j,k.
\ee
Firstly the associativity equations were found in topological string models
(where they follow from the crossing relations) but later it turned out that
they show up in much more vast class of effective theories, for example in
the Seiberg-Witten theory.

\item {\bf "Quasiclassical" integrable hierarchies}. These hierarchies
usually arise on attempts to find exactly the tree-level or spherical
contributions $\F({\bf T})\equiv \F_0({\bf T})$. They are usually reduced to
well-known dispersionless analogs of the hierarchies of
Kadomtsev-Petviashvili or Toda lattice types. In a wider sense the
quasiclassical hierarchies are applicable, say, to the description of the
Seiberg-Witten theory: the prepotential $\F$ is logarithm of the
tau-function of some nontrivial solution to quasiclassical hierarchy. The
known solutions to quasiclassical hierarchies are related mostly to geometry
of complex manifolds. One of the consequences of such a relation is the
existence of so called "localization" or the residue formulas of the form
\be
\label{res}
{\d^3\F\over\d T_i\,\d T_j\,\d T_k} =\ \res\ \left({dH_idH_jdH_k\over
\Omega}\right)
\ee
where $dH_i$ are one-forms related to the variables $T_i$, and $\Omega$ is
some "symplectic" two-form. One of the possible consequences of the residue
formulas is the existence of associativity equations (\ref{WDVV}).
\end{itemize}
\end{footnotesize}

\setcounter{equation}0
\section{Strings and Duality between Gauge Theories and Gravity
\label{ss:holography}}

\subsection{Holography and Strings}

One of the most interesting recent physical ideas in string theory is
applying the "holographic principle" which allows to describe theory in
full $D$-dimensional space-time (or in some part of this space-time) -- in
the so called bulk -- in terms of the information encoded on its
{\em boundary}. Such a possibility exists far from everywhere, since the bulk
theory contains, in general, much more information than the theory on the
boundary -- the number of degrees of freedom of the bulk theory is much
larger. Roughly  speaking, the ratio of the number of degrees of freedom in
the bulk of dimension $D$ and on the boundary of co-dimension $\delta $
(usually $\delta=1$) under the growth of characteristic size of the system
$L$ grows as $L^D/L^{D-\delta} = L^\delta$. Besides this fact, in
traditional quantum field theory the field theory "inside" (say, the Green
functions) is completely determined by the boundary theory only in
quadratic or free case.

\begin{figure}[tb]
\epsfysize=7cm
\centerline{\epsfbox{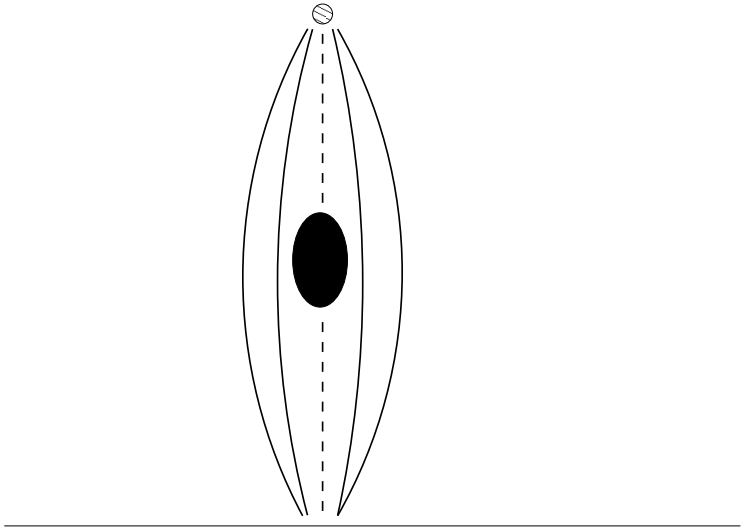}}
\caption{\sl Holographic 't~Hooft principle. A point which cannot be
naively projected to the boundary due to the presence of some
material "screen", is nevertheless projected due to deviation of
rays by the gravitational field, induced by this "screen".}
\label{fi:hologr}
\end{figure}

In contrast to quantum field theory, string theory necessarily contains
gravity, in which the relation between the bulk and boundary theories seems
to be completely different. One of the manifestations of this fact is the
well-known {\em linear} connection between the entropy of the black hole and
the area of the horizon, demonstrating that the number of degrees of freedom in
gravity is proportional not to the volume, as one would expect from quantum
field theory. Another side of the same phenomenon is known as the 't~Hooft
holographic principle \cite{thooftholo}. According to this principle due to
deviation of rays in gravitational field any point from the
bulk can be "independently" projected to the boundary
(see fig.~\ref{fi:hologr}).

String theory unifies "matter" (open strings) and gravity (closed).
Moreover, as was already discussed in sect.~\ref{ss:dbr}, there are
natural vacua in string theory where matter is localized on some
hypersurfaces in space-time, while gravitons or closed strings can propagate
everywhere in bulk. A necessary production of closed strings in the theory
of open strings (see fig.~\ref{fi:opclo}) leads to the possibility
establishing some {\em holographic} (in the above sense) analogy between the
theory of matter or open strings on a D-brane (on the 'boundary") and the
theory of closed strings or gravity in the bulk.

In other words, the same effects can be formulated both in terms of open
strings or the Yang-Mills theory as well as in the language of the
closed string theory or gravity. In this chapter we will try to discuss some
consequences of this duality, in the modern parlance usually called
"AdS/CFT-correspondence", since the most well-known example of this
phenomenon is the duality between \4N supersymmetric conformal field theory
of the Yang-Mills fields (conformal field theory -- "CFT") and gravity in
five-dimensional anti-de-Sitter space ("AdS") \cite{Malda}, see
sect.~\ref{ss:ads} below. The most physically interesting effect which can
hopefully be better understood in the framework of such correspondence is
the parallel between two very important
phenomena in modern theoretical physics proposed by Polyakov \cite{Polholo}
-- the confinement of quarks in
non-Abelian gauge theories and the confinement of matter beyond the
horizon of the black hole.

Another interesting aspect of this picture is adding to the physical picture
of the world so-called "extra dimensions". In contrast to already
traditional Kaluza-Klein ideas \cite{KaKle} (see, also,
\cite{Scherk}) about additional {\em small} dimensions, responsible for the
internal symmetries in the theory, in the new proposed physical picture the
extra dimensions should not necessarily be small (and can in general be
even non compact). The problems of the theories with extra dimensions
(although not in the context of string theory) were considered recently in
\cite{rubrane}.

\subsection{Duality of Open and Closed Strings}

\begin{figure}[tb]
\epsfysize=5cm
\centerline{\epsfbox{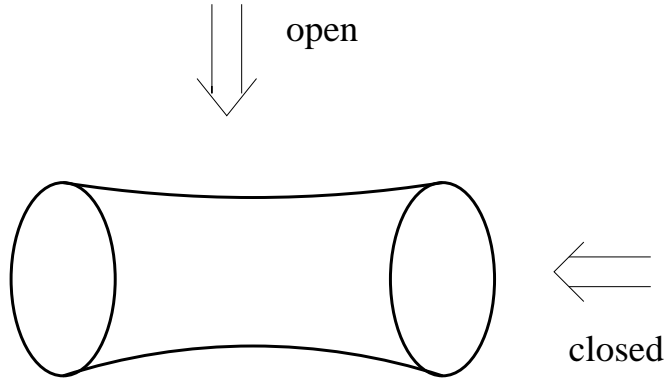}}
\caption{\sl One-loop diagram in the theory of open strings is equivalent to
a tree diagram in the theory of closed strings.}
\label{fi:cyl}
\end{figure}

As we already discussed in sect.~\ref{ss:ym-f-str}, string theory is the
only reasonable candidate for the role of the unifying theory of the vector
fields and gravity since it naturally unifies the carriers of these
interactions as excitations of open and closed strings. One of the
consequences of this relation is the possible interpretation of closed
strings as bound states in the theory of open strings
(see fig.~\ref{fi:opclo}). Another rather natural conclusion comes out if
one considers the one-loop diagram in the open string theory corresponding
to the world-sheet with topology of a cylinder (see fig.~\ref{fi:cyl}).
Looking at the same diagram from the perspective of closed string theory, it
is clear that it corresponds just to a tree-level propagator
(cf. with fig.~\ref{fi:prop}). Thus, it says that the one-loop (i.e.
quantum) effects in the open string theory may have a dual formulation in
terms of tree-level (i.e. classical) gravity -- the massless part of the
closed string spectrum.

This purely string duality can in principle be realized as a duality between
the gauge theories and gravity and this leads to the already mentioned parallels
between the confinement of quarks inside hadrons and keeping matter beyond
the horizon of black holes. This idea has become very popular due to the
more or less explicit example of the "holographic" duality between the
${\cal N}=4$ supersymmetric gauge theory and geometry
$AdS_5\times {\bf S}^5$, or the direct product of five-dimensional
Lobachevsky space or anti-de-Sitter space and five-dimensional sphere, see
sect.~\ref{ss:ads}.
Such duality is often called "holographic" since from the point of view
of nonperturbative string theory one may consider it as a consequence of the
holographic principle or, in more simple terms, of the fact that bulk
gravity can be described in terms of some effective theory on the boundary
of its volume. In more detail, the hypothetical scenario of such duality is
based on the following assumptions:

\begin{itemize}

\item Matter, described in terms of gauge fields and their superpartners,
or, generally, by open strings
is confined to certain hypersurfaces in multidimensional (e.g. $D=10$ or
$D=11$)
space-time, since open strings are allowed to have their ends {\em only}
on these Dirichlet or D-branes
\footnote{At least in the context
of type II string theory.}, see fig.~\ref{fi:dbrane}.

\item In contrast to matter, gravity corresponding to the massless
excitations of {\em closed} strings, is allowed to propagate everywhere
in the bulk of ten-dimensional space-time, i.e. is indeed (at least)
ten-dimensional theory, as any consistent {\em quantum} gravity should be.

\item The matter branes (D-branes) themselves induce a gravitational
field, which, at the level of the classical ($\alpha'\to 0$) approximation could
be considered just as a solution of the {\em bulk} equations of motion with
the boundary terms arising from effective theories on branes. Hence,
on the one hand one
may look at the boundary terms induced by matter as at the (localized) sources
for gravitational field, on the other hand deeper correspondence implies
that gravitational boundary action may play the role of a generating
function for the correlators in matter theory on brane.

\item In most nowadays popular concrete models, the bulk geometry is
"reducible" i.e. has form of a direct product like
$AdS_5\times S^5$, where the compact $S^5$ part is kept
to be "fixed" while the real physics takes place within the other
part, so that four co-ordinates $\{ x_\mu\}$ play the role of
"visible" space-time, while the rest, the fifth co-ordinate $y$
(which the background metric nontrivially depends on), serves as
a {\em scale} of observable space-time
\footnote{In this context the five-dimensional geometry plays the role
of the five-dimensional gravitational "bulk",
restricted by "boundary" branes of
codimension $\delta=1$.}. In other words the metric can be written in the
distinguished in string theory form of the "Friedman universe"
\be
\label{RSanz}
ds^2 = dy^2 + a(y)(dx_\mu)^2
\ee

\item The scale factor of matter theory or the position of the matter
brane in the auxiliary (fifth) dimension can be found as solution to the
five-dimensional equations of motion (on the "gravitational side"), or by the
renormalization group equations (on the "matter" or gauge theory side).
Since equations of motion are differential
equations of the second order (while conventional renormalization group
contains only the first order equations in scale parameter), the relation
between them is rather nontrivial. An interesting existing proposal is that
of \cite{ver} -- to use the Hamiltonian formalism \cite{FadUFN} in
five-dimensional gravity theory. Going along this way one should come to
a direct description of the effective boundary action in terms of a
tau-function of some integrable system, see sect.~\ref{ss:is}.

\end{itemize}

Most of these ideas about the relations between the gauge theories and
theory of gravity arose \cite{Polholo} as a direct generalization of the
well-studied correspondence between zero-dimensional (or one-dimensional)
gauge theories -- the so called matrix models (\ref{mm}) (or matrix quantum
mechanics) and theory of two-dimensional gravity or $D\leq 1$
string models \cite{mamo,ds,dsBK,dsDS,dsGM}.

\subsection{Confinement and Black Holes
\label{ss:confbh}}

One of the oldest problems in string theory, moreover being in a sense its
main origin is the description of one-dimensional extended objects in the
theory of strong interactions. Multiple attempts to formulate string
theory adequate for the description of the Wilson loops in gauge theories
and QCD has led to the idea \cite{Polholo} that such a theory should be
necessarily noncritical in the sense that the effective tension must depend on
auxiliary string co-ordinates playing the role of the scale factor and at
some point this tension should vanish or become infinite. All that means
that the string action in such model has the general form
\be
\label{polact}
\int_\Sigma\left( \d\varphi\bar\d\varphi + a(\varphi)\d{\bf X}\bar\d{\bf X}
+ \dots\right)
\ee
in order to be able to coincide with the theory of gauge fields at critical
point. The main problem then is to identify the action (\ref{polact}) with
some exactly solvable two-dimensional conformal field theory with the
necessary spectrum and other properties.
\begin{figure}[tb]
\epsfysize=8.5cm
\centerline{\epsfbox{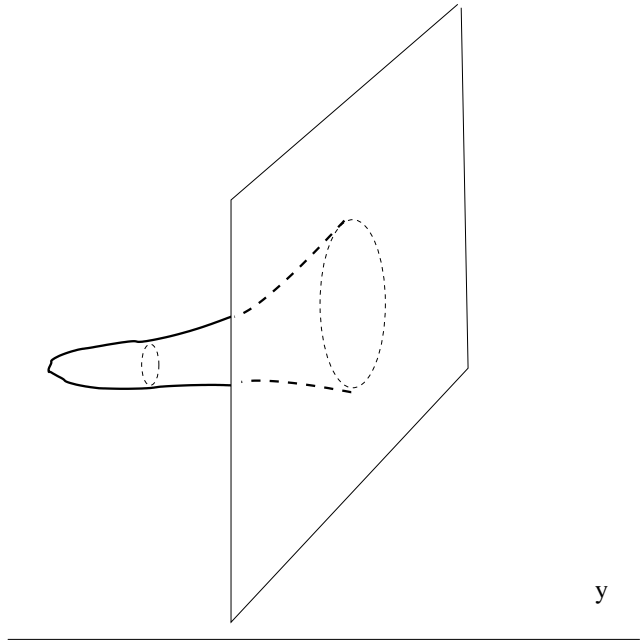}}
\caption{\sl Gravitational analog of confinement -- the metric similar to
the metric of a "black hole", which is almost flat far from the
"horizon" and near
the horizon turns into narrow throat with strong gravitational fields,
confining "quarks".}
\label{fi:ads}
\end{figure}
In its main features the "gravitational picture" of confinement is depicted
in fig.~\ref{fi:ads}.
The action (\ref{polact}) in gravitational approximation
corresponds to the "Friedman metric" (\ref{RSanz}), the co-ordinates
$\{ x_\mu\}$ are the zero modes of "two-dimensional fields"
$\{ X_\mu(\sigma,\tau)\}$, while the co-ordinate $y$ is the zero mode of the
"two-dimensional field" $\varphi(\sigma,\tau)$. The function $a(y)$
qualitatively behaves in the following way: on one side of the
$y$-axis it grows and the space-time becomes the macroscopic
five-dimensional space. On the other side of the $y$-axis, contrarily
$a(y)\to 0$, and one gets a "throat" with a strong gravitational field
confining the matter.

The essential part of this picture is the "nonstandard" nature of gravity,
compared to "ordinary" gravity of the observable (macroscopic) space-time.
First, the effects of this "hadronic" gravity \cite{Polholo} should become
essential not at the Planck scale but already at the scale of strong
interaction of the order of $10^3$ MeV. Second, metric (\ref{RSanz}) is not
observable at least in the sense that the co-ordinate $y$ is not a real
co-ordinate or co-ordinate of "visible" space-time, but rather plays the role
of a scale in the theory. Moreover, it is necessary to point out that the
gravitational description is applicable only in the situation when string
corrections are suppressed. It happens, for example, in the planar limit
$N\to\infty$ \cite{thooft}, which corresponds to the tree-level Feynman
diagrams of spherical topology or the spherical (i.e. tree-level) limit in
dual closed string theory.

Thus, the existing examples of duality between gauge theory and gravity are
implied to be correct at least in the phase where $N \gg 1$ and
$g_{YM}^2N > 1$. The first requirement is the well-known large $N$ limit
\cite{thooft} and this means that in gauge theory only the planar diagrams
survive, or that the loop corrections of the {\em closed} strings are
suppressed. In contrast to this transparent limit of large $N$
(which literally means $N\to\infty$ for the properly normalized quantities),
while the second constraint onto the coupling constant is absolutely
nontrivial. In order to compare it with the boundary action in
the theory of gravity one should first sum up the contribution of all loops
of the gauge theory or theory of open strings. Hence, the theory of gravity
should predict the {\em nonperturbative} results in gauge theories which are
not analytic in coupling constant. It is especially necessary to stress this
circumstance in order to avoid mixing between the nontrivial string duality,
relating the classical bulk theory with the boundary theory at strong
coupling and rather trivial "continuation" of the (free) Green functions
from the boundary. Such "continuation" is well-defined for conformal theory
at the boundary and metric of constant negative curvature in the bulk.

\subsection{AdS/CFT Correspondence
\label{ss:ads}}

The most well-known example of duality between the gauge theory and gravity
is the so called AdS/CFT correspondence -- the correspondence of gravity in
anti-de-Sitter space and the conformal field theory, or more exactly the
\4N supersymmetric Yang-Mills theory which is the four-dimensional (do not
mix with two-dimensional) conformal field theory with the vanishing
beta-function (\ref{rg}), (\ref{beta}) (at least in the perturbation
theory). Such gauge theory can be represented directly by the picture at
fig.~\ref{fi:dbrane}, i.e. for the $SU(N)$ gauge theory -- by a "stack" of
$N$ (completely coinciding!) D$3$-branes. The dual gravitational picture
can be constructed as a solution to supergravity equations with
corresponding boundary conditions. Such solution is well-known
(see, for example, references in \cite{Malda}), and its metric has the form
\be
\label{ads5s5}
ds^2 = U^{-1/2}\left(dx_\mu\right)^2 +
U^{1/2}\left(dr^2 + r^2d\Omega^2\right)
\ee
while the source of this metric is the Ramond-Ramond $4$-form
\be
\label{cads5}
C_{\mu\nu\lambda\rho} = \epsilon_{\mu\nu\lambda\rho}\left({1\over U}-1\right)
\ee
with the D$3$-branes being charged with respect to this form. In formulas
(\ref{ads5s5}) and (\ref{cads5}) the function $U=U(r)$ depends only on the
distance$r$ from the "stack" of branes
\be
\label{uma}
U(r) = 1 + {g^2 N \alpha'^2\over r^4}
\ee
where $N$ is the number of D-branes and $g$ is the coupling of
\4N gauge theory. Metric (\ref{ads5s5}) is a metric of manifold consisting
of the five-dimensional sphere (the second term in the r.h.s. of
(\ref{ads5s5})) and some five-dimensional manifold with the metric similar
to (\ref{RSanz}), where the role of distinguished co-ordinate
$y$ is played by the distance $r$ to D-branes.
Since
\be
U(r) \stackreb{r\to 0}{\sim} {g^2 N \alpha'^2\over r^4}
\ee
in the vicinity of the horizon $r\to 0$, the first term in the r.h.s. of
(\ref{ads5s5}) turns into the anti-de-Sitter metric
\be
\label{ads}
ds^2 = \alpha'\sqrt{g^2N}\left({dr^2\over r^2} + a(r)dx_\mu^2 +
d\Omega^2\right)
\ee
where
\be
a(r)={\alpha'\over g^2N}\left(r\over\alpha'\right)^2
\ee
From (\ref{ads}) it follows that the squared radius of the five-dimensional
sphere $R_{\rm sphere}^2=\alpha'\sqrt{g^2N}$ is equal to the so called
't~Hooft constant in units of $\alpha'$. As is was mentioned above, the
string
corrections are suppressed as $N\to\infty$, besides this requirement, metric
(\ref{ads}) is close to the exact solution at large 't~Hooft coupling,
i.e. when $g^2N \gg 1$.

This example is in fact the only {\em explicit} example of a relation
between the gauge theory and gravity, which allows in particular to study
the correlation functions and anomalous dimensions of composite operators
\cite{GKP}. Unfortunately this example cannot really be "deformed" into more
sensible physical theories, i.e. all the construction is rather rigid. Some
attempts of the dual gravitational description of the gauge theories with
less supersymmetry were made in \cite{Polstra}, though by now without any
striking success.

From the more general point of view the AdS/CFT correspondence in the framework
of string theory can be divided into two, generally speaking, different
parts
\be
\label{cor}
\log \int DA_\mu \exp\left({-S_{YM}[A_\mu;\phi_0] +
\sum\int_{d^4x}\phi_iO_i^{YM}
(F_{\mu\nu})}\right) =
\\ =
\sum\int D\varphi DX \exp\left({-\int_{\Sigma}
\left(G_{MN}\d X^M\bar\d X^N + R^{(2)}
\Phi (X) + \sum \phi_i (X)V_i(X)\right)}\right) =
\\ =
\int_{dx} \sqrt{G}e^{-2\Phi}\left(R(G) + V(\phi_i)+{1\over 2}(\nabla\Phi)^2 +
{1\over 2}(\nabla\phi_i)^2 + \dots\right)
\ee
which are "labeled" by two different equality signs in the formula
(\ref{cor}). This formula deserves further explanations which are now in
order:

\begin{itemize}
\item The l.h.s. contains the logarithm of the generating function of the
(supersymmetric, omitted for simplicity) Yang-Mills {\em matrix} field
theory, which is considered in the sense of
't~Hooft ${1\over N}$-expansion, reproducing the perturbative
expansion in string theory with both holes (open string loops) and handles
(closed string loops, see fig.~~\ref{fi:perturb}). One adds in this part
the sum of the {\em gauge-invariant} operators
$O_i^{YM}(F_{\mu\nu})$ \cite{GKP} to
the Yang-Mills action, depending on the (covariant
derivatives of the) Yang-Mills field-strength with
the {\em external} sources $\phi_i(x)$.

\item The middle part is literally the string theory generating functional.
As it should be in the first-quantized theory, there is the sum over
topologies and number of "holes" (the Yang-Mills expansion we noted above).
The integration is performed over all embeddings $X^M = (X^\mu,\varphi)$ of a
two-dimensional world-sheet parameterized by $(\sigma_1,\sigma_2)$
into the bulk space-time. By definition,
the world-sheets may have holes only "attached" to the boundary in space-time,
i.e. the Dirichlet boundary conditions have to be imposed on $\varphi$. The
gauge invariant operators coupled to $\phi_i$ are now represented by the
{\em closed-string} background fields $\phi_i(X)$, interacting with the
string over the whole world-sheet surface.

\item The requirement of two-dimensional conformal invariance
(see sect.~\ref{ss:conf}) is equivalent to the condition
that external background fields $\phi_i(X)$ (including the specially
singled out background metric $G_{MN}(X)$ and dilaton $\Phi(X)$)
should be on a mass-shell, i.e. satisfy the equations of motion.
This is an important point, because the equations of motion
should be "supplemented" by boundary conditions, which are not explicitly
mentioned in (\ref{cor}); nevertheless one should remember them and
{\em add} to the "middle" part of (\ref{cor}) that the boundary conditions
are imposed at $\left.\varphi\right|_0=y=y_\ast$ and the couplings in the
Yang-Mills part (the l.h.s.) are exactly the {\em boundary} values of the
string couplings $\phi_i(x) = \phi_i(\left.X\right|_0,\left.\varphi\right|_0=
y_\ast)$
\footnote{We are now discussing this correspondence at a relatively "rough"
level, forgetting more delicate questions, like the relation of
the basis of gauge-invariant operators in the Yang-Mills theory and the
basis of the corresponding vertex operator in string theory. This is a
nontrivial issue, since there is no way to adjust these basises
{\em a priori} in the first and second part of equality in the formula
(\ref{cor}). This can be seen already for the simplest example of the AdS/CFT
correspondence -- the matrix model (\ref{mm}) and the dual theory of {\em
two-dimensional} gravity.}.

\item The equality between the middle part and the r.h.s. requires even more
additional detailed explanations. The r.h.s. contains what is called the
string theory effective action (see sect.~\ref{ss:fratse}; in particular
eq.~(\ref{effgr})). Literally as is written in (\ref{cor}) it looks like
an ordinary low-energy effective action in quantum field theory. However,
things are not
so simple since one should remember that the middle part of the
equality and, thus,
the r.h.s. is defined {\em only} on mass shell. In fact the last part of
formula (\ref{cor}) contains a {\em non-local} expression, arising if one
substitutes into the action solutions to the equations of motion as
functionals of the boundary conditions! Thus, despite it seems that formula
(\ref{cor}) reformulates the quantum problem of computation of the
generating function (taking into account all loop contributions) as some
classical problem, the last one -- the classical problem of finding the
effective action as a functional of the boundary conditions -- is not in
fact simpler. An exception is the case of dilaton field with vanishing
potential, where the comparison between the gauge theory and gravity was
indeed performed in \cite{GKP}.

\end{itemize}

\subsection{Life on a brane}

Interpretation of the scale factor as an auxiliary co-ordinate of
{\em space-time} allows one to consider the problems of confinement in the
theory of elementary particles and the problems of gravity and cosmology on
equal footing. In analogy to the previous section in the theory of gravity
already at the level of simplest
{\em classical} consideration it is easy to demonstrate \cite{RS2} that
\begin{itemize}
\item It is easy to get a vanishing {\em effective} cosmological constant
of the four-dimensional matter theory;
\item It is also easy to get a massless four-dimensional graviton, non
propagating to the bulk at least in the
linear approximation.
\end{itemize}
These two statements arise without any additional information from
solving the Einstein equations of motion for bulk gravity with
certain boundary conditions, induced by brane sources.

The most general classical action in this approach includes only two terms
(the rest of contributions to the action are marked by dots)
\be
\label{RSact}
\int_{d^5x}\sqrt{G_5}
\left( {R_5\over 2\gamma_{\rm N}^{(5)}}+\Lambda_5 \right) +
\int_{d^4x}\sqrt{G_4}\Lambda_4 + \dots
\ee
where, according to the accepted rules, we consider only the nontrivial
five-dimensional part of $D$-dimensional theory and
write down two terms corresponding to the bulk five-dimensional contribution
(with metric $G^{(5)}_{MN}\equiv G_{MN}$ and its curvature $R_5 = R_5(G)$;
$\gamma_{\rm N}^{(5)}$ is the five-dimensional Newton constant) and the
boundary four-dimensional contribution (where $G_4$ denotes the determinant
of metric on the brane world volume, induced by the five-dimensional metric
with the determinant $G_5\equiv G$). The terms, omitted in (\ref{RSact}),
are
generally non-local or contain higher derivatives; they however should
necessarily be taken into account in an exact string formulation of the
problem.

It is remarkable that the action (\ref{RSact}) written in the simplest approximation,
does not really depend on {\em any} details of the model.
In the simplest case, the second term can be chosen as a $\delta$-function
along the fifth co-ordinate $x_5=y$ and the "potentials"
$\Lambda_5$ and $\Lambda_4$ can be considered as constants -- the five-dimensional
bulk cosmological constant and "bare" four-dimensional cosmological
constants or tension of the correspondent brane.
Nobody forbids, however, considering them as nontrivial functions
of co-ordinates, being, say, the values of the matter (scalar) fields
potentials --
then the simplest picture is easily generalized to the case of several thin
branes or a  thick brane. The analysis in any case does not differ from
the simplest examples of localization \cite{RS1,RS2}, when the second
term represents
the only thin brane sitting at $y=0$ with no other sources, or, better to say,
the contribution of {\em all} other sources is encoded in the non-vanishing
five-dimensional cosmological constant $\Lambda_5=const<0$ giving rise to
the anti-de-Sitter $AdS_5$ geometry far outside the brane.

The appropriate solutions to the equations of motion, following from
(\ref{RSact})
\be
\label{RSeqmo}
{1\over\gamma_{\rm N}^{(5)}}\left(R^{(5)}_{MN}-
{1\over 2}G_{MN}R_5\right) =
{1\over 2}\Lambda_5G_{MN} + T_{MN}^{(4)}
\ee
(in this section large indices run over five values
$M,N=1,\dots,5$ while the small indices over the four values
$\mu,\nu=1,\dots,4$) can be found in a very simple way,
using the symmetries of the problem. Since $T_{MN}^{(4)}\sim\delta(y)
t_{\mu\nu}^{(4)}(x)\delta_M^\mu\delta_N^\nu$ one can first solve
eqs.~(\ref{RSeqmo}) for $y\neq 0$, which naturally suggest the anzatz of
a "Friedman universe" (\ref{RSanz}). Substitution of
(\ref{RSanz}) into (\ref{RSeqmo}) gives
\be
\label{eq2}
a''(y) +{\Lambda_5\gamma_{\rm N}^{(5)}\over 3}\ a(y) = 0, \ \ \ \ \ y\neq 0
\ee
with the solution
\be
\label{AB}
a(y) = A\exp({ky})+B\exp({-ky})
\\
\Lambda_5\gamma_{\rm N}^{(5)}=-3k^2 < 0
\ee
(the cosmological constant of five-dimensional space is negative).
A natural choice would be
$A=0$ for $y>0$ and $B=0$ for $y<0$, then we have an AdS horizon as
$|y|\to\infty$. On the brane surface at $y=0$ one has to "glue" two
exponents with different signs,
then $a(y)=e^{-k|y|}$, but this would bring us to an extra contribution into
(\ref{RSeqmo}) at $y=0$, i.e. proportional to $\delta(y)$.
However, {\em tuning}
$\Lambda_4\gamma_{\rm N}^{(5)}=3k$ one exactly cancels this term by the
contribution
of the variation of the second term in (\ref{RSact}) so that (\ref{RSeqmo})
also holds at $y=0$. Thus, the solution is finally
\be
\label{RSback}
ds^2 = \exp({-k|y|})(dx_\mu)^2 + dy^2
\ee
so that the {\em effective} cosmological constant in four-dimensional theory
\be
\Lambda_4^{\rm eff} = \Lambda_4 + \int dy\sqrt{G_5}\Lambda_5 = \Lambda_4 +
{\Lambda_5\over k} = 0
\ee
vanishes. Thus, in this scenario the "observable" cosmological
constant $\Lambda_4^{\rm eff}$ classically vanishes independently of any
particular details of the model in a given class.

One of the {\em very} important immediate consequences we got in this
context is that the boundary
conditions (here -- gluing on the brane) reduce exactly half of the bulk modes
existing in the theory. In a more general context this condition could be
different if speaking about its exact form, but one may always express in
(\ref{AB}) $B$ as a function of $A$ or vice versa.

Next question to study is the spectrum of small fluctuations
of the (linearized) action (\ref{RSact}) in the vicinity of the
background (\ref{RSback}). It is easy to see that for the perturbation
$g_{\mu\nu} = a(y)\eta_{\mu\nu}+h_{\mu\nu}(x,y)=a(y)\eta_{\mu\nu}+
\psi^{(p)}_{\mu\nu}(y)e^{ipx}$ one gets an equation
\be
\label{Schro}
\left( -\d_y^2 + p_\mu^2\exp({k|y|})-2k\delta(y)+k^2\right)\psi^{(p)}(y)=0
\ee
rather similar to the Schr\"odinger equation in a $\delta$-function well with a
coefficient $-2k$. From elementary quantum mechanics it is well-known
that there always exists a {\em single} level, localized to this
well (here at $y=0$) with the energy
$E=-k^2$. This immediately gives rise to $p_\mu^2=0$ in (\ref{Schro}),
or to the four-dimensional {\em massless}
graviton which is forbidden to propagate into the fifth direction (to the
bulk) by the exponential wave function $\psi^{(p^2=0)}\sim e^{-k|y|}$.

This is, in fact, a generic phenomenon -- for {\em any} metric of the form
(\ref{RSanz}) with $a(y)=e^{-\alpha(y)}$ with suitable
$$
a(y)\stackreb
{|y|\to\infty}{\to}0
$$
there exists a solution to (\ref{RSeqmo}) with
{\em non} constant bulk "potential" $\Lambda_5(y)$ and $\Lambda_4(y)$,
corresponding in general to some thick brane, satisfying
\footnote{Notice that the expression
${\cal T}(y) = \Lambda_5(y)+\Lambda_4(y)$ has exactly the form of the Miura
stress-energy tensor, widely
appearing in two-dimensional conformal theory, in particular in the
procedure of bosonization or in the
Liouville theory. Such "Virasoro" properties of the
conformal mode of the space-time metric
may serve as a possible origin for the target-space Virasoro symmetries
(\ref{lntau}), often appearing when describing the effective string theory
actions in terms of integrable systems.}
\be
\Lambda_5(y) = -3 \alpha'(y)^2, \ \ \ \ \ \ \ \
\Lambda_4(y) = {3\over 2}\alpha''(y)
\\
\Lambda_5(y) + \Lambda_4(y) = 3\left(-\alpha'(y)^2+{\alpha''(y)\over 2}\right)
\\
\int dy(\Lambda_5 + \Lambda_4)\exp({-2\alpha(y)}) =
{3\over 2}\int dy {d\over dy}
\left(\alpha'\exp({-2\alpha(y)})\right) =
\\ = -{3\over 4}\int dy {d^2\over dy^2}
\exp({-2\alpha(y)}) = 0
\ee
Of course, the "gravity description" presented above
answers almost all simple questions but cannot pretend to
be complete
\footnote{For example, within pure gravity theory it is not clear why the
classical vanishing of cosmological constant is not violated by quantum effects,
say, by contribution of graviton tadpoles etc. This is just one more
manifestation of the main concept of this review: the only way to "quantize"
gravity is to consider it as low-energy limit of string theory.}.
The massive modes $\psi(y)$ can be expressed in terms of the
Bessel functions and their contribution to the deviation from the Newton
law in a four-dimensional world seems to be consistent with
 {\em any} one-loop contribution to the graviton propagator
$\langle h_{\mu\nu}(x)h_{\alpha\beta}(0)\rangle$ which should be of the form
$\int d^4q e^{iqx}k^4\log{q^2\over\mu^2}$ giving rise to ${1\over r^3}$
correction to the potential of four-dimensional gravity.

Now, let us remember that gravity arises only as an effective description of
string theory and in the string theory picture the previous formulas
can be understood in the following way.
Consider the generating functional of
string theory in the background (\ref{RSanz}), (\ref{polact})
\be
\label{polstr}
\int D\varphi DX \exp\left({-\int_{\Sigma}a(\varphi)\d X_\mu\bar\d X_\mu +
\d\varphi\bar\d\varphi + {\cal R}^{(2)}\Phi(\varphi) + \dots}\right)
\ee
so that the zero modes of $\left.X_\mu(\sigma)\right|_0=x_\mu$
play the role of four-dimensional co-ordinates in (\ref{RSanz})
while the zero mode of
the Liouville field $\left.\varphi(\sigma)\right|_0=y$ is the extra
bulk co-ordinate. The action (\ref{polstr}) should be
consistent in the sense of string theory, in particular after the
integration over co-ordinates $X_\mu$, the arising correction
действию
\be
\label{det}
\int DX \exp\left({-\int_{\Sigma}a(\varphi)\d X_\mu\bar\d X_\mu }\right) =
\det \left({\bar\d} a(\varphi)\d\right) ^{-D/2}= \\ =
\exp\left({-\int_{\Sigma}\d\alpha\bar\d\alpha
+ {\cal R}^{(2)}\alpha + \dots}\right)
\ee
should not break the conformal invariance (independence of the macroscopic
theory of the choice of the world-sheet co-ordinates). In the last formula,
which is a particular case of a general anomaly formula from \cite{KaMo},
$\alpha=\alpha(\varphi)=\log a(\varphi)$, and the anomaly contributions
depending only on metric are marked by dots.
We see, that, identifying the Liouville or dilaton field with the
fifth co-ordinate, eq.~(\ref{det}) gives rise to a
reparameterization in the fifth dimension $\varphi\to\varphi+\alpha(\varphi)$
and $\Phi(\varphi)\to\Phi(\varphi)+\alpha(\varphi)$.
For the particular background (\ref{RSback})
one gets just a trivial {\em renormalization} of the string action for
the Liouville component. In particular, this means that the
background (\ref{RSback}) is {\em stable} against string corrections. The
integration over $X_\mu$-coordinates is effectively equivalent to the
study of
nontrivial dependence only upon fifth coordinate in the bulk theory, taking
four-dimensional branes as effective boundary sources and this is quite
similar to what we have considered above in the classical gravity
approximation. Moreover, the
solution $\alpha=\alpha(\varphi)=\log a(\varphi)$ is the only one naively
consistent with the requirement of world-sheet conformal invariance.

\setcounter{equation}0
\section{Some New Directions in String Theory
\label{ss:new}}

Finally in this review let us say a few words about the
directions which have begun development only in recent years. We will
discuss only few such
directions and let us note immediately that the understanding of most of the
problems considered in this chapter deserves to be better.

\subsection{M(atrix) Theory}

M(atrix) theory \cite{Mmatrix} is one of the most interesting (though not
very successful) attempts to
construct an alternative to strings formalism in M-theory. For
the role
of such formalism some particular {\em matrix} quantum mechanics is
proposed. This origins already in its name and special attention to the
first letter can be considered as a rather transparent hint that this letter
should be identified with "M" in M-theory and the rest of the word "matrix"
can be omitted.

As a building blocks m(atrix) theory uses the $N\times N$ matrices
$X_i$, $i=1,\dots,9$, whose diagonal elements can be interpreted as the
transverse co-ordinates of the D$0$-branes (their number is equal to
$N$) in the light-cone co-ordinates in the eleven-dimensional
compactified M-theory. The Lagrangian of such a theory can be written in the
form
\be
\label{mmatrix}
\int_{dt}\Tr\left({1\over 2R}{\dot X}_i^2 + M_{\rm pl}^6R\sum_{i<j}[X_i,X_j]^2
+ \dots\right)
\ee
where the dots correspond to omitted fermionic terms. Eq.~(\ref{mmatrix})
explicitly contains the eleven-dimensional Planck mass $M_{\rm pl}$
(cf. with formulas (\ref{mpl11}) and (\ref{relM})), together with the radius
of the compact dimension $R$, which in the formalism of matrix theory somewhat
artificially corresponds to the light-cone co-ordinate $X_-$. Hence, nine
transverse co-ordinates and two light-cone co-ordinates -- time and
compactified $X_-$, corresponding to the trace over matrices in
(\ref{mmatrix}), together form the eleven-dimensional
target space of M-theory.

The quantum mechanical action (\ref{mmatrix}) can be interpreted in the
following way. If $N=1$, action (\ref{mmatrix}) corresponds to the
Hamiltonian $H\sim P^2$ and the ground state is degenerate with respect to
all auxiliary (absent explicitly in (\ref{mmatrix})) Grassmann variables
$\theta_\alpha$. Simple counting of all states shows (see, for example
\cite{BiSumma}), that their total number is $2^8=256$, so that half of them
are bosonic: $44={9(9+1)\over 2}-1$ gravitons and $84$ of antisymmetric
tensor field, and half of them are fermionic.

Thus, the "vacuum" of m(atrix)
theory corresponds to the supergraviton, or, better to say, the supergraviton
multiplet of eleven-dimensional supergravity \cite{11SUGRA}, in which the
only bosonic fields are metric and three-form. It is also claimed that
nontrivial solutions to the equations of motion in m(atrix) theory can be
identified with a membrane, fivebrane etc. For example, in the
"quasiclassical"
$N\to\infty$ action (\ref{mmatrix}) can be rewritten, replacing the
commutator with the Poisson bracket in auxiliary variables
$(\sigma_1,\sigma_2)$
\be
\label{membrane}
\int_{dt}\int_{d^2\sigma}\left({1\over 2R}{\dot X}(\sigma_1,\sigma_2)_i^2 +
M_{\rm pl}^6R\sum_{i<j}\left\{X(\sigma_1,\sigma_2)_i,
X(\sigma_1,\sigma_2)_j\right\}_{PB}^2
+ \dots\right)
\ee
This action can be identified with the action of membrane in the light-cone
gauge.

The hope for a matrix formalism in nonperturbative string theory has not been
justified in the sense that the new formalism appeared to be not very
effective in solving essential problems. Nevertheless, it can be already
considered as a "relative success" that at least some properties of string
theory and eleven-dimensional M-theory can be extracted from this at
first glance totally
absurd concept. To finish this section let us note, that
some problems of the m(atrix) formalism were discussed in \cite{MaZa}.

\subsection{Non-commutative Field Theories}

{\bf Non-commuting co-ordinates}. The fact that the co-ordinates of the
"stack" of D-branes from the point of view of effective field theory become
eigenvalues of the matrix of scalar field in the adjoint representation is
sometimes interpreted as appearance of {\em non-commuting coordinates}. In
the first-quantized formalism one may consider this as a relatively simple
and formal representation for the effective theories in terms of
D$0$-branes, D-strings etc, studying the corresponding matrix quantum
mechanics or two-dimensional non-Abelian gauge theory.

\noindent
{\bf Non-vanishing background $B$-field}. Another manifestation of
non-commutativity shows up (see \cite{SWNC}) if we consider string theory in the nontrivial
background $B$-field (\ref{B}), for example
\be
\label{Bconst}
B_{\mu\nu} = B\epsilon_{\mu\nu}
\\
B = {\rm const}
\ee
This case can be clearly understood by analogy with the well-known
example of a charged particle in a constant magnetic field. Indeed, the
interaction, say, with the constant $B$-field (\ref{Bconst}), is performed
over the whole surface of the world sheet
\be
\label{Bfo}
\int_{\Sigma} B_{\mu\nu}dX^{\mu}\wedge dX^{\nu} =
\int_{\d\Sigma} B_{\mu\nu}X^{\mu}dX^{\nu}
\ee
and by the Stocks formula it can be rewritten as a boundary term, equivalent
to the interaction of a string with the vector-potential
$A_\mu(X) = B_{\mu\nu}X^{\nu}$, corresponding to the constant magnetic
field. If the value of the $B$-field is large enough the contribution of
the term (\ref{Bfo}) to the two-dimensional correlator of the fields
$X(t)= \left.X\right|_{\d\Sigma}$ dominates
\be
\langle X_\mu(t)X_\nu(t')\rangle \propto \epsilon_{\mu\nu}{\rm sign}(t-t')
\ee
and in the field-theory limit this corresponds to non-commuting
coordinates
\be
\label{commX}
[X_\mu,X_\nu] = \zeta\epsilon_{\mu\nu}
\ee
where $\zeta\sim {1\over B}$. This reasoning is in fact a rather rough
illustration of the well-known effect when the role of non-commutative
variables is played by the centers of (small) circles -- the
trajectories of particles in a magnetic field.

The corresponding effective field theory can be described by a Lagrangian,
where all the products are replaced with the so called Moyal products
\be
\label{moyal}
f(x)*g(x) = \left.\exp\left(\epsilon_{\mu\nu}{\d\over\d x_\mu}
{\d\over\d y_\nu} \right)f(x)g(y)\right|_{x=y} = f(x)g(y) +
\{ f(x),g(x)\} + {\cal O}(\d^2)
\ee
where $f$ и $g$ are any two functions (local functionals)
of "ordinary" fields $\phi(x)$, and
\be
\{ f(x),g(x)\} = \epsilon_{\mu\nu}{\d f\over\d x_\mu}
{\d g\over\d x_\nu}
\ee
is the Poisson bracket, corresponding to the "quasiclassical" limit of the
commutator (\ref{commX}). The Lagrangians where the fields are multiplied by
the law (\ref{moyal}), obviously contain infinitely many derivatives
\footnote{Despite this, their ultraviolet properties are not better than
the corresponding properties of ordinary, i.e. commutative quantum field
theories.}. Examples of non-commutative field theories usually include the
theories of scalar fields
\be
\label{scalnc}
S = \int_{dx}\left(\2\d_\mu\phi*\d_\mu\phi + V(\phi)\right) =
\int_{dx}\left(\2\d_\mu\phi\d_\mu\phi + V(\phi)\right)
\ee
where $*$-multiplication (\ref{moyal}) is essential only in the interaction
terms, and the gauge theories
\be
\label{noncym}
F_{\mu\nu} = \d_\mu A_\nu - \d_\nu A_\mu + A_\mu*A_\nu - A_\nu*A_\mu
\\
S = {1\over g^2}\int_{dx} F_{\mu\nu}*F_{\mu\nu}
\ee
which are rather natural generalization of the Yang-Mills theories. Notice,
that in contrast to commutative case already the Abelian variant of
(\ref{noncym}) is a nontrivial interacting theory. Practically without any
changes (just considering $A_\mu(x)$ as matrix-valued functions of
non-commuting variables and adding the trace over matrix indices) formula
(\ref{noncym}) defines also the noncommutative Yang-Mills theories.

The most interesting by now applications of the non-commutative field
theories are their classical solutions.

\noindent
{\bf Solitons and instantons in non-commutative theories}. In contrast to
common scalar field theories where the existence of localized classical
solutions is forbidden by scaling arguments in almost all dimensions
(starting with $D\geq 2$), such solutions can arise in non-commutative field
theories where the scaling is much less trivial due to an extra dimensional
parameter ($\zeta$ in the formula
(\ref{commX})) \cite{GMS}. The simplest is the two-dimensional case. After the
scale transformation of co-ordinates $X\to\sqrt{\zeta}X$
in the action (\ref{scalnc}), one gets for the two-dimensional (or static
three-dimensional) case
\be
E = \int_{d^2x}\left(\2(\d\phi)^2 + \zeta V(\phi)\right)
\ee
and as $\zeta\to\infty$ the solution and its energy is completely determined
by potential terms. The stationarity equation is reduced in such case, for
example, for the potential
$V(\phi) = {m^2\over 2}\phi^2 + {\lambda\over 3}\phi^3$, to
\be
\label{soleqn}
m^2\phi + \lambda\phi*\phi = 0
\ee
With normal multiplication, the solutions to (\ref{soleqn}) would be "maps
into the set of points" $\phi(x) = 0$ and $\phi(x) = -{m^2\over\lambda}$,
however non-commutativity "washes away" these points in the space of fields.
Indeed, formally a solution to
(\ref{soleqn}) can be written as
$\phi = -{m^2\over\lambda}{\hat P}$, where ${\hat P}$ is the projector,
i.e., generally, any operator with the property
${\hat P}^2={\hat P}$. In two-dimensional non-commutative space (isomorphic
to the phase space of quantum mechanics with the only degree of freedom)
projectors can easily be constructed in terms of, say, the Fock space
operators. For example, one can take
${\hat P}_n \sim |n\rangle\langle n|$, where $|n\rangle$ -- is the state of
$n$-th energy level of harmonic oscillator. One can write correspondingly
their representation in (non-commutative) $x$-space, the simplest solution
will have a form of "bell" $\phi_0(x) = -{2m^2\over\lambda}\
\exp({-(x_1^2+x_2^2)})$.

In the non-commutative gauge theories the main interest is caused by the
instanton solutions \cite{NeSch}. In contrast to commutative theory, the
nontrivial solutions to the self-duality equations arise already in the case
of Abelian (noncommutative) group $U(1)$. From the physical point of view their
main attraction is that they do not contain the singularities of the
"zero-size" $1/x^4$ any longer (for example, in the expression for the field-strength
at $\rho = 0$ in formula (\ref{bpst})), the parameter of non-commutativity
turns the non-integrable singularity in four dimensions
$1/x^4$ into the integrable expression $1/x^2(x^2+\zeta)$. Construction of
the solutions is almost the same as in the commutative case with the only
distinction being replacement, as much as possible, of ordinary multiplication
by the Moyal $*$-multiplication (\ref{moyal}).

The detailed discussion of different aspects of the non-commutative theories
can be found, for example, in the review \cite{DouNe}.

\subsection{Tachyon Potential
\label{ss:tachpot}}

One of the main problems of many well-known string models is the presence of
tachyons or states with negative squared masses. The tachyons lead, in
particular, to infrared divergences in string amplitudes and since the
infrared and ultraviolet regions are identified by two-dimensional geometry
this problem "screens" the ultraviolet finiteness of string theory.

\begin{figure}[tb]
\epsfysize=8.5cm
\centerline{\epsfbox{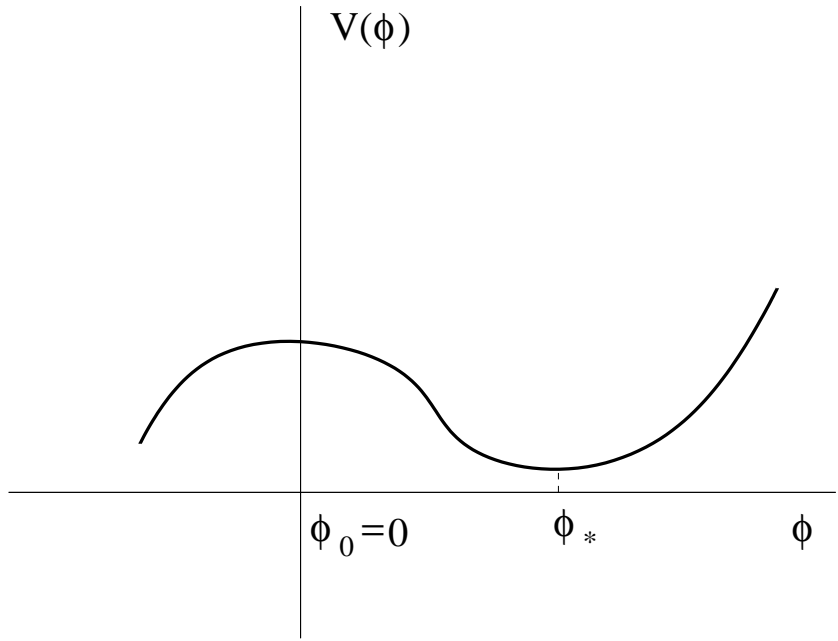}}
\caption{\sl Effective potential with minimum at
$\phi=\phi_*$ and extremum at $\phi=\phi_0=0$. The value
$\phi_0=0$ is an extremum point for the potential
$V'(\phi_0)=0$, but the second derivative is negative
$V''(\phi_0)<0$, what corresponds to the presence of the
tachyon in the vicinity of this point.}
\label{fi:tachpot}
\end{figure}

The interpretation of negative masses is absolutely clear in field theory
(in particular, in the effective field theories for string models with
tachyons) and it causes the {\em instability} of the corresponding vacuum.
Indeed, drawing the effective potential with the requirement
$m^2 = V''(\phi_0) <0$, we immediately see
(see fig.~\ref{fi:tachpot}),
that corresponding point (in the space of fields) is a local extremum but
not a minimum, and under any perturbation the theory "runs" in the "true"
vacuum at $\phi=\phi_*$.

Unfortunately, string theory by now does not have any self-consistent
second-quantized formalism or string field theory
\footnote{The problems of constructing string field theory go beyond the
scope of this review. Notice only that there exists a huge amount of
literature, devoted to this problem, whose total volume can be
easily comparable with amount of literature devoted to all other problems of
string theory, taken altogether. The
possibility of construction of string field theory is seriously restricted
at least by absence of "universal variables" which allow to see
{\em all} (and not a single!) string vacua, since the first-quantized string
theory is described in terms of {\em different} variables -- two dimensional
conformal field theories -- in the vicinity of different vacua. Another
problem is that at least in the closed string theory the counting of states
in the loops is different for a different number of loops (or for Riemann
surfaces with different genera) and therefore, in contrast to quantum field
theory, it is very hard (if possible!?)
to write down a Lagrangian, taking into account
this obstacle.}
at least in the form, like the second-quantized approach exists in quantum
field theory. Say, any field theoretical Lagrangian with the potential
depicted on fig.~\ref{fi:tachpot}, allows one immediately to see both stable
$\phi=\phi_*$, and non-stable $\phi=\phi_0$ vacua. This effect cannot be
really seen in string theory since there is no formalism (yet?), which
would allow one to consider the points $\phi=\phi_0$ and
$\phi=\phi_*$ simultaneously.

In the bosonic string theory the existing formalism allows to compute
amplitudes in the vicinity of vacuum of $\phi=\phi_0$ type, generally
with two tachyons -- from the open and closed spectrum. A.Sen
\cite{sentach} has proposed a nice D-brane interpretation, which allows
partially to get rid of the tachyon of the open spectrum. It is based on the
fact that the bosonic open string theory may be interpreted as D$25$-brane
(the Dirichlet brane of dimension $p=25$), whose world volume fills in the
whole twenty-six-dimensional space-time. Equally the ten-dimensional
superstring can be seen as a D$9$-brane. The standard way to get rid of the
tachyon in ten-dimensional superstring -- the GSO-projection
\cite{GSO}, which was already discussed in sect.~\ref{ss:susy} --
in fact corresponds to fig.~\ref{fi:dbrane} with parallel BPS D-branes. From
some perspective this may even be considered as a definition of what is
drawn on fig.~\ref{fi:dbrane}.

Sen proposed to interpret the tachyon as a ground
state of string, stretched between the Dirichlet and anti-Dirichlet branes,
defining such a configuration as corresponding to the "opposite sign" in
the GSO
projection. It should be noted here that it corresponds only to the
"non-diagonal" or "non-Abelian" tachyon of the open-string spectrum, since
it corresponds to a string stretched between two different branes. Such
a situation, in contrast to non-interacting parallel D-branes, is unstable.
The Dirichlet and anti-Dirichlet branes tend towards each other and want to
annihilate. From the energy conservation it follows that
(see fig.~\ref{fi:tachpot})
\be
V(\phi_0) - V(\phi_*) = 2T_D
\ee
where $T_D$ -- is the D-brane or anti-D-brane tension.

Moreover, since it is possible to stretch two strings between the D-brane
and the anti-D-brane, different by orientation, the corresponding tachyon
field becomes complex, and the potential from fig.~\ref{fi:tachpot} should be
"complexified" by rotation around the vertical axis. Then it becomes similar
to "bottom of a bottle" well-known in the framework of the Standard Model. The
effective theory in such potential possesses "kink"-like solutions depending
on some space-time co-ordinate $x$. For such solution one may take
$\phi(x)\stackreb{x\to +\infty}
{\to}|\phi_*|\exp({i\theta_1})$ and $\phi(x)\stackreb{x\to -\infty}
{\to}|\phi_*|\exp({i\theta_2})$, with $\theta_1\neq \theta_2$.

Hence, if
the tachyon under discussion corresponds to the pair of D$p$- and
anti-D$p$-branes, the arising kink is very similar to an extended object of
a dimension less by unity, i.e. to a D$(p-1)$-brane. This kink is also
unstable and it exists together with an "anti-kink" -- a solution running along
the co-ordinate $x$ to the opposite direction. It is natural to interpret
the anti-kink as an anti-D$(p-1)$-brane, and continue this procedure by
induction. Such qualitative reasoning leads to the idea, that "falling down"
along the tachyon potential depicted at fig.~\ref{fi:tachpot},
from the point $\phi_0=0$ to the point
$\phi=\phi_*$, and starting with a pair of D$p$- and anti-D$p$-branes,
where $p=D-1$ -- is the dimension of our space (without time), we will find on
our way many local extrema corresponding to the branes of smaller
dimensions and finally will arrive at the "true" vacuum $\phi=\phi_*$,
where the open string excitations are simply absent.

Unfortunately this sort of reasoning does not allow to compute the exact
tachyon potential, even for restricted class of tachyonic fields. The only
way to calculate such quantities is to use the effective actions which were
discussed in sect.~\ref{ss:fratse}. Literally this method can be applied
only in the vicinity of "false" vacuum $\phi_0=0$ of the tachyon potential,
where corresponding two-dimensional conformal theory is a theory of free
fields. However, there have been many attempts to "extrapolate" the results of
such computations towards the direction of "real vacuum"
$\phi=\phi_*$ (see, for example, \cite{tsef75}). Moreover, one can
even find
claims that the tachyon potential in tree-level approximation can be
computed {\em exactly} \cite{gesha}, and equals to the rather simple
expression (for the canonical kinetic term)
\be
V(\tilde\phi) = -\2\tilde\phi^2\log\tilde\phi
\ee
with $\tilde\phi\sim \exp({-\phi})$. Despite the arguments in favor of this
formula deserve to be more strict, qualitatively this means that in "true
vacuum" $\tilde\phi=0$ or $\phi\to\infty$ the mass of tachyon field becomes
infinite, and it is consistent with the Sen hypothesis about the disappearance
of all excitations of the open string spectrum.

\section{Conclusion. String Theory or Field Theory?}

In this review we have tried to discuss the main aspects of string theory in
the form, as it exists at present. Certainly, as any physical theory
detached from experiment it looks like it is "flying in the air" and the
only excuses for such theory may come from new ideas, which have shown up
inside string theory and, very slowly, affect the modern scientific paradigm
of what is quantum field theory.

It becomes more and more evident that microworld physics cannot be
simply reduced to an infinite set or "media" made of harmonic oscillators.
Such theories arise only as the low-energy effective description of phenomena in
the weak-coupling regime, which however finds lots of applications both in
elementary particle and condensed matter physics. However, the main
physical problems, which are not now understood, are contrarilyy related to
the strong-coupling phase or strong filed regime, or exactly where the
traditional quantum field theory or "theory of oscillator" does not have new
successes. The very popular attempts thirty or even twenty years ago to
develop "correct" or "general" formalism in quantum field theory, such that
its computations can be "prolonged" towards the strong coupling look less and
less promising. String theory in contrast implies (and originally implied)
the existence of a principally new perspective on the problems of strong
coupling.

Having
appeared almost phenomenologically in the theory of strong interactions, the
theory of one-di\-men\-si\-o\-nal extended objects gained huge popularity
because, at variance with many other languages, it proposed a reformulation of
many problems in terms of extremely simple two-dimensional conformal field
theory, where the structure of computations is under the rigid control of
infinite-dimensional symmetry and complex geometry, in particular by the
language of complex analytic functions. Despite the observable world being
multidimensional, the string scattering amplitudes are expressed through the
correlation functions in two-dimensional conformal theories with
well-defined operator product expansions etc. Moreover, the majority of
target-space multidimensional symmetries are in this or that way related to
the two-dimensional symmetries of the world-sheet theories.

In string theory the approach based on a dual description of the strong
coupling effects was proposed and developed. Rather soon, this approach led
to a certain hypothesis about nonperturbative results in supersymmetric gauge
theories. These results are beyond the framework of traditional
field-theoretical methods and allow one to get a deeper understanding of the
problem of quark confinement.

String theory seems to be the only natural continuation of General
Relativity to the region of strong fields and small distances. The (almost
obvious) idea that there can be no quantum gravity in the framework of
quantum field theory since these are two totally different theories
is becoming
more and more widespread. The appearance of time as a scale factor together
with the distinguished role of solutions similar to the "Friedman universe"
demonstrate deep internal relations between gravity and string theory.

Thus, the experience of development of string theory brought lots of rich new
ideas into modern science. The only trouble is that string theory now
does not possess not only a well-developed, but even any fixed
formalism, allowing to perform computations of physical effects without
applying of some "intuition". All these problems exist on the background of
enforced development of connections with different spheres of mathematics
and mathematical physics, and it allows to think that these problems have
a temporary and mathematical, but not physical character. On the other hand,
it is very nice to believe that it is a necessity to apply continuously physical
intuition is called Theoretical Physics.

\vspace{0.5 cm}
\noindent
I am grateful to V.L.Ginzburg who proposed to collect several lectures and
write a review to {\em Physics Uspekhi} and to
S.Apenko, I.M.Dremin and L.B.Okun for reading the manuscript and
useful comments. I am also grateful to E.Akhmedov, A.Barvinsky, I.Batalin,
V.Ya.Fainberg, V.Fock, A.Gerasimov, A.Gorsky, Vl.Dotsenko, V.Kazakov,
S.Kharchev, I.Kri\-che\-ver, A.Losev, V.Losyakov, Yu.Makeenko, R.Metsaev,
A.Mironov,
N.Nekrasov, A.Polyakov, A.Rosly, V.Rubakov, K.Saraikin, J.H.Schwarz,
K.Selivanov, A.Tseytlin, I.Tyutin, A.Vainshtein, A.Yung,
A.Zabrodin
and, especially to B.Voronov and A.Morozov for lots of useful discussions of
various questions considered in this text. I am also indebted to S.Apenko,
H.Braden and E.Corrigan for useful advice concerning the English version of this
review.

The work was partially supported by the RFBR grant
No.~00-02-16477, the INTAS grant No.~00-00561
and the grant of support of scientific schools No.~00-15-96566. I would like
to thank Laboratoire de Physique
Th\'eorique de l'\'Ecole Normale Sup\'erieure and Institut des Hautes
\'Etudes Scientifiques, where this work was completed, for the warm
hospitality


\newpage
\begin{thebibliography}{799}


\bibitem{Pol}
A.~Polyakov, {\sl Gauge Fields and Strings}, Harwood Academic Publishers, 1987.

\bibitem{GSW}
M.~Green, J.~Schwarz and E.~Witten, {\sl Superstring Theory}, Cambridge
University Press, 1987.

\bibitem{Polch}
J.~Polchinski, {\sl String Theory}, Cambridge University Press, 1998.

\bibitem{SlavFad}
A.~Slavnov and L.~Faddeev, {\sl An Introduction to the Quantum Theory of
Gauge Fields}, Moscow, Nauka, 1978.

\bibitem{Okun}
L.~Okun, {\sl Leptons and Quarks}, Nauka 1990

\bibitem{Andreev}
I.~Andreev, {\sl Chromodynamics and hard processes at high energies}, Nauka
1981.

\bibitem{SUSYWest}
P.~West, {\sl Introduction to Supersymmetry and Supergravity}, WS 1986.

\bibitem{Mbook}
A.~Marshakov, {\sl Seiberg-Witten Theory and Integrable Systems},
World Scientific, 1999.

\bibitem{Kni}
V.~Knizhnik, Physics Uspekhi {\bf 32} (1989) 945.

\bibitem{Mor}
A.~Morozov, Physics Uspekhi, {\bf 162} (1992) 84-176.

\bibitem{Mormat}
A.~Marshakov,
Int.\ J.\ Mod.\ Phys.\ A {\bf 8}, 3831 (1993), hep-th/9303101;\\
A.~Morozov, Physics Uspekhi, {\bf 164} (1994) 3-62;\\
M.~Adler and P.~van Moerbeke,
Commun.\ Pure Appl.\ Math.\  {\bf 50}, 241 (1997), hep-th/9706182.



\bibitem{MaZa}
K.~Zarembo and Yu.~Makeenko, Physics Uspekhi, {\bf 168} N 1 (1998) 3.

\bibitem{AkhmedovStrR}
E.~Akhmedov, Physics Uspekhi, {\bf 171} (2001) 1005-1024; hep-th/9911095.

\bibitem{SchwarzStrR}
J.~H.~Schwarz,
hep-ex/0008017; hep-th/0011078;
hep-th/9812037; hep-th/9807135; hep-th/9711029.


\bibitem{SenStrR}
A.~Sen,
Nucl.~Phys.~Proc.~Suppl. {\bf 94} (2001) 35-48, hep-lat/0011073;
hep-th/9904207;
hep-th/9802051.

\bibitem{PolchST}
J.~Polchinski,
hep-th/9411028.

\bibitem{PolchDbr}
J.~Polchinski,
hep-th/9611050.

\bibitem{VafaStrR}
C.~Vafa,
hep-th/9810149;
hep-th/9702201.

\bibitem{TownsendMR}
P.~Townsend, hep-th/ 9612121.

\bibitem{SchwarzMR}
J.~H.~Schwarz, hep-th/ 9607201.

\bibitem{LosevMR}
A.~Losev, M-theory for pedestrians, preprint ITEP, unpublished.

\bibitem{loopsstR}
E.~Verlinde and H.~Verlinde,
``Lectures On String Perturbation Theory,''
Published in Trieste School 1988: Superstrings:189 (QCD161:T7322:1988);
\\
E.~D'Hoker and D.~Phong, Rev.~Mod.~Phys. {\bf 60} (1988) 917-1065\\
and references therein;\\
E.~D'Hoker and D.~Phong,
hep-th/0110247, hep-th/0110283, hep-th/0111016, hep-th/0111040.

\bibitem{mirrorVafa}
C.~Vafa,
hep-th/9111017.

\bibitem{mirrorWitten}
E.~Witten,
hep-th/9112056.

\bibitem{mirrorAspinwall}
P.~Aspinwall,
hep-th/0001001.

\bibitem{mirrorMorrison}
D.~Morrison,
math.ag/0007090.

\bibitem{Mtmf}
A.~Marshakov,
Theor.~\& Math.~Phys. {\bf 112} (1997) 3-46, hep-th/9702083;
Theor.~\& Math.~Phys. {\bf 121} (1999) 179-243

\bibitem{BiSumma}
D.~Bigatti and L.~Susskind,
hep-th/9712072.

\bibitem{BiSuTAHo}
D.~Bigatti and L.~Susskind,
hep-th/0002044.

\bibitem{branezoo}
A.~Giveon and D.~Kutasov,
Rev.~Mod.~Phys. 71 (1999) 983-1084, hep-th/9802067;\\
S.~Kachru,
hep-th/0009247;\\
P.~Argyres and K.~Narayan,
JHEP {\bf 0103}, 047 (2001), hep-th/0101114.

\bibitem{SUSYOM}
V.~Ogievetsky and L.~Mesincescu, Physics Uspekhi {\bf 117} (1975) 637.

\bibitem{VZNSUFN}
A.~Vainshtein, V.~Zakharov, V.~Novikov and M.~Shifman,
Physics Uspekhi {\bf 136} (1982) 553-591;
see also
M.~Shifman and A.~Vainshtein,
``Instantons versus supersymmetry: Fifteen years later,''
hep-th/9902018.



\bibitem{FadUFN}
L.~Faddeev, Physics Uspekhi {\bf 25} 130 (1982)

\bibitem{MorAn}
A.~Morozov, Physics Uspekhi {\bf 150} (1986) 337-416.

\bibitem{NeOsc}
S.~Gershtein, E.~Kuznetsov and V.~Ryabov, Physics Uspekhi
{\bf 40} (1997) 773-806.


\bibitem{Rubufn}
V.~Rubakov, Physics Uspekhi {\bf 169} (1999) 1299.

\bibitem{tHoVeufn}
G.~'t~Hooft, Physics Uspekhi {\bf 170} (2000) 1218-1224;\\
M.~Veltman, Physics Uspekhi {\bf 170} (2000) 1225-1234.

\bibitem{rubrane}
V.~Rubakov, Physics Uspekhi, {\bf 171} (2001) 913-938; hep-ph/0104152.

\bibitem{VyNe}
M.~Vysotsky and R.~Nevzorov, Physics Uspekhi, {\bf 171} N 9 (2001) 939-950.

\bibitem{SUSYL}
E.~Likhtman, Physics Uspekhi, {\bf 171} (2001) 1025-1032.


\bibitem{Wils}
K.~Wilson and J.~Kogut,
Phys.~Rep. {\bf 12C} (1974) 75-199.

\bibitem{Scherk}
J.~Scherk, in {\sl Recent Developments in Gravitation}, eds. M.~Levy and
S.~Deser, Plenum, 1979.

\bibitem{Stra}
M.~Strassler, Prog. Theor. Phys. Suppl. {\bf 131} (1998) 439;
hep-th/9803009;
{\em On confinement and duality}, Lecture notes for Trieste
Spring School 2001.

\bibitem{Yung}
A.~Yung,
hep-th/0005088.

\bibitem{DouNe}
M.~Douglas and N.~Nekrasov,
hep-th/0106048.

\bibitem{Ven}
G.~Veneziano, Nuovo Cim. {\bf 57A} (1968) 190.

\bibitem{NG}
Y.~Nambu, Proc.~Int.~Conf. on Symmetries and Quark Models, Wayne State Univ.,
Gordon and Beach 1970, 269-275;\\
T.~Goto, Progr.~Theor.~Phys. {\bf 46} (1971) 1560.

\bibitem{NS}
A.~Neveu and J.~Schwarz, Phys.~Lett. {\bf 34B} (1971) 86.

\bibitem{R}
P.~Ramond, Phys.~Rev. {\bf D3} (1971) 2415.

\bibitem{ShSch}
J.~Sсherk and J.~Schwarz, Nucl.~Phys. {\bf B81} (1974) 118.

\bibitem{GSO}
F.~Gliozzi, J.~Scherk and D.~Olive, Phys.~Lett. {\bf 65B} (1976) 282.

\bibitem{Pol81}
A.~Polyakov,
Phys.~Lett. {\bf B103} (1981) 207;
Phys.~Lett. {\bf B103} 211.

\bibitem{BPZ}
A.~Belavin, A.~Polyakov and A.~Zamolodchikov, Nucl. Phys. {\bf B241} (1984)
333.

\bibitem{WZW}
E.~Witten, Commun. Math. Phys. {\bf 92} (1984) 455-472.

\bibitem{KZ}
V.~Knizhnik and A.~Zamolodchikov, Nucl. Phys. {\bf B247} (1984) 83-103.

\bibitem{GS84}
M.~Green and J.~Schwarz, Phys.~Lett. {\bf 149B} (1984) 117.

\bibitem{FraTse}
E.~Fradkin and A.~Tseytlin, Nucl.~Phys. {\bf 261} (1985) 1-27.

\bibitem{BKni}
A.~Belavin and V.~Knizhnik, JETP {\bf 91} (1986) 364-390.

\bibitem{mamo}
V.~Kazakov,
Phys.\ Lett.\ B {\bf 150} (1985) 282.
\\
F.~David, Nucl. Phys. {\bf 257} (1985) 45.

\bibitem{ds}
V.~Kazakov, Mod. Phys. Lett., {\bf A4} (1989) 2125.

\bibitem{duadbr}
J.~Polchinski, Phys. Rev. Lett. {\bf 75} (1995) 4724, hep-th/ 9510017;
Prog. Theor. Phys. Suppl. {\bf 123} (1996) 9, hep-th/ 9511157;
Rev. Mod. Phys. {\bf 68} (1996) 1245,  hep-th/ 9607050.

\bibitem{KaKle}
T.~Kaluza, Sitzungber. Preuss. Akad. Wiss. Berlin, Math.-Phys. KA, 1966
(1921);\\
O.~Klein, Zs. Phys., {\bf 37} (1926) 895.

\bibitem{SUSYGL}
Yu.~Golfand and E.~Likhtman, JETP Lett. {\bf 13} (1971) 452.

\bibitem{BeHaw}
J.~Bekenstein, Lett. Nuovo Cim., {\bf 4} (1972) 737;\\
S.~Hawking, Comm.~Math.~Phys. {\bf 43} (1972) 737.

\bibitem{GVP}
D.~Gross and F.~Wilczek, Phys.~Rev.~Lett. {\bf 30} (1973) 1343-1346;\\
H.~Politzer, Phys.~Rev.~Lett. {\bf 30} (1973) 1346-1349.

\bibitem{SUSYWZ}
J.~Wess and B.~Zumino, Nucl. Phys. {\bf B70} (1974) 39.

\bibitem{tHoPo}
G.~'t~Hooft, Nucl.~Phys. {\bf B79} (1974) 276;\\
A.~Polyakov, JETP Lett. {\bf 20} (1974) 194.

\bibitem{thooft}
G.~'t~Hooft, Nucl.~Phys. {\bf B72} (1974) 461.

\bibitem{Pol75}
A.~Polyakov, Phys.~Lett. {\bf B59}(1975) 79;
A.~Belavin and A.~Polyakov, JETP Lett. {\bf 22} (1975) 503-506.

\bibitem{BPST}
A.~Belavin, A.~Polyakov, A.~Schwarz and Yu.~Tyupkin,
Phys.~Lett. {\bf B59} (1975) 85-87.

\bibitem{BPS}
E.~Bogomolny, Sov.~J.~Nucl.~Phys. {\bf 24} (1976) 449;\\
M.~Prasad and C.~Sommerfield, Phys.~Rev.~Lett. {\bf 35} (1975) 760.

\bibitem{11SUGRA}
E.~Cremmer, B.~Julia and J.~Scherk,
Phys.\ Lett.\  {\bf 76B} (1978) 409.

\bibitem{SW}
N.~Seiberg and E.~Witten, Nucl. Phys. {\bf B426} (1994) 19,
hep-th/ 9407087;
Nucl. Phys. {\bf B431} (1994) 484,
hep-th/ 9408099.

\bibitem{ChaPa}
J.~Paton and H.~Chan, Nucl.~Phys. {\bf B10} (1969) 516.

\bibitem{GeSa}
J.~-L.~Gervais and B.~Sakita, Nucl.~Phys. {\bf B34} (1971) 632.

\bibitem{SUSYVA}
D.~Volkov and V.~Akulov, Phys. Lett. {\bf 46B} (1973) 109.

\bibitem{BDHDZ}
L.~Brink, P.~Di Vecchia, P.~Howe, S.~Deser and B.~Zumino,
Phys.~Lett. {\bf B64} (1976) 435-438.

\bibitem{BDH&DZ}
S.~Deser and B.~Zumino,
Phys. Lett. {\bf B65} (1976) 369-373;\\
L.~Brink, P.~Di Vecchia and P.~Howe,
Phys. Lett. {\bf B65} (1976) 471-474.

\bibitem{GSsst}
M.~Green and J.~Schwarz, Nucl.~Phys. {\bf B181} (1981) 502; Phys.~Lett. {\bf
109B} (1982) 444; Phys.~Lett. {\bf 136B} (1984) 367.

\bibitem{bosonizaFF}
B.~Feigin and D.~Fuchs, Func.~An. \& Appl. {\bf 16} (1982) 47.

\bibitem{bosonizaDF}
Vl.~Dotsenko and V.~Fateev, Nucl.~Phys. {\bf B240} (1984) 312-348;
{\bf 251} (1985) 691.

\bibitem{heter}
D.~Gross, J.~Harvey, E.~Martinec and R.~Rohm, Phys.~Rev.~Lett. {\bf 54} (1985)
502.

\bibitem{FTDBI}
E.~Fradkin and A.~Tseytlin,
Phys.~Lett. {\bf B163} (1985) 123.

\bibitem{bosonizaFMS}
D.~Friedan, E.~Martinec  and  S.~Shenker, Nucl.~Phys. {\bf B271} (1986)
93-150.

\bibitem{SWss}
N.~Seiberg and E.~Witten,
Nucl.\ Phys.\ {\bf B276} (1986) 272.

\bibitem{ShiVa}
M.~Shifman and A.~Vainshtein, Nucl. Phys. {\bf B277} (1986) 456.

\bibitem{bosonizaVV}
E.~Verlinde and H.~Verlinde, Nucl.~Phys. {\bf B288} (1987) 357-379.

\bibitem{loopsst}
E.~Verlinde and H.~Verlinde, Phys. Lett. {\bf 192B} (1987) 95;\\
J.~Atick, G.~Moore and A.~Sen,
Nucl.~Phys. {\bf B307}
(1988) 221-273;\\
H.~La and P.~Nelson, Phys.~Rev.~Lett. {\bf 63} (1989) 24-27.

\bibitem{AtWi}
J.~Atick and E.~Witten,
Nucl. Phys. {\bf B310} (1988) 291-334.

\bibitem{FM2}
V.~Fainberg and A.~Marshakov,
Phys.~Lett. {\bf B211} (1988) 81-85.

\bibitem{KaMo}
R.~Kallosh and A.~Morozov, JETP {\bf 94} (1988) 42-56.

\bibitem{gauss}
R.~Dijkgraaf, E.~Verlinde and H.~Verlinde,
Comm. Math. Phys. {\bf 115} (1988) 649;\\
P.~Ginsparg, Nucl. Phys. {\bf B295\ [FS21]} (1988) 153.

\bibitem{NiRo}
H.~B.~Nielsen and D.~Rohrlich,
Nucl.\ Phys.\  {\bf B299} (1988) 471.

\bibitem{AFS}
A.~Alekseev, L.~Faddeev and S.~Shatashvili,
Journ. Geom \& Phys. {\bf 5} (1989) 391-406.

\bibitem{Mprop}
A.~Marshakov,
Nucl.~Phys. {\bf B312} (1989) 178-196.

\bibitem{bosonizaGMMOS}
A.~Gerasimov, A.~Marshakov, A.~Morozov, M.~Olshanetsky and S.~Shatashvili,
Int. J. Mod. Phys. {\bf A5} (1990) 2495-2589.

\bibitem{dsBK}
E.~Br\'ezin and V.~Kazakov, Phys. Lett. {\bf B236} (1990) 144.

\bibitem{dsDS}
M.~Douglas and S.~Shenker, Nucl. Phys. {\bf B335} (1990) 635.

\bibitem{dsGM}
D.~Gross and A.~Migdal, Phys. Rev. Lett. {\bf 64} (1990) 127.

\bibitem{WDVV}
E.~Witten, Nucl.\ Phys.\ {\bf B340} (1990) 281;\\
R.~Dijkgraaf, H.~Verlinde and E.~Verlinde,
Nucl.\ Phys.\ {\bf B352} (1991) 59.

\bibitem{vir}
M.~Fukuma, H.~Kawai and R.~Nakayama, Int. J. Mod. Phys., {\bf A6} (1991) 1385;\\
R.~Dijkgraaf, E.~Verlinde and H.~Verlinde, Nucl. Phys. {\bf B348} (1991) 435.

\bibitem{virmamo}
A.~Gerasimov, A.~Marshakov, A.~Mironov, A.~Morozov and A.~Orlov, Nucl. Phys.
{\bf B357} (1991) 565.

\bibitem{virGKM}
E.~Witten, in {\it New York 1991, Proceedings,
Differential geometric methods in theoretical physics, vol. 1} 176-216;\\
A.~Marshakov, A.~Mironov and A.~Morozov, Phys. Lett. {\bf B274} (1992) 280;\\
S.~Kharchev, A.~Marshakov, A.~Mironov, A.~Morozov and A.~Zabrodin,
Nucl. Phys. {\bf B380} (1992) 181, hep-th/ 9201013;
see also \cite{WDVV}.

\bibitem{LGGKM}
S.~Kharchev, A.~Marshakov, A.~Mironov and A.~Morozov,
Mod. Phys. Lett. {\bf A8} (1993) 1047, hep-th/9208046;
see also \cite{RGWhi}.

\bibitem{thooftholo}
G.~'t~Hooft,
gr-qc/9310026; see also \cite{BiSuTAHo}.

\bibitem{Seib94}
N.~Seiberg, hep-th/ 9408013.

\bibitem{sun}
A.~Klemm, W.~Lerche, S.~Theisen and S.~Yankielowicz,
Phys.~Lett. {\bf 344B} (1995) 169; hepth/9411048;\\
P.~Argyres and A.~Faraggi, Phys.~Rev.~Lett. {\bf 73}
(1995) 3931, hepth/9411057.

\bibitem{GKMMM}
A.~Gorsky, I.~Krichever, A.~Marshakov, A.~Mironov and A.~Morozov, Phys. Lett.
{\bf B355} (1995) 466; hep-th/ 9505035.

\bibitem{MtheoryT}
P.~Townsend, Phys. Lett. {\bf B350} (1995) 184, hep-th/ 9501068.

\bibitem{MtheoryWi}
E.~Witten, Nucl. Phys. {\bf B443} (1995) 85, hep-th/ 9503124.

\bibitem{MMM}
A.~Marshakov, A.~Mironov and A.~Morozov,
Phys. Lett. {\bf B389} (1996) 43, hep-th/9607109.

\bibitem{VafaC}
S. Kachru, A. Klemm, W. Lerche, P. Mayr and C. Vafa,
Nucl.~Phys. {\bf B459} (1996) 537-558, hep-th/9508155.

\bibitem{WittD}
E.~Witten, Nucl. Phys. {\bf B460} (1996) 335, hep-th/ 9510135.

\bibitem{DHW}
D.~-E.~Diaconescu, hep-th/ 9608163;\\
A.~Hanany and E.~Witten, hep-th/ 9611230.

\bibitem{SW3}
N.~Seiberg and E.~Witten, hep-th/ 9609219.

\bibitem{WittM97}
E.~Witten, hep-th/ 9703166.

\bibitem{Polholo}
A.~Polyakov,
hep-th/9711002;
hep-th/9809057;
hep-th/0006132;
hep-th/0110196.

\bibitem{Mmatrix}
T.~Banks, W.~Fischler, S.~Shenker and L.~Susskind, Phys. Rev. {\bf D55} (1997)
5112-5128;
see also
T.~Banks,
Nucl.\ Phys.\ Proc.\ Suppl.\  {\bf 67} (1998) 180, hep-th/9710231 and
references therein.


\bibitem{MMaM}
A.~Marshakov, M.~Martellini and A.~Morozov,
Phys. Lett. {\bf B418} (1998) 294, hep-th/ 9706050;\\
A. Marshakov, in
{\sl New Developments in Quantum Field The\-o\-ry}, Plenum Press 1999, NATO ASI
Series B: Physics {\bf 366} (Eds. P.~H.~Damgaard and J.~Jurkiewicz), 279;
hep-th/ 9709001.

\bibitem{LaxCo}
B.~Dubrovn, I.~Krichever and S.~Novikov, {\sl Integrable systems - I},
in VINITI, {\it Dynamical systems - IV} (1985) 179;
\\
N.~Hitchin,
Duke. Math. Journ. {\bf 54} (1987) 91;
\\
A.~Gorsky and N.~Nekrasov,
hep-th/9401021.

\bibitem{RGWhi}
A.~Gorsky, A.~Marshakov, A.~Mironov and A.~Morozov,
Nucl.~Phys. {\bf B527} (1998) 690-716; hep-th/9802007.

\bibitem{Malda}
J.~Maldacena, Adv.~Theor.~Math.~Phys. {\bf 2} (1998) 231, hep-th/9711200.

\bibitem{GKP}
S.~Gubser, I.~Klebanov and A.~Polyakov,
Phys.~Lett. B428 (1998) 105-114, hep-th/9802109.

\bibitem{NeSch}
N. Nekrasov and A. Schwarz,
Commun.~Math.~Phys. {\bf 198} (1998) 689-703, hep-th/9802068.

\bibitem{sentach}
A.~Sen,
hep-th/9805170;
hep-th/9911116;
hep-th/9902105.

\bibitem{ver}
J.~ de Boer, E. Verlinde and H. Verlinde,
JHEP {\bf 0008} (2000) 003, hep-th/9912012;\\
E. Verlinde and H. Verlinde,
JHEP {\bf 0005} (2000) 034, hep-th/9912018;\\
E.Verlinde,
Class.~Quant.~Grav. {\bf 17} (2000) 1277-1285, hep-th/9912058.

\bibitem{Polstra}
J.~Polchinski and M.~Strassler,
hep-th/0003136;\\
J.~Polchinski,
Int.~J.~Mod.~Phys. A16 (2001) 707-718, hep-th/0011193.

\bibitem{RS1}
V.~Rubakov and M.~Shaposhnikov, Phys.~Lett. {\bf B125} (1983) 136.

\bibitem{RS2}
L.~Randall and R.~Sundrum,
Phys.~Rev.~Lett. {\bf 83} (1999) 4690-4693, hep-th/9906064.

\bibitem{SWNC}
N.~Seiberg and E.~Witten,
JHEP {\bf 9909} (1999) 032,
hep-th/9908142.

\bibitem{GMS}
R.~Gopakumar, S.~Minwalla and A.~Strominger, JHEP {\bf 5} (2000) 20,
hep-th/0003150.

\bibitem{tsef75}
A.~Tseytlin, Theor.\& Math.~Phys., {\bf 128} (2001) 540-560.

\bibitem{gesha}
A.~Gerasimov and S.~Shatashvili,
hep-th/0009103.

\end{thebibliography}
\end{document}